\documentclass[letter,11pt]{article}%
\usepackage{amsmath,bm}
\usepackage{amsfonts}
\usepackage{amsthm}
\usepackage{amssymb}

\usepackage{comment}
\usepackage{lscape}
\usepackage{multibib}
\usepackage{float}
\usepackage{graphicx}
\usepackage{rotating}
\usepackage[]{subfigure}
\usepackage[height=9.5in,left=0.75in,right=0.75in,bottom=1in]{geometry}
\usepackage[colorlinks=true,citecolor=blue, linkcolor=black]{hyperref}
\usepackage{natbib}
\usepackage{color}
\usepackage{colortbl}
\usepackage{appendix}
\usepackage{paralist}
\usepackage{verbatim}
\usepackage{threeparttable}
\usepackage{multirow}
\usepackage{amsmath}%
\usepackage{makecell}

\providecommand{\U}[1]{\protect\rule{.1in}{.1in}}
\usepackage{titlecaps}
\Addlcwords{for a an and with in of to the for}

\RequirePackage{hypernat}

\RequirePackage{hypernat}

\theoremstyle{plain}

\newtheorem{theorem}{Theorem}

\newtheorem{assumption}{Assumption}

\newtheorem{corollary}[theorem]{Corollary}

\newtheorem{lemma}[theorem]{Lemma}

\newtheorem{proposition}[theorem]{Proposition}

\catcode`~=11
\newcommand{\urltilde}{\kern -.15em\lower .7ex\hbox{~}\kern .04em}
\catcode`~=13
\def \@seccntformat#1{\csname the#1\endcsname.\quad}
\makeatother
\numberwithin{equation}{section}
\hypersetup{pdftitle={JMPFA}, pdfsubject={}, pdfauthor={Facundo Argañaraz}, pdfkeywords={} }

\usepackage{array}
\newcolumntype{P}[1]{>{\centering\arraybackslash}p{#1}}

\begin{document}
	
	\title{
		Automatic Debiased Machine Learning of Structural Parameters with General Conditional Moments}
	\author{Facundo Arga\~{n}araz\thanks{Department of Economics. E-mail:
			\href{mailto:facundo.arganaraz@sciencespo.fr}{facundo.arganaraz@sciencespo.fr}. Web-page: \href{https://argafacu.github.io}{https://argafacu.github.io}. The author appreciates comments by Adam Lee, Christian Hansen, Clément de Chaisemartin, Jesús Carro, Juan Carlos Escanciano, Martin Weidner, and Stéphane Bonhomme.  The author is also thankful to participants of  multiple Econometrics seminars. All errors that remain are the author’s own.}\\\textit{Sciences Po}}
	\date{\today}
	
	\maketitle
	
	\begin{abstract}
		{This paper proposes a method to automatically construct or estimate Neyman-orthogonal moments in general models defined by a finite number of conditional moment restrictions (CMRs), with possibly different conditioning variables and endogenous regressors. CMRs are allowed to depend on non-parametric components, which might be flexibly modeled using Machine Learning tools, and non-linearly on finite-dimensional parameters. The key step in this construction is the estimation of Orthogonal Instrumental Variables (OR-IVs)—``residualized" functions of the conditioning variables, which are then combined to obtain a debiased moment. We argue that computing OR-IVs necessarily requires solving potentially complicated functional equations, which depend on unknown terms. However, by imposing an approximate sparsity condition, our method finds the solutions to those equations using a Lasso-type program and can then be implemented straightforwardly. Based on this, we introduce a GMM estimator of finite-dimensional parameters (structural parameters) in a two-step framework. We derive theoretical guarantees for our construction of OR-IVs and show $\sqrt{n}$-consistency and asymptotic normality for the estimator of the structural parameters. Our Monte Carlo experiments and an empirical application on estimating firm-level production functions highlight the importance of relying on inference methods like the one proposed.} \vspace{1mm}
		
		\begin{description}
			\item[Keywords:] Conditional Moment Restrictions; Debiased Inference; Machine Learning.
			
			\item[\emph{JEL classification:}] C14; C31; C36.\newpage
			
		\end{description}
	\end{abstract}
	
	\newgeometry{height=9in,left=1in,right=0.75in,bottom=1in}
	
	\section{Introduction}
	\label{intro}

	Models defined by conditional moment restrictions (CMRs) are ubiquitous in economics, appearing in a wide range of settings such as regressions, quantile models, dynamic discrete choice models, simultaneous equations models, demand estimation, auctions,  missing data, and production functions, to name a few \citep[see][]{chen2016methods}. These settings are often considered semiparametric as they involve parameters belonging to an infinite dimensional space, for example, the policy function that characterizes a firm's optimal investment decision given its unobserved productivity and inputs. In addition to these infinite-dimensional components, these models also include a finite-dimensional vector of structural parameters, such as the marginal products of capital and labor, which we are interested in. While our focus is on estimating these structural parameters, it is necessary to estimate the nuisance parameters as well, since they are unknown and play a role in the estimation of the parameters of interest.
	
	 Machine Learning tools, e.g., Boosting, Random Forest, Lasso, and Neural Networks, have proven to be suitable algorithms for handling such nonparametric estimations due to their strong prediction capabilities. Consequently, there is great potential to leverage these tools in a wide range of economic models. However, the use of machine learning typically introduces statistical challenges. The strength of these algorithms lies in their specific regularization strategies, which aim to achieve an optimal bias-variance trade-off---crucial for reliable estimation of unknown functions. This, nonetheless, typically leads to biased estimations of the parameters of interest and the impossibility of attaining $\sqrt{n}-$consistency. A practical implication of this issue is poor coverage of the commonly used confidence intervals and tests with sizes exceeding nominal levels.
	
	This paper introduces an algorithm that restores $\sqrt{n}-$consistency when machine learning is used to estimate nuisance parameters in a broad class of models defined by semiparametric CMRs. Thus, it enables the application of  modern estimation techniques in various structural economic settings. Specifically, this work presents a general estimation method for Locally Robust (LR)/(Neyman-)Orthogonal/ Debiased Moments \citep[][]{chernozhukov2022locally, neyman1959} in these contexts.\footnote{Throughout the paper, we will use these terms interchangeably.} These moments are  particularly appealing as estimation based on them is less affected by the regularization involved in learning high-dimensional parameters.   
	
	Our work contributes to the existing literature on debiased machine learning by developing a general algorithm for constructing debiased moments in a data-driven way within models characterized by a finite number of CMRs, with possibly different conditioning variables and endogenous regressors (in the sense that the nuisance parameters may be a function depending on variables other than those used for conditioning). To be specific, this paper develops an estimation technique for obtaining Orthogonal Instrumental Variables (OR-IVs) which are the key components to obtain LR moments in models with CMRs, as pointed out by \cite{chernozhukovetaldebiased2016}, \cite{chernozhukov2018double}, and, more recently, by \cite{arganaraz2025machine}. OR-IVs can be interpreted as functions of the conditioning variables, with finite second moments, where the effect of estimating the non-parametric component has been partialed out.

	\cite{chernozhukovetaldebiased2016}, Section 7, first proposed a definition of OR-IVs in models defined by a finite number of semiparametric CMRs, with different conditioning information sets, when the researcher is interested in a finite-dimensional vector of parameters appearing in the residual functions. They focused on a specific OR-IV that minimizes the asymptotic variance of GMM estimators in the class of GMM estimators based on orthogonal instruments. The authors did not discuss how to estimate such optimal OR-IV or other OR-IVs in general.  \cite{chernozhukov2018double} considered the special case with only one conditioning information set under exogeneity (i.e., when the nonparametric component is a function of a sub-vector of the conditioning variable). \cite{arganaraz2025machine} characterized OR-IVs in the same model of \cite{chernozhukovetaldebiased2016}, but for the case where the parameter of interest is a general smooth functional of finite-dimensional and nonparametric components, which might not necessarily appear in the residual functions of the model. \cite{chernozhukov2018double} and \cite{arganaraz2025machine} worked out estimation of particular OR-IVs in specific applications such as the partially linear model with and without an endogenous treatment.  Differently to these works, the goal of the current paper is to introduce a strategy to estimate \textit{any} OR-IV, in a computationally convenient way, in a \textit{general} model with CMRs. For simplicity, we construct LR moments for a finite-dimensional vector of parameters appearing in the residual functions; nevertheless, the general principles developed in this paper can be extended to more general functionals. 
	
	Building on this strategy, we introduce a Debiased GMM estimator (DGMM) that is locally insensitive to regularization techniques. In our context, the parameter of interest must have a finite efficiency bound, but it can map into the CMRs non-linearly, which is a common situation in structural models in economics.\footnote{Recall that having a finite efficiency bound is equivalent to the (efficient) Fisher information matrix being positive definite, which is a necessary condition for $\sqrt{n}-$ estimability of the parameters of interests in general; see \cite{van1991differentiable}.} For instance, the object of interest could be interpreted as the parameters of a production function on which the CMRs depend in a non-linear way, as in the semiparametric production function setting considered by \cite{ackerberg2014asymptotic}.  Furthermore, we obtain a rate of consistency of our OR-IV estimators, which is essential to show $\sqrt{n}-$consistency and asymptotic normality of DGMM, as we do in this paper. 
	
	Our approach to obtaining debiased moments leverages a characterization of such moments in our general setting for general smooth functionals,  previously derived by \cite{arganaraz2025machine}. They prove that an orthogonal moment is \textit{necessarily} a sum of the products between the residual functions and the OR-IVs. Since the residual functions are primitives of the economic model, known up to the unknown parameters, the key step in computing an LR moment is the calculation of OR-IVs. This will allow us to construct a moment that is locally insensitive to estimation of the infinite-dimensional parameters, including the OR-IVs, and then proceed with a suitable estimation of the parameters of interest. Unfortunately, OR-IVs are necessarily solutions to functional equations, which might potentially be challenging to solve analytically. 

    The key idea is based on constructing a linear operator derived from the CMRs.\footnote{See \cite{luenberger1997optimization} and \cite{CARRASCO20075633} for key results on linear operators.} The range of this operator collects all the possible Fréchet derivatives of the initial CMRs with respect to the high-dimensional parameters. OR-IVs must be orthogonal to the range of this operator, under a suitable notion of orthogonality. Our work builds on the observation that this orthogonalization can be accomplished by ``residualizing" a known function of the conditional variables, i.e., an IV, from the orthogonal projection onto the range of the operator. This exercise requires obtaining this orthogonal projection, which is challenging as the operator itself is unknown, and thus its orthogonal projection is unknown.  
	
We show that computing such an orthogonal projection can be seen as minimizing a mean squared error, resulting from projecting a vector of known starting IVs onto the linear span of suitable regressors. The part of the starting functions that is not explained by the regressors, i.e., the residual, is shown to be a valid OR-IV.  We allow the dimension of the regressors involved in that minimization to be greater than the sample size (and depend on it) by adding an $\ell_1$- norm penalization term. The resulting objective function is then familiar from standard Lasso problems. Nonetheless,  this Lasso program is special  in that it is based on unknown regressors. This is a by-product of not knowing the linear operator that we mentioned in the previous paragraph. As such, these regressors must be estimated prior to minimizing the corresponding Lasso objective function.

 Conveniently, we show that those regressors take the form of conditional expectations (also known as reduced forms), given the conditioning variables of the model. Consequently, they can be estimated by suitable machine learning tools, useful for predicting regression functions. Once these regressors are estimated, the problem reduces to a Lasso problem with generated regressors, which can be implemented straightforwardly by well-known algorithms.  We provide theoretical guarantees for our estimation procedure, assuming that some sparsity condition holds, which implies that the orthogonal projection can be well-approximated by a ``small" number of terms.

 This paper argues that for inference in high-dimensional settings, researchers should use OR-IVs only, as moments based on them are debiased. Hence, our algorithm justifies the use of OR-IVs over other choices of instruments that do not necessarily yield a debiased moment and that are currently employed in applied work. Our approach converts any suitable function of the conditioning variables (e.g., $Z$ or $Z^2$) into an OR-IV that can be used in estimation as all these functions can be ``residualized" using the approach that we described above. If multiple such functions are provided, multiple orthogonal moments can be estimated and combined as usual in GMM.  A natural approach is to start with the standard choices in applied work  (e.g., $Z$ or $Z^2$) and apply our method to transform them into OR-IVs. 
	
We propose estimating the nuisance parameters with machine learning and cross-fitting \citep[as in][]{chernozhukov2018double}, and then estimate the parameters of interest with GMM . We develop this estimator in a two-step framework, where the nuisance parameters are estimated in a first step, and the estimation of the finite-dimensional parameter is conducted in a second step, using the first-step estimations. The rationale for employing GMM stems from our theoretical findings, which assure that standard inference remains valid for the finite-dimensional parameter, despite using machine learning tools to estimate the nuisance parameters in a first stage. Specifically, leveraging the asymptotic properties established for the OR-IV estimators, we demonstrate that the estimator we introduce is $\sqrt{n}-$consistent and asymptotically normal. For this, we only require that the nuisance parameters being estimated at a rate faster than $n^{-1/4}$, which can be achieved by a variety of machine learners.  When orthogonal moments are not used, standard errors must be correctly computed to account for first-stage estimation \citep[][]{newey1994asymptotic}. Nonetheless, we show that standard errors based on our debiased moments directly account for such a first-stage estimation by using the usual ``sandwich" formula, simplifying their calculation and avoiding using bootstrap methods. 

Additionally, we apply our general theory to a semiparametric version of production functions estimation using the proxy variable approach, studied by \cite{olleypakes96}, \cite{levinsohn2003estimating}, and \cite{ackerberg2015identification}, which might be of independent interest.\footnote{Production functions have been at the core of economics since the early 1800's; see the introduction in \cite{chambers1988applied} for an historical review on production function and productivity.} Production functions and measures of productivity play a central role in several empirical settings in economics. For instance, they have been used to study the effects of trade liberalization, exporting, foreign ownership, competition, importing intermediate goods, investment climate, learning by doing; see \cite{ackerberg2007econometric}, \cite{ackerberg2015identification}, and references therein.  We apply debiased machine learning in this context using off-the-shelf machine learning tools to conduct inference on the parameters of all the inputs simultaneously.  \cite{cha2023inference} explore the use of the Double Debiased Machine Learning (DML) estimator of \cite{chernozhukov2018double} in the context of production functions. Since the authors based estimation on the orthogonal score function from \cite{robinson1988root}, they conduct inference on only one parameter at a time and treat the parameters of the rest of inputs, including capital, as nuisance ones. As \cite{ackerberg2015identification} pointed out, conducted inference using the Robinson's transformation might not be ideal as it could lead to identification failures from functional form dependence. Moreover, \cite{cha2023inference} focus on a particular strategy, introduced by the authors,  to estimate the nonparametric components. Independently, \cite{doraszelski2025productionfunctionestimationinvertibility} have derived a Neyman-orthogonal moment for the proxy variable approach by explicitly obtaining the first-step influence function (accounting for regularization bias in the first stage) and imposing restrictions among the conditioning variables, which our more general approach does not require.

	We have assessed the finite sample performance of the introduced estimator through several Monte Carlo experiments, in the context of production functions. Despite this setting is not ``of high dimension" (as there are only a few inputs), it entails an unknown nonparametric object which is potentially highly nonlinear. It arises from an unknown policy function---a solution to a potentially complex firm's dynamic problem---that links productivity (observed by the firm but unobserved to the econometrician) and a proxy variable, e.g., investment or intermediate inputs, given the other inputs in the production function. We aim to estimate the marginal products of these inputs in the presence of the unknown link function. Estimation is performed using machine learning, and we employ our OR-IV construction to correct for potential regularization bias so that standard inference remains valid. Our results show that DGMM exhibit smaller bias and coverage closer to the nominal level compared to a naive estimator based on the same first stage but not addressing regularization bias, consistent with our theoretical findings.
	
	In our empirical application, we used data from a panel of Chilean firms, which has been extensively studied in the production function literature. We aimed to assess whether using machine learning could lead to different conclusions compared to other estimators that model first stages less flexibly and are commonly used in applied work. The results suggest that the choice of approach for modeling the first stage should be made with caution, as it can significantly influence empirically relevant conclusions, particularly regarding production function parameters. Consequently, inference methods like the one introduced in this paper are potentially appealing because they allow researchers to adopt less parametric specifications in the first stage. This is important for two reasons. First, it provides an additional empirical tool for scholars who wish to avoid imposing restrictive parametric assumptions. Second, it helps those more confident in traditional approaches to evaluate the sensitivity of their results to less flexible specifications.

	The rest of the paper is organized as follows. Section \ref{literature} revises the most related works to ours, emphasizing what our contribution is. Section \ref{motivatingexamples} works out a production functions example to motivate our results and describe the problem we face. Section \ref{orthogonality} defines LR moments in our context and discusses our approach heuristically. Section \ref{estimationoivs}  is the core of the paper as it introduces our approach to estimate OR-IVs in a general setting. Section \ref{sthealgorithm}  proposes an algorithm that can be followed in practice. Section \ref{asymptoticskappa} obtains a convergence rate for the estimators of the OR-IVs. In Section \ref{estimationparint}, we present the estimator of the finite-dimensional parameter in a two-step setting, and Section \ref{apparameterinterest} provides conditions under which it is $\sqrt{n}-$consistent and asymptotically normal distributed. Section \ref{montecarlo}  studies the performance of our estimator through different Monte Carlo experiments. Section \ref{empiricalapp} contains our empirical results. Finally,  Section \ref{conclusion} provides final remarks. A Supplementary Appendix collects the proofs of all the theoretical results, states additional formal findings, applies our general asymptotic theory to our example, elaborates on implementation details of the Lasso-type program we introduce and its justification, and provides additional details on the Monte Carlo experiments.

	\section{Related Literature}
	\label{literature}
	
	Our work relates to various strands of the literature. First, it contributes to the body of research on CMRs. Our general model encompasses several settings that have been extensively studied in this literature. A seminal contribution in this regard is made by \cite{ai2012semiparametric}, who establish the semiparametric bound (and propose an efficient estimator) for structural parameters in semiparametric models defined by nested CMRs, and thus involving different conditioning variables. Their work extends previous results by \cite{chamberlain1987asymptotic} for the purely parametric case with one conditioning variable, as well as by \cite{chamberlain1992efficiency} and \cite{ai2003efficient}, who study semiparametric models with a single conditioning variable. It also relates to the findings in \cite{chamberlain1992cond} and \cite{brown1998efficient}, who consider parametric cases with nested CMRs; see also \cite{ai2007estimation}. Furthermore, our model closely relates to the nonparametric setting with general CMRs studied by \cite{chen2018overidentification}.
	
	As discussed by \cite{chen2016methods}, nuisance parameters in general CMRs settings have typically been estimated using sieve-based methods \citep[][]{chen2007large}, with finite-dimensional linear sieve approximations being a popular choice. Estimation of the parameters of interest can be performed simultaneously or in a two-step fashion. While the sieve approach is computationally attractive for both estimation and inference (for example, they already lead to debiased estimators when first steps are efficiently estimated) it can be restrictive. For instance, sieve methods often assume knowledge of the important approximating terms that should be included in estimation, an assumption that might be strong in complex, high-dimensional settings. In contrast, Lasso over a dictionary does not require such an assumption, as it can learn the relevant terms from the data, as discussed by \cite{bradic2022minimax}.\footnote{A prominent exception is neural networks, which can be seen as a non-linear sieve approach. When neural networks are used to model the first-stage, LR moments and cross-fitting might not be needed, and inference can be alternatively conducted following, e.g., \cite{ai2007estimation}, \cite{ai2012semiparametric}, \cite{chen2023efficient}, and \cite{chen2024inference}.}
	
	We build upon this literature by allowing for the utilization of a wide array of machine learning tools to estimate nuisance parameters in a first stage, and by automatically constructing LR moments. This provides researchers with additional tools to undertake their analyses using general semiparametric CMRs, invoking more general conditions on the space of functions to be approximated. In this paper, we follow a two-step approach for estimating the parameters of interest. A two-step framework is often more computationally attractive than a one-step estimation \citep[see][]{ackerberg2014asymptotic}.

	Second, this paper is connected to the recently developed literature on debiased moments. The general construction and asymptotic theory of these moments have been established by \citet*{chernozhukov2022locally} (CEINR) for high-dimensional settings where machine learners are employed in a first stage. Typically, a debiased moment function consists of two components: the initial identifying moment function and an adjustment term, known as the first-step influence function. This function characterizes the local effect of the first step on the starting (non-debiased) moment and its existence is generally equivalent to the starting moment having a finite semiparametric variance bound. When the functional form of this adjustment term is known, CEINR provide a general method to estimate the components in such functional form.\footnote{Typically, the first-step influence function depends on the data, the finite-dimensional and nonparametric components, and an extra nuisance parameter. CEINR propose a general method to estimate this nuisance object.} In this paper, we specifically examine semiparametric models defined by CMRs that involve varying conditioning variables and we do not need to obtain the functional form of the adjustment term, as it will be entirely estimated from the data, as we explain below. Focusing on CMRs with different conditioning variables allows us to extend the automatic construction of debiased models to new interesting applications such as learning firm-level production functions and productivity measures.\footnote{As CMRs imply unconditional moments, an alternative approach to construct LR moments in our setting, is to first convert conditional restrictions into unconditional ones, and then apply CEINR's approach. This, however, will entail deriving the first-step influence function, which might not be trivial in several structural models in economics. Our method, in contrast, avoids this.}
	
	As we stated in the Introduction, OR-IVs have been introduced by the first version of CEINR, \cite{chernozhukovetaldebiased2016}, within models with semiparametric CMRs and varying conditioning variables, which leads to debiased moments for a finite-dimensional vector of parameters appearing in the residual functions. \cite{chernozhukov2018double} consider the special case where there is a common conditioning information set and the nonparametric component is a function of a subset of the IVs of the model. More generally, \cite{arganaraz2025machine} characterize OR-IVs for general smooth functionals of finite-dimensional and nonparametric unknowns, where the functionals might not index the CMRs. \cite{chernozhukov2018double} and \cite{arganaraz2025machine}  consider estimation of OR-IVs in specific contexts such as the partially linear model with and without an exogenous treatment, respectively.  We complement this line of research by developing an algorithm to estimate any OR-IVs in general contexts with CMRs and providing its asymptotic guarantees. This procedure allows for valid inference on the finite-dimensional components in the residual functions, where there are varying conditioning variables, and  potential endogeneity. While we focus on these parameters, our theory can be extended to more general functions, similarly to the algorithms developed by \cite{arganaraz2025existence} for models with unobserved heterogeneity. However, this is beyond the scope of the current paper and will be explore in future research.     
	
	Third, our paper contributes to the literature on the automatic construction of debiased moments, which have typically developed specific algorithms based on that in CEINR. In many applications, the first-step influence function can be expressed as the product of a residual function and the Riesz representer of the initial moment function that identifies the parameter of interest \citep[][]{ichimura2022influence}. \citet{chernozhukov2022automatic} propose a Lasso-type minimum distance estimator for such Riesz representer, allowing the nuisance parameter to be estimated using machine learning tools; see also \citet{chernozhukov2020adversarial}, \citet{chernozhukov2021automatic}, \citet{chernozhukov2022debiaseda}, and \citet{chernozhukov2022debiasedb} for related work. While these papers focus on exogenous cases, where the nuisance parameter depends on conditioning variables, \citet{bakhitov2022automatic} extends this framework to endogenous settings, proposing a penalized GMM approach for estimating the Riesz representer. Additionally, \citet{farrell2021deep} offer an automatic method for constructing adjustment terms that produces debiased moments, even when the first-stage parameter is not simply a prediction but any well-defined parameter in an economic model. In these references, the parameter of interest is the average of some smooth function of the first-stage parameter.\footnote{While \citet{farrell2021deep} focus on parameters that are averages, their first-step correction  can potentially be extended to situations where the object of interest solves a general moment condition, see their Remark 3.1.}

	Our paper complements these works by working with CMRs and not focusing on situations where the parameter of interest represents an expectation, but on parameters of economic models ``with more structure". Specifically, we address the more general case where the model implies conditional moments of arbitrary (but smooth) functions that depend on both finite and infinite-dimensional parameters. Thus, our paper paves the way for applying the automatic construction of debiased moments developed in these previous works to nonlinear and highly complex models with CMRs. This might be especially relevant in semiparametric models prevalent in structural economics such as firm-level production functions \citep[][]{olleypakes96, ackerberg2014asymptotic, levinsohn2003estimating}, general missing data problems \citep[][]{hristache2016semiparametric, hristache2017conditional, graham2011efficiency}, auctions \citep[][]{kawai2011auction}, and simultaneous equations models \citep[][]{newey1999nonparametric}. Further examples can be found in Section 5 of \cite{chen2016methods}. In the production function setting, for instance, the parameter of interest are marginal productivities which cannot be written as simple means.

	Lastly, our paper develops a robust method for estimation of production functions and productivity at the firm level. Estimation and inference in this context has been studied by extremely popular papers in applied work, including \cite{olleypakes96}, \cite{levinsohn2003estimating}, \cite{wooldridge2009estimating}, \cite{ackerberg2015identification}, and \cite{gandhi2020identification}. We contribute to this literature by allowing for the use of off-the-shelf nonparametric estimation routines to handle first-steps in a semiparametric version of this model, while maintaining the ability to perform standard inference on \textit{all} marginal productivities and related parameters. \cite{cha2023inference} have recently explored the use of debiased machine learning in this setting. We departure from this paper in two important ways. First, we do not exploit the score in \cite{robinson1988root} and this allows us to estimate all the parameters of the production function simultaneously without treating the capital marginal product as a nuisance component. In contrast, \cite{cha2023inference} can only run inference for the labor coefficient with their strategy. More importantly, the Robinson's score might lead to identification failures in this context, as discussed by \cite{ackerberg2015identification}. Second, we simplify the implementation of machine learning procedures to estimate production functions by invoking several well-known routines already implemented in R or Python, subject to some rate conditions. Differently,  \cite{cha2023inference} estimate the first step using their specific algorithm, but they are able to relax exact sparsity conditions.  \cite{doraszelski2025productionfunctionestimationinvertibility} have independently obtained a Neyman-orthogonal moment in this context by deriving the first-step influence function and imposing restrictions among the conditioning variables. Our approach is more general as we rely on automatic Neyman-orthogonalization, without restricting the conditioning information sets of the model.

	\bigskip \noindent \textbf{Notation:}  For an arbitrary vector $x \in \mathbb{R}^r$, let $\left|\left|x\right|\right|_1$ and $\left|\left|x\right|\right|$ be the $\ell_1$ and $\ell_2$ norm, respectively. If $A$ is a matrix, $\left|\left|A\right|\right|$ denotes spectral norm. For a measurable function $a$ of an arbitrary random variable $H$, let $\left|\left|a\right|\right|_2 = \sqrt{\mathbb{E}\left[a(H)^2\right]}$. Let $P$ be a probability distribution. Let $L^2$ be the space of functions of $W$ that are square-integrable, when $ W \sim P$, where the precise meaning of $W$ will be established below.\footnote{Technically, we should index $L^2$ by some $\sigma-$finite measure. We avoid this to simplify our notation.} In addition, let $L^2_0$ be a subset of $L^2$ with the additional mean-zero restriction. Similar definitions apply to objects such as $L^2(V)$ and $L^2_0(V)$ for functions of $V$. For a $n \times r$ matrix $A = \left[a_{ik}\right]$, let $\left|\left|A\right|\right|_{\infty} = \underset{i,k}{\operatorname{max}} \left|a_{ik}\right|$, and $\left|\left|A\right|\right|_{\ell_{\infty}} = \underset{i}{\operatorname{max}} \sum^r_{k=1}\left|a_{ik}\right|$. For a set of indexes $S \subseteq \left\{1,\cdots,r\right\}$, let $x_S$ be the modification of $x$ that places zeros in all entries of $x$ whose indexes do not belong to $S$. Moreover, let $S_x$ be a subset of $\left\{1,\cdots,r\right\}$ such that $x_k \neq 0$ for all $k \in S_x$, and $S^c_{x}$ be the complement of $S_x$ in $\left\{1,\cdots,r\right\}$. For a bounded set $Q$, $\# Q$ denotes the cardinality of $Q$. For $a, b \in \mathbb{R}$, $a \vee b = \max\left\{a,b\right\}$. For any arbitrary subset $\mathcal{K}$, let $\overline{\mathcal{K}}$ denote the closure of $\mathcal{K}$ and $\overline{\mathcal{K}}^{\perp}$ be its orthocomplement, when a topology and inner product $\left<\cdot,\cdot\right>$ is defined. Moreover, $\Pi_{\overline{\mathcal{K}}}$ is the orthogonal projection operator onto $\overline{\mathcal{K}}$. Let $J < \infty$ and  $Z = \left(Z_1, Z_2, \cdots,Z_J\right)^{'}$, we say that a vector-valued function $f(Z) = \left(f_1(Z_1), \cdots, f_J(Z_J)\right)^{'}$ is in $L^2(Z) \equiv \bigotimes^J_{j=1} L^2(Z_j)$ when each of the elements in a such vector belongs to the corresponding $L^2\left(Z_j\right)$, $j=1,\cdots,J$, and we define $\left|\left|f\right|\right|_{L^2(Z)} := \sqrt{\sum^J_{j=1}\left|\left|f_j \right|\right|^2_{2}}$. Finally, all the CMRs in the sequel are satisfied almost surely (a.s.), but we will not make it explicit often to simplify the exposition.
	
	\section{Production Functions at the Firm Level: A Motivating Example}
	\label{motivatingexamples}
	To illustrate the key ideas of this paper we will consider an interesting setting throughout. We emphasize that this is purely for the sake of clarity in exposition. Our theoretical framework is broad and applicable to models defined by general semiparametric CMRs, as we will show in Section \ref{estimationoivs}.
	
	\bigskip \noindent \textsc{Example: Production Functions at the Firm Level.} We observe a panel of $n$ firms across $T$ periods, where $i$ and $t$ index firms and periods, respectively. Let $Y_{it}$ be the output of firm $i$ at time $t$, and $X_{it}$ be a vector of inputs, e.g., capital and labor.  Output is determined by the following equation:\footnote{Unless otherwise stated, all the variables are expressed in logarithms.}
	\begin{equation}
		\label{production}
		Y_{it} = F\left(X_{it},\theta_{0p}\right) +  \omega_{it} + \epsilon_{it},
	\end{equation}
	where $F$ is assumed to be known up to $\theta_{0p}$, where $\theta_{0p} \in \mathbb{R}^{p_1}$, $\omega_{it}$ is firm $i$'s productivity shock (known as anticipated productivity) in period $t$, which is allowed to be correlated with inputs, and $\epsilon_{it}$ is noise in output, which is independent and identically distributed (iid), is assumed to be independent of the current and previous optimal decisions of the firm and anticipated productivities, and with zero mean.  Since $\omega_{it}$ is not observed and is correlated with inputs, OLS will provide inconsistent estimates.\footnote{In the case where $F$ is assumed to be linear in inputs.} To address this, we follow the so-called proxy variable approach, started by \cite{olleypakes96}; see also \cite{levinsohn2003estimating} and \cite{wooldridge2009estimating}. We assume that there exists some firm's choice, $I_{it}$, at $t$ that is linked to $\omega_{it}$:  
	$$
	I_{it} = I_t\left(\omega_{it}, X_{it}\right).
	$$
	The precise meaning of the variable $I_{it}$, a ``proxy", differs across different formulations of the model.\footnote{For instance, \cite{olleypakes96} consider in $I_{it}$ the firm's current investment towards future physical capital. In \cite{levinsohn2003estimating}, $I_{it}$ is the firm's choice of an intermediate input, e.g., electricity or material input.} In addition, let us assume that $I_t$ is strictly monotonic in $\omega_{it}$. No parametric assumptions are imposed on $I_t$. Then, we shall write 
	$$
	\omega_{it} = \omega_t\left(I_{it}, X_{it}\right), 
	$$
	where $\omega_t$ is also non-parametric. Hence, we were able to express the unobservable productivity in terms of observable inputs. It is immediate that Equation \eqref{production} becomes 
	\begin{equation}
		\label{production2}
		Y_{it} = F\left(X_{it},\theta_{0p}\right) +  \omega_t\left(I_{it}, X_{it}\right) + \epsilon_{it}.
	\end{equation}
	Let 
	$$
	\eta_{0t}\left(I_{it}, X_{it}\right) = F\left(X_{it},\theta_{0p}\right) +  \omega_t\left(I_{it}, X_{it}\right).
	$$
	Then, by the independence and zero mean assumptions, 
	\begin{equation}
		\label{firstmoment}
		\mathbb{E}\left[\left.Y_{it} - \eta_{0t}\left(I_{it}, X_{it}\right)\right|I_{it}, X_{it}\right] = 0.
	\end{equation}
	Note that \eqref{firstmoment} does not necessarily exhaust all the information in the model, e.g., we can also condition on $I_{i-t-1}$ and $X_{i,t-1}$. We write this CMR in this way as it is what researchers typically do in applied work, presumably for simplicity; see \cite{ackerberg2014asymptotic}, Section 4.2.  While Equation \eqref{firstmoment} identifies $\eta_{0t}$, it is not enough to identify all the parameters of the production function. This is true since $X_{it}$ enters parametrically and non-parametrically in \eqref{production2}. The last element of the model is the evolution of $\omega_{it}$. Often, this is also treated non-parametrically, as in \cite{olleypakes96} and \cite{levinsohn2003estimating}. We, nonetheless, follow \cite{ackerberg2014asymptotic} and work with a more ``natural" semiparametric model. Let us assume that $\omega_{it}$ follows a First-Order Markov process in the sense that 
	\begin{equation}
		\label{fom}
		\mathbb{E}\left[\left.\omega_{it}\right|X_{i,t-1}, I_{i,t-1}\cdots,\omega_{i,t-1}, \omega_{i,t-2}, \dots,  \omega_{i,0}\right] = \mathbb{E}\left[\left.\omega_{it}\right|\omega_{i,t-1}\right]. 
	\end{equation}
	Equation \eqref{fom} is indeed assumed by \cite{olleypakes96} and \cite{levinsohn2003estimating}. What \cite{ackerberg2014asymptotic} suggests is to parameterize \eqref{fom}. To keep things simple, let us consider 
	\begin{equation}
		\label{fomp}
		\mathbb{E}\left[\left.\omega_{it}\right|\omega_{i,t-1}\right] = \theta_{0\omega} \omega_{i,t-1},
	\end{equation}
	where $-1 < \theta_{0\omega} < 1$. Then, letting $\Omega_{it}$ be the firm information set at $t$, with $\left(X_{it}, I_{it}\right)$ being a (measurable) function of $\Omega_{it}$, and using the independence assumption, equations \eqref{production2}, \eqref{fom}, and \eqref{fomp}, it is not difficult to show that 
	\begin{equation}
		\label{secondmoment}
		\mathbb{E}\left[\left.Y_{it} - F\left(X_{it},\theta_{0p}\right) - \theta_{0\omega}\left(\eta_{0,t-1}\left(Z_{i,t-1}\right) - F\left(X_{i,t-1},\theta_{0p}\right)\right)\right|\Omega_{i,t-1}\right] = 0.
	\end{equation}
	Suppose that $T=3$, i.e., firms are observed during three periods. Ignoring subscript $i$, and denoting $\eta_{0} = \left(\eta_{01}, \eta_{02}\right)^{\prime}$, the model can be defined by the following CMRs: 
	\begin{align}
		\mathbb{E}\left[\left.Y_{1} - \eta_{01}\left(I_{1}, X_{1}\right)\right|I_{1}, X_{1}\right] & = 0, \label{prod1} \\ \mathbb{E}\left[\left.Y_{2} - F\left(X_{2},\theta_{0p}\right) - \theta_{0\omega}\left(\eta_{01}\left(I_1,X_1\right) - F\left(X_1,\theta_{0p}\right)\right)\right|\Omega_{1}\right] & = 0, \label{prod2} \\  \mathbb{E}\left[\left.Y_{2} - \eta_{02}\left(I_{2}, X_{2}\right)\right|I_{2}, X_{2}\right] & = 0, \label{prod3} \\ \mathbb{E}\left[\left.Y_{3} - F\left(X_{3},\theta_{0p}\right) - \theta_{0\omega}\left(\eta_{02}\left(I_2,X_2\right) - F\left(X_2,\theta_{0p}\right)\right)\right|\Omega_{2}\right] & = 0.\label{prod4}
	\end{align}
 Our goal is to learn $\theta_0 = \left(\theta_{0p}^{'},\theta_{0\omega}\right)^{'}$, the parameter of interest, in the presence of an unknown $\eta_0$. To this end, we might exploit \eqref{prod1}-\eqref{prod3} to estimate $\eta_{0}$, and then plug the resulting estimator into moments based on \eqref{prod2}-\eqref{prod4} for estimation of $\theta_0$.  $\square$

	\bigskip  In the example above, $\eta_0$ is unknown and ultimately needs to be estimated to estimate $\theta_0$.  
 The traditional literature on production functions estimation estimates the nonparametric component of the model using sieves methods \cite[][]{chen2007large}, especially linear sieves, as done in \cite{olleypakes96}, \cite{levinsohn2003estimating}, and \cite{ackerberg2015identification}.\footnote{In \cite{olleypakes96} and \cite{levinsohn2003estimating}, the vector of inputs $X_t$ is such that $X_t = \left(X^{\prime}_{1t}, X^{\prime}_{2t}\right)^{\prime}$, and the link function depends on $I_t$ and the subvector $X_{1t}$, so that  $\eta_{0t}$ depends on $I_{it}$ and $X_{1it}$ only.} We instead seek to model $\eta_0$ using machine learning tools, which allow us to exploit a wide range of algorithms to estimate $\eta_0$, different from that of the sieves approach. Those tools (e.g., Lasso, random forest, neural networks, and boosting) have proven useful for approximating functions (where the leading case involves conditional expectations) without imposing stringent parametric assumptions. While the sieves approach provides a convenient theoretical and empirical tool, it might be less flexible in modeling unknown nuisance components than machine learning tools. For example, the sieves approach requires knowledge of all the terms necessary to approximate an unknown function, a requirement that might not always be plausible. In contrast, tools like Lasso aim to identify those terms from the data itself---similar notions apply to other machine learning techniques. Therefore, we aim to equip empirical researchers with an additional inference procedure, based on a less parametric modeling for first stages.  This is important for two reasons. First, it provides a valuable tool for scholars who prefer to avoid imposing restrictive assumptions. Second, it offers a way to conduct robustness checks for those more confident in traditional (somewhat less flexible) approaches.

\begin{figure}[H]%
		\caption{Comparison of Naive and DGMM Estimators}
	\label{fig:histbias}
		\begin{threeparttable}
			\centering
			\subfigure[Naive]{\includegraphics[width=0.4\textwidth]{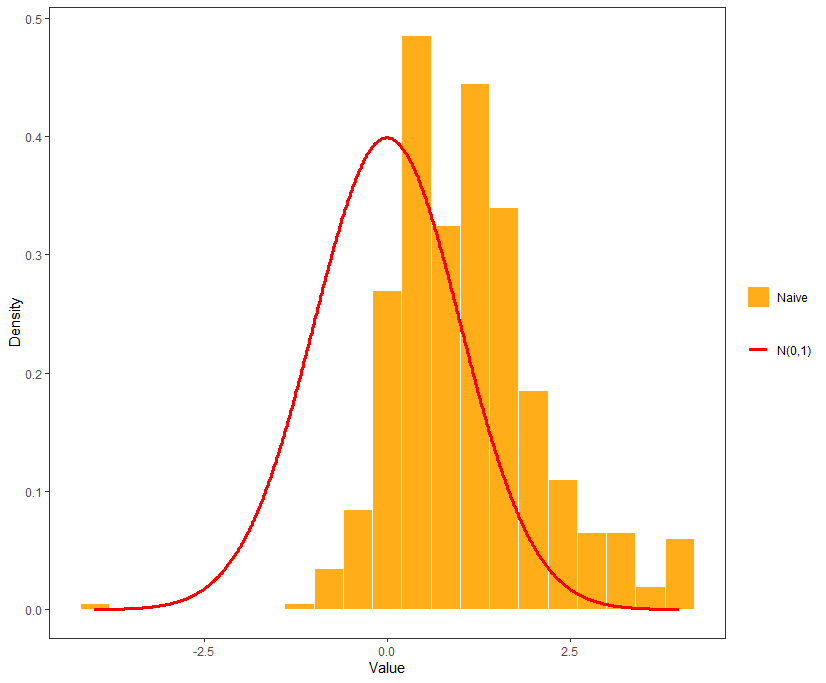}}\qquad
			\subfigure[DGMM]{\includegraphics[width=0.4\textwidth]{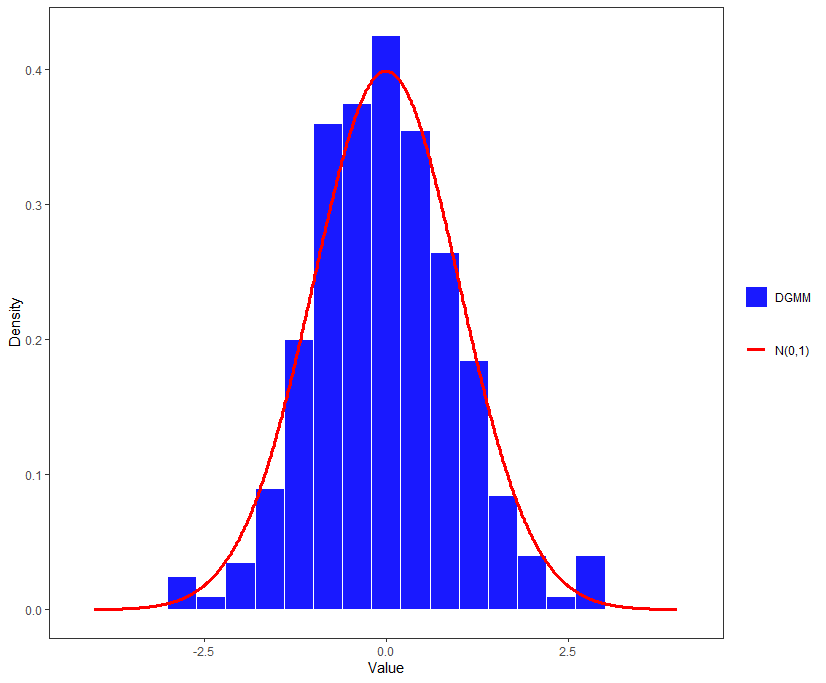}}
			\begin{tablenotes}
				\scriptsize
				\item NOTE: The figure shows the histogram of the standardized estimator $\left(\hat{\theta} - \theta_0\right)$ for one of the parameters of the production function in our example, using the naive plug-in approach (Naive) and the corrected one based on debiased moments (DGMM), based on the construction we propose. The density of the standard normal distribution is also displayed in red. The sample size is $n = 1,000$. Results are based on $500$ Monte Carlo repetitions.
			\end{tablenotes} 
		\end{threeparttable}
	\end{figure}

	Nevertheless, by employing those algorithms, it would be difficult to estimate $\theta_0$ without bias. This is a by-product of the regularization techniques that all those methods impose in estimation. More importantly, this bias typically decays at  a rate slower than $\sqrt{n}$ \citep[][]{chernozhukov2022locally}. Hence, a naive estimator that directly plugs the estimated $\hat{\eta}$ in estimating $\theta_0$, could lead to a biased estimator, say $\hat{\theta}_{Naive}$. Then it would not hold that $\sqrt{n}\left(\hat{\theta}_{Naive}- \theta_0\right )$ is asymptotically normally distributed. Indeed, $\sqrt{n}\left(\hat{\theta}_{Naive}- \theta_0\right )$ would not be $O_p(1)$, invalidating standard inference on $\hat{\theta}_{Naive}$.  To illustrate the point, Figure \ref{fig:histbias} displays (in orange) the empirical distribution of the standardized $\left(\hat{\theta}_{Naive} - \theta_0\right)$ for one of the parameters of the production function in our example obtained from simulated data on 1,000 firms and 500 Monte Carlo repetitions. This is computed by estimating $\eta_0$ using boosting in the first stage and then estimating $\theta_0$ using GMM.\footnote{For this illustration, we have used the R function \texttt{gbmt} and its default options.} The figure illustrates that this estimator suffers from bias, as its mean is not centered around zero. Furthermore, the shape of the distribution is certainly different from the standard normal distribution (depicted by the red curve), which would be similar to the expected asymptotic distribution if the bias were absent. The situation would likely become worse as the sample size increases.

	Our goal is to construct moments that, under regularity conditions, can be used to estimate $\theta_0$, and that will lead to an estimator that remains $\sqrt{n}-$consistent, despite $\eta_{0}$ being estimated flexibly. Such moments, known as LR/orthogonal/debiased moments, play a crucial role in achieving this goal. A key object in this framework will be a function that we denote Orthogonal Instrumental Variables (OR-IVs) depending on the conditioning variables, $\left(I_1,X_1,\Omega_1,I_2,X_2,\Omega_2\right)$ in the example, that satisfies an important property.  We will argue that to compute OR-IVs the researcher needs to solve certain functional equations (cf. Equation \eqref{funceqprod}). The terms involved in it as well as the plausibility of solving these equations would be extremely linked to the particular setting that one is working on. Our aim is to construct LR moments in a data-driven manner that can be uniformly applied across various settings, e.g., production functions. This is the main contribution of this work. Hence, the researcher does not have to derive explicit expressions for such functions. The algorithm will provide her with a suitable adjustment term that can subsequently be employed in the estimation of $\theta_0$. This approach results in an estimator $\hat{\theta}_{DGMM}$, which is debiased and  yields a  $\sqrt{n}\left(\hat{\theta}_{DGMM}- \theta_0\right)$ normally distributed. Figure \ref{fig:histbias} displays the standardized empirical distribution of such an estimator (in blue). Notably, it closely resembles the expected asymptotic distribution and, crucially, is centered around zero, implying that our construction is less affected by the first-stage bias.

	\section{Debiased Moments and Heuristics}
	\label{orthogonality}
	
	We provide a precise notion of an LR moment for $\theta_0$ in our setting. Let $\kappa_0 \in L^2(Z)$ be some vector-valued function, which will be characterized below. Let $\eta_{0} \in \mathbf{B}$, where $\mathbf{B}$ is a possibly infinite dimensional vector space. A debiased moment in our setting is a moment based on a function $\psi: \mathcal{W} \times \Theta \times \mathbf{B} \times L^2(Z) \mapsto \mathbb{R}$ satisfying the following two restrictions: 
	\begin{align}
		\frac{d}{d \tau} \mathbb{E}\left[\psi\left(W,\theta_0,\eta_0 + \tau b,\kappa_0\right)\right] & = 0, \;\;\; \text{for all}\; b \in\mathbf{B}, \label{derivlr} \\
		\mathbb{E}\left[\psi\left(W,\theta_0,\eta_0,\kappa\right)\right] & = 0, \;\;\;\; \text{for all}\; \kappa \in L^2(Z), \label{zeromeankappa}
	\end{align}
	where $\frac{d}{d\tau}$ denotes derivatives from the right (i.e., from non-negatives values of $\tau$), at $\tau = 0$ . Equation \eqref{derivlr} implies that the Gateaux derivative at $\eta_0$ of a moment based on $\psi$ is zero. Here $b$ represents a possible direction of deviation from $\eta_0$. Intuitively, local perturbations to the nuisance parameter do not affect the moment. This is an appealing property as it would be hard to learn $\eta_0$ exactly, which is particularly true in high-dimensional contexts. This is the reason why estimation of $\theta_0$ based on a moment that is orthogonal is less affected by first-stage bias, typically present in estimating $\eta_0$, and will be a key attribute to establish standard inference results on an estimator of $\theta_0$. As we will show below, we need to introduce the nuisance parameter $\kappa_0$ to achieve \eqref{derivlr}. Nonetheless, this would not affect the moment itself as Equation \eqref{zeromeankappa} implies that the expectation of $\psi$ is \textit{globally} insensitive to deviations from $\kappa_0$. 
	
	Now, how can we construct moments that satisfy the key properties \eqref{derivlr}-\eqref{zeromeankappa} in the context of our production function example? The basic idea will be to obtain a function $\kappa_0$ that is orthogonal to a set that contains all the possible derivatives of the CMRs with respect to $\eta$. This corresponds to a more general result of orthogonality for general functionals in models defined by several CMRs, studied by \cite{arganaraz2025machine}.
	
	\bigskip \noindent \textsc{Example:} Let $W = \left(Y_1,Y_2,Y_3,X_1,X_2,X_3,I_2,I_2\right)$ be the vector of outputs, inputs, and proxy variables.  For this case, given some smoothness conditions, we can obtain a debiased moment by means of 
	\begin{equation}
		\label{lrfuncproduction}
		\begin{split}
			\psi\left(W,\theta_0,\eta_0, \kappa_0\right)  = &  \left(Y_1 - \eta_{01}\left(Z_1\right)\right)\kappa_{01}\left(Z_1\right) + \left(Y_2 - F\left(X_2,\theta_{0p}\right) - \theta_{0\omega}\left(\eta_{01}\left(Z_1\right) - F\left(X_1,\theta_{0p}\right)\right)\right)\kappa_{02}\left(Z_1\right) \\ & + \left(Y_2 - \eta_{02}\left(Z_2\right)\right)\kappa_{03}\left(Z_2\right) \\ & + \left(Y_3 - F\left(X_3,\theta_{0p}\right) - \theta_{0\omega}\left(\eta_{02}\left(Z_2\right) - F\left(X_2,\theta_{0p}\right)\right)\right)\kappa_{04}\left(Z_2\right)
			, 
		\end{split}
	\end{equation}
	where $Z_1 = \left(I_1,X_1\right)$, $Z_2 = \left(I_2,X_2\right)$, $Z = (Z_1,Z_2)$, $\kappa_0 = \left(\kappa_{01}, \kappa_{02}, \kappa_{03}, \kappa_{04}\right)\in L^2\left(Z_1\right) \times L^2\left(Z_1\right) \times L^2\left(Z_2\right) \times L^2\left(Z_2\right)$ is such that 
	\begin{equation}
		\begin{split}
			\label{funceqprod}
			\frac{d}{d\tau}\mathbb{E}\left[\psi\left(W,\theta_0,\eta_0 + \tau b, \kappa_0\right)\right] & = \mathbb{E}\left[b_1\left(Z_1\right)\left(-\kappa_{01}\left(Z_1\right) - \theta_{0\omega}\kappa_{02}\left(Z_1\right)\right) + b_2\left(Z_2\right)\left(-\kappa_{03}\left(Z_2\right) - \theta_{0\omega}\kappa_{04}\left(Z_2\right)\right)\right] \\ & = 0,\;\;\; \text{for all}\;\; b \in \mathbf{B}. 
		\end{split}
	\end{equation}
	Then, we can verify that \eqref{lrfuncproduction} indeed satisfies \eqref{derivlr}-\eqref{zeromeankappa}. $\square$

	\bigskip Expression \eqref{lrfuncproduction} indicates that we can obtain debiased moments as the sum of the products between the residual functions and the elements in $\kappa_0$. This linear combination is special in the sense that we need to find $\kappa_0 \in L^2(Z)$ such that the moment is invariant to local perturbations to $\eta$ around $\eta = \eta_0$. Therefore, by construction, the resulting moment \eqref{lrfuncproduction} is debiased.  Since $\kappa_0$ yields an orthogonal moment by properly combining the initial residual functions, \cite{chernozhukovetaldebiased2016} refer to these as Orthogonal Instrumental Variables (OR-IVs).    
	
	Obtaining a valid $\kappa_0$ entails solving the functional equation \eqref{funceqprod}. Various approaches can be pursued to achieve this goal. One option is to directly solve \eqref{funceqprod} and find a closed from expression for $\kappa_0$, as \cite{arganaraz2025machine} did for the partially linear model with endogeneity. This strategy, although convenient, may not always be ideal, as its plausibility depends on the specific expressions involved in the functional equations defining $\kappa_0$. Consequently, there is no guarantee that a procedure for solving for $\kappa_0$ in one model can be readily applied in other contexts. For instance, solving Equation \eqref{funceqprod} requires finding four terms, and each pair involves functions of the same random variables. In more complex settings, additional terms may emerge. Instead of relying on the tractability of the equations defining the OR-IVs, we propose an approach that can be applied generally. In addition, as it is evident from  \eqref{funceqprod}, these equations might involve unknown quantities, making direct computations potentially infeasible. Our goal is to design a \textit{feasible} algorithm that addresses unknown quantities intelligently.

	 Our algorithm converts any suitable vector of instruments $f \in L^2(Z)$, where $Z$ collects all the distinct random elements in the conditioning variables, into a unique $\kappa_0$. In other words, our approach automatically transforms a function of the conditioning variables such that it satisfies \eqref{funceqprod}. To illustrate the core idea of our procedure, let us again focus on our example. We start with a known function $f = \left(f_1, f_2, f_3, f_4\right) \in L^2\left(Z\right)$, chosen by the researcher. For instance, $f(Z)=\left(K_{1}, K_{1}, I_{2}, I_{2}\right)^{\prime}$---where $K$ denotes capital---provided that these entries have finite second moments. Under some regularity conditions and for suitable vectors $\hat{M}_1$, $\hat{M}_2$, $\hat{M}_3$, and $\hat{M}_4$, our algorithm finds some finite-dimensional vector $\hat{\beta}$ such that an estimator $\hat{\kappa} = \left(\hat{\kappa}_1, \hat{\kappa}_2, \hat{\kappa}_3, \hat{\kappa}_4\right)$ is constructed as follows: 
	\begin{align}
		\hat{\kappa}_{1}\left(Z\right) & = K_1 - \hat{M}^{'}_1\hat{\beta}, \label{betas11}\\ 
		\hat{\kappa}_{2}\left(Z\right) & = K_1 - \hat{M}^{'}_2\hat{\beta}, \label{betas12} \\
		\hat{\kappa}_{3}\left(Z\right) & = I_2 - \hat{M}^{'}_1\hat{\beta}, \label{betas13} \\ 
		\hat{\kappa}_{4}\left(Z\right) & = I_2 - \hat{M}^{'}_2\hat{\beta}, \label{betas14}
	\end{align}
	and satisfies \eqref{funceqprod}, with probability approaching one, where we replace population objects with their sample counterparts. In particular, $\hat{\beta}$ is interpreted as the solution to a Lasso-type problem with generated regressors $\hat{M}_1$-$\hat{M}_4$. Essentially, the vector $\hat{\beta}$ makes $\hat{\kappa}$, constructed as in \eqref{betas11}-\eqref{betas14},  orthogonal to a linear operator whose range is equal to all the possible Gateux derivatives of a moment based on $\psi$ with respect to $\eta$ at $\eta = \eta_0$. Obtaining such orthogonalization amounts to minimizing a mean squared error. As we allow the dimension of each of the regressors $\hat{M}_1$-$\hat{M}_4$ to be greater than the sample size, we add a $\ell_1-$norm penalization term. Hence, our problem of finding $\hat{\beta}$ can be seen as solving as a Lasso problem. Under an approximate sparsity condition, we show that if this $\hat{\beta}$ is plugged into \eqref{betas11}-\eqref{betas14}, a valid  $\hat{\kappa}$ and thus a debiased moment can be obtained.

	Based on this construction, we derive below a convergence rate for our OR-IV estimator, $\hat{\kappa}$. Let $\hat{\eta}$ be a suitable machine learning estimator of $\eta_0 \in \Xi \subseteq \mathbf{B}$.\footnote{While establishing asymptotic results, the distinction between $\Xi$ and $\mathbf{B}$ is important in endogenous settings, as we clarify below.}  We provide below conditions under which
	$$
	\sqrt{n}\left|\left|\hat{\eta} - \eta_0\right|\right|_{\Xi}\left|\left|\hat{\kappa} - \kappa_0\right|\right|_{L^2(Z)} \rightarrow 0, 
	$$
	where $\left|\left|\cdot \right|\right|_{\Xi}$ is a suitable norm in $\Xi$. Note that the previous display allows for rates slower than $n^{-1/2}$ for $\hat{\eta}$ and $\hat{\kappa}$. Under this key condition, we prove asymptotic normality of a two-step GMM estimator $\hat{\theta}$, implying that standard inference can be conducted straightforwardly. Debiasing plays a pivotal role in this derivation. The following sections develop these ideas in a general framework, derive the exact program that $\hat{\beta}$ solves, and provides the technical conditions required. Readers less interested in these details can skip these sections, except for Section \ref{estimationparint}, where we present our estimator. 
	
	\section{Computation of the OR-IVs}
	\label{estimationoivs}
	
	\subsection{General Setting}
	
	 The data $W_i = \left(Y_i, V_i, Z_i\right)$, $i = 1,\cdots,n$, is iid with the same distribution as $W = \left(Y,V,Z\right)$, where  $Y$ has support $\mathcal{Y} \subseteq \mathbb{R}^{d_Y}$. Let $\theta \in \Theta \subset \mathbb{R}^{d_\theta}$ denote a finite-dimensional parameter vector. Let $\eta  \in \Xi \subseteq \mathbf{B}$ be a vector of real-valued measurable functions of $V$ that might depend on additional unknown parameters that we do not specify, and $\mathbf{B}  \subseteq \bigotimes^{d_\eta} L^2\left(V\right)$ is a Hilbert space with inner product $\left<\cdot,\cdot \right>_{\mathbf{B}}$. To be specific,  $\eta = \left(\eta_{1}, \cdots, \eta_{d_\eta}\right)$ with $\eta_{s} \equiv \eta_{s}\left(V\right)$.\footnote{With this notation, we also allow for cases where $\eta_s$ depends on a subvector of $V$.} We assume that there is a vector of residual functions $m_j: \mathcal{Y} \times \Theta \times \mathbf{B} \mapsto \mathbb{R}$ such that: 
	\begin{equation}
		\label{moment}
		\mathbb{E}\left[\left.m_j\left(Y,\theta_0,\eta_{0}\right)\right|Z_j\right] = 0,\;\;\; \text{a.s.},\;\;\; j = 1,2,\cdots,J,
	\end{equation}     
	where $\mathbb{E}[\cdot]$ is expectation under the distribution of $Y$ given $Z_j$, $Z$ denotes the union of distinct random elements of the conditioning variables $\left(Z_1,\cdots,Z_J\right)$, and each $m_j$ is known up to the parameters $(\theta_0,\eta_0)$, with finite second moment. To be precise, $m_j$ might depend on $\theta_0$ arbitrarily and  the entire function $\eta_0$, not only on its evaluation at a particular realization of $V$.  Observe that we are not imposing anything regarding how the conditioning variables relate. These might have all or some elements in common. Note that in the case where $Z_j$ is a constant, for some $j$, we have an unconditional moment.  Hereafter, we assume that there exists a unique $\left(\theta_0,\eta_0\right) \in \Theta \times \mathbf{B}$ such that  \eqref{moment} holds. We say the vector $Z$ gathers the exogenous variables in the model, which might include $V$ or not. 
	
	Let $\kappa = \left(\kappa_1, \cdots, \kappa_J\right)$, where $\kappa_j \equiv \kappa_j\left(Z_j\right)$, and $\kappa_j \in L^2(Z_j)$, $1 \leq j \leq J$. Hence, $\kappa \in L^2(Z)$, where $L^2(Z) = \bigotimes^J_{j=1} L^2(Z_j)$. We define
	$$
	h_j\left(Z_j, \theta, \eta\right) := \mathbb{E}\left[\left.m_j\left(Y,\theta,\eta\right) \right|Z_j\right].
	$$
	Our results will rely heavily on smoothness conditions of the functions $h_j$'s. In particular, we maintain the key assumption throughout: 
	\begin{assumption}
		\label{gateaux}
	For all $j$, $h_j\left(Z_j,\theta_0,\cdot\right): \mathbf{B} \mapsto L^2(Z_j)$ is Fréchet differentiable in a neighborhood of $\eta_0$, where the derivative $S^{(j)}_{\theta_0,\eta_0}b$ is given by 
		\begin{equation}
			\begin{split}
				\label{fderivativegral}
				\left[\nabla h_j\left(Z_j,\theta_0,\eta_0\right)\right](b) & \equiv  \frac{d}{d \tau }h_j\left(Z_j,\theta_0,\eta_0 + \tau b\right) \\ & = \left[S^{(j)}_{\theta_0,\eta_0}b\right]\left(Z_j\right),
			\end{split}
		\end{equation}
		for all $b \in \mathbf{B}$.
	\end{assumption}
	In our notation above, $b$ denotes a specific direction of deviation and $\tau \in \left[0, \varepsilon\right)$ is the associated length. We next provide conditions in our specific example under which Assumption \ref{gateaux} holds. 
	
	 \bigskip 	\noindent \textsc{Example:} The following lemma states sufficient conditions for the CMRs to be Fréchet differentiable. 
	 \begin{lemma}
	 	\label{fde2}
	 	Let $b_s = \left(b_{s1}, b_{
	 	s2}\right)^{\prime}$, $s=1,2$,  $\mathbf{B} \subseteq L^2(Z_1) \times L^2(Z_2)$, $\left<b_1, b_2 \right>_{\mathbf{B}} = \mathbb{E}\left[b_{11}(Z_1)b_{21}(Z_1)\right] + \mathbb{E}\left[b_{12}(Z_2)b_{22}(Z_2)\right] $.  Then, the CMRs are Fréchet differentiable. $\square$
	 \end{lemma}
	
	Remark that \eqref{fderivativegral} defines a linear operator $S^{(j)}_{\theta_0,\eta_0}: \mathbf{B} \mapsto L^2(Z_j)$; see \cite{luenberger1997optimization} and \cite{carrasco2000generalization} for a theory on linear operators. In addition, let us define
	$$
	S_{\theta_0,\eta_0}b := \left(S^{(1)}_{\theta_0,\eta_0}b, \cdots, S^{(J)}_{\theta_0,\eta_0}b\right).
	$$
	Then, $S_{\theta_0,\eta_0}: \mathbf{B} \mapsto L^2(Z)$ is also a linear  operator. We equip $L^2(Z)$  with the inner product 
	$$
	\left<f_1,f_2\right>_{L^2(Z)} = \sum^J_{j=1} \mathbb{E}\left[f_{1j}(Z_j)f_{2j}(Z_j)\right],
	$$
	where $f_s = \left(f_{s1},\cdots,f_{sJ}\right)$, $s=1,2$. Therefore, $L^2(Z)$ is a Hilbert space. The range of that operator can be defined as follows: 
	$$
	\mathcal{R}\left(S_{\theta_0,\eta_0}\right) := \left\{f \in L^2\left(Z\right): f = S_{\theta_0,\eta_0}b\;\; \text{for some}\;\; b \in \mathbf{B}\right\}. 
	$$
	A key object for us is $\overline{\mathcal{R}\left(S_{\theta_0,\eta_0}\right)}^{\perp}$, i.e., the orthogonal complement of the closure of the range of $S_{\theta_0,\eta_0}$ in $L^2(Z)$, which can be defined as 
	$$
	\overline{\mathcal{R}\left(S_{\theta_0,\eta_0}\right)}^{\perp} :=\left\{f \in L^2\left(Z\right): \sum^J_{j=1}\mathbb{E}\left[f_j\left(Z_j\right)h_j\left(Z_j\right)\right] = 0, \;\;\;\text{for all}\;\;\; h = \left(h_1,\cdots,h_J\right) \in \overline{\mathcal{R}\left(S_{\theta_0,\eta_0}\right)}\right\}. 
	$$
	\cite{arganaraz2025machine} show that a debiased moment in model \eqref{moment} can be constructed as follows: 
	\begin{equation}
		\label{lrmomenteq}
		\psi\left(W,\theta_0, \eta_0\right) = \sum^J_{j=1} m_j\left(Y,\theta_0,\eta_0\right)\kappa_{0j}\left(Z_j\right), \;\;\;  \kappa_0 \in \overline{\mathcal{R}\left(S_{\theta_0,\eta_0}\right)}^{\perp}.
	\end{equation}
	
	We point out two important observations. First, $\kappa_0$ might not be unique. In fact, there can potentially exist an infinite number of such  $\kappa_0$'s. Despite this, not every IV, i.e., functions in $L^2\left(Z\right)$, belongs to $\overline{\mathcal{R}\left(S_{\theta_0,\eta_0}\right)}^{\perp}$, and thus not every IV may serve as a valid  $\kappa_0$. Consequently, choices of instruments commonly made in applied work might not lead to orthogonal moments. For example, $\left(K_{1}, K_{1}, I_{2}, I_{2}\right)^{\prime}$ might not be  OR-IVs in our example. Nevertheless, we will demonstrate how to convert suitable functions in $L^2\left(Z\right)$, e.g.,$\left(K_{1}, K_{1}, I_{2}, I_{2}\right)^{\prime}$, into a $\kappa_0$. Therefore, one may begin with the common choices of instrument functions and then apply our transformation to directly obtain an orthogonal moment.
	
	Second, a $\kappa_{0}$ does not necessarily exist. This situation occurs when $\overline{\mathcal{R}\left(S_{\theta_0,\eta_0}\right)}^{\perp} = \left\{0\right\}$.  Following the terminology in \cite{arganaraz2025existence}, when $\overline{\mathcal{R}\left(S_{\theta_0,\eta_0}\right)}^{\perp} = \left\{0\right\}$  we say the model satisfies a \textit{local surjectivity} property.\footnote{Similar notions have appeared elsewhere for certain panel data models, e.g., in \cite{bonhomme2012functional}.} A practical implication of this situation is that this would result in a trivial orthogonal moment, i.e., $\psi = 0$. Unfortunately, the plausibility of encountering this situation depends on the particular model at hand.\footnote{Using our approach for estimating OR-IVs, we may check for local surjectivity in a practical way. Let $\hat{\kappa}$ be an estimator of $\kappa_0$. Local surjectivity implies $\hat{\kappa} \approx  0$ for each individual in the sample for a large class of initial functions $f$'s. An alternative (and more rigorous) approach consists of using duality theory and exploiting the fact that $\overline{\mathcal{R}\left(S_{\theta_0,\eta_0}\right)}^{\perp} = \mathcal{N}\left(S^{*}_{\theta_0,\eta_0}\right)$, where $S^{*}_{\theta_0,\eta_0}$ is the adjoint operator of $S_{\theta_0,\eta_0}$ and $\mathcal{N}\left(\cdot\right)$ denotes the null space of an operator \citep[][Theorem 6.6.3]{luenberger1997optimization}. Theoretically checking that $\mathcal{N}\left(S^{*}_{\theta_0,\eta_0}\right) = \left\{0\right\}$ might be easier than studying if  $\overline{\mathcal{R}\left(S_{\theta_0,\eta_0}\right)}^{\perp} = \left\{0\right\}$.}
	
	\subsection{Estimation of OR-IVs}
	\label{estimationOIV}
	We next explain how to obtain OR-IVs automatically in the general model \eqref{moment}. As we explained above, the key step to construct an orthogonal moment is to find a function $\kappa_{0} \in L^2\left(Z\right)$  such that it is orthogonal to $\overline{\mathcal{R}\left(S_{\theta_0,\eta_0} \right)}$. This implies that
	$$
	\sum^J_{j=1}\mathbb{E}\left[ \left[S^{(j)}_{\theta_0,\eta_0}b\right](Z_j)\kappa_{0j}\left(Z_j\right)\right] = 0, 
	$$
	for all $b \in \mathbf{B}$. We can obtain such $\kappa_0$ by selecting some function $f \in L^2\left(Z\right)$, and then computing 
	$$\kappa_{0} = f - \Pi_{\overline{\mathcal{R}\left(S_{\theta_0,\eta_0}\right)}}f,
	$$ where recall that $\Pi_{\overline{\mathcal{R}\left(S_{\theta_0,\eta_0}\right)}}$ denotes the orthogonal projection operator onto $\overline{\mathcal{R}\left(S_{\theta_0,\eta_0}\right)}$, and is defined as follows
	\begin{equation}
		\label{orthproject}
		\Pi_{\overline{\mathcal{R}\left(S_{\theta_0,\eta_0}\right)}}f := \underset{\tilde{f} \in \overline{\mathcal{R}\left(S_{\theta_0,\eta_0}\right)}}{\operatorname{arg\;min}} \; \sum^J_{j=1}\mathbb{E}\left[\left(f_j(Z_j) - \tilde{f}_j(Z_j)\right)^2\right].
	\end{equation}
	By the Projection Theorem, \eqref{orthproject} exists and is unique \citep[see][Theorem 3.3.2]{luenberger1997optimization}. Hence, for a given $f$, we have a unique $\kappa_0$ a.s. Next, we exploit the following facts. Notice that $\Pi_{\overline{\mathcal{R}\left(S_{\theta,\eta_0}\right)}}f \in \overline{\mathcal{R}\left(S_{\theta_0,\eta_0}\right)}$, and that the range of $S_{\theta_0,\eta_0}S^{*}_{\theta_0,\eta_0}$ is dense in $\overline{\mathcal{R}\left(S_{\theta_0,\eta_0}\right)}$, where $S^{*}_{\theta_0,\eta_0}$ is the adjoint operator of $S_{\theta_0,\eta_0}$. Since  $S_{\theta_0,\eta_0}$ is bounded, $S^{*}_{\theta_0,\eta_0}: L^2(Z) \mapsto \mathbf{B}$ exists, is linear, and bounded \citep[see Theorem 2.21 in][]{CARRASCO20075633}. Moreover, it satisfies, for all $g \in L^2(Z)$ and $b \in \mathbf{B}$, 
	$$
	\left<S_{\theta_0,\eta_0}b, g\right>_{L^2(Z)} = \left<b, S^{*}_{\theta_0,\eta_0}g\right>_{\mathbf{B}}.
	$$
   Let $\Pi_{\overline{\mathcal{R}\left(S_{\theta_0,\eta_0}\right)}}f = f^{*}$, then the previous observations imply that $f^{*} \in \overline{\mathcal{R}\left(S_{\theta_0,\eta_0}S^{*}_{\theta_0,\eta_0}\right)}$. This is potentially useful as $S_{\theta_0,\eta_0}S^{*}_{\theta_0,\eta_0}$ is an operator that maps elements in $L^2(Z)$ into $L^2(Z)$, spaces of exogenous variables in the data. Moreover, $S_{\theta_0,\eta_0}S^{*}_{\theta_0,\eta_0}$ can be characterized by exploiting the CMRs given by the model in a general way, as we show below. In general, we are interested in solving 
	\begin{equation}
		\label{restprogram}
		\underset{g \in L^2\left(Z\right)}{\operatorname{inf}} \; \sum^J_{j=1}\mathbb{E}\left[\left(f_j(Z_j) - S^{(j)}_{\theta_0,\eta_0}S^{*}_{\theta_0,\eta_0}g\right)^2\right].
	\end{equation}

From a practical standpoint, problem \eqref{restprogram} presents two difficulties. First, we might not able to directly compute $S^{(j)}_{\theta_0,\eta_0}S^{*}_{\theta_0,\eta_0}$ as it might involve unknown objects. Hence, we shall construct an estimator for it. 
	
	Second, the problem implied by \eqref{restprogram} requires a search over an infinite-dimensional space, which would be challenging to directly accomplish in practice. To tackle this issue, we exploit a particular regularization strategy. We will look for a solution to \eqref{restprogram} in a set $\mathcal{G}_n$ defined as follows. Let $\gamma\left(Z\right) = \left(\gamma_1\left(Z_1\right)^{'}, \cdots, \gamma_J\left(Z_J\right)^{'}\right)^{'} \in L^2(Z)$ be a vector of basis functions, where $\gamma_j\left(Z_j\right)$ is a $r_j-$dimensional vector of known real-valued functions, with $\mathbb{E}\left[\gamma_{jk}\left(Z_j\right)\right] = 0$ and $\mathbb{E}\left[\gamma^2_{jk}\left(Z_j\right)\right] = 1$, $k=2,\cdots,r_j$.\footnote{We let the first element of $\gamma_j$ be 1.} Let $\beta = \left(\beta_{11}, \cdots, \beta_{1r_1},\cdots,\beta_{1r_J}, \cdots,\beta_{Jr_J}\right)^{'}$ be a $r-$dimensional vector, where $r = \sum^J_{j=1}r_j$. Then, we define $\mathcal{G}_n$ as\footnote{Extending our results to other functional classes such as Reproducing Kernel Hilbert Spaces or Neural Networks is an interesting avenue of future research.} 
	$$
	\mathcal{G}_n := \left\{g \in L^2(Z): g_j\left(Z_j\right) = \gamma_j\left(Z_j\right)^{'}\beta_j, \;\;\left|\left|\beta\right|\right|_{\infty} < c \right\}.
	$$
	Remark that since we will look for a solution in $\mathcal{G}_n$ and $\gamma$ is known, we only need to focus on finding an optimal finite-dimensional vector $\beta$. We let $\beta$ be general in the sense that its dimension, $r$, can grow with $n$ and be much larger than it. To this end, we augment the program in \eqref{restprogram} by including an $\ell_1-$norm penalization term on $\beta$. In sum, the need for estimating $S^{(j)}_{\theta_0,\eta_0}S^{*}_{\theta_0,\eta_0}$ and our regularization strategy yields writing a feasible version of \eqref{restprogram} as 
	
	\begin{equation}
		\label{restprogram3}
		\underset{\beta \in \mathbb{R}^r}{\operatorname{min}} \; \sum^J_{j=1}\mathbb{E}_n\left[\left(f_j(Z_{ji}) - \hat{S}^{(j)}_{\hat{\theta},\hat{\eta}}\hat{S^{*}}_{\hat{\theta},\hat{\eta}}\gamma(Z_i)^{'}\beta\right)^2\right] + 2\lambda_n\left|\left|\beta\right|\right|_1,
	\end{equation}
	where $\gamma(Z)^{'}\beta \equiv \left(\gamma(Z_1)^{'}\beta_1,\cdots,\gamma(Z_J)^{'}\beta_J\right)^{\prime}$, $\hat{S}^{(j)}_{\hat{\theta},\hat{\eta}}\hat{S^{*}}_{\hat{\theta},\hat{\eta}}$ is a suitable estimator of $S^{(j)}_{\theta_0,\eta_0}S^{*}_{\theta_0,\eta_0}$, $\lambda_n \geq 0$ is a regularization term (or tuning parameter) to be determined below, and $\mathbb{E}_n\left[h(Z_i)\right] = \sum^n_{i=1}h(Z_i)/n$. The solution to \eqref{restprogram3} is denoted by $\hat{\beta}$. 
	
	Let us be more precise about the estimation of $S^{(j)}_{\theta_0,\eta_0}S^{*}_{\theta_0,\eta_0}$. During its estimation, we use cross-fitting, following \cite{chernozhukov2018double}.\footnote{Cross-fitting has a long tradition in the semiparametric literature; see, e.g., \cite{bickel1982adaptive}, \cite{klaassen1987consistent}, \cite{vandervaart98}, \cite{robins2008higher}, and \cite{Zheng2010AsymptoticTF}.} This entails randomly partitioning the sample into $L$ groups, $I_1, \cdots, I_L$, of the same size. Let $I^c_\ell$ be the complement of $I_\ell$.  We compute $\hat{\beta}_\ell$, using observations in  $I^c_\ell$ only. In other words, we compute, for each $I_{\ell}$, $\hat{\beta}_\ell$ using $\hat{S}^{(j)}_{\hat{\theta}_\ell,\hat{\eta}_\ell}\hat{S^{*}}_{\hat{\theta}_\ell,\hat{\eta}_\ell}$, which is constructed from observations in $I^c_\ell$. Let $n_\ell$ denote the number of observations in $I_\ell$. Hence, 
		\begin{equation*}
		\hat{\beta}_\ell = \underset{\beta \in \mathbb{R}^r}{\operatorname{argmin}} \; \sum^J_{j=1}\frac{1}{n-n_\ell}\sum_{i \notin I_\ell}\left[\left(f_j(Z_{ji}) - \hat{S}^{(j)}_{\hat{\theta}_\ell,\hat{\eta}_\ell}\hat{S^{*}}_{\hat{\theta}_\ell,\hat{\eta}_\ell}\gamma(Z_i)^{'}\beta\right)^2\right] + 2\lambda_n\left|\left|\beta\right|\right|_1.
	\end{equation*}
		Ultimately, we propose to estimate each component in $\kappa_0$, for each individual $i \in I_\ell$, by means of
	\begin{equation}
		\label{estkappa}
		\begin{split}
			\hat{\kappa}_{j\ell}\left(Z_{ji}\right) & = f_j\left(Z_j\right) - \hat{f^{*}_{j\ell}}\left(Z_{ji}\right) \\ & = f_j\left(Z_{ji}\right) - \hat{S}^{(j)}_{\hat{\theta}_\ell,\hat{\eta}_\ell}\hat{S^{*}}_{\hat{\theta}_\ell,\hat{\eta}_\ell}\gamma(Z_i)^{'}\hat{\beta}_\ell.    
		\end{split}
	\end{equation}
	
	\section{The Algorithm}
	\label{sthealgorithm}
	The approach previously developed provide researchers with a general guidance for estimating OR-IVs. We now want to propose a specific algorithm, which can be followed in a wide range of settings that share a common structure in the operator $S_{\theta_0,\eta_0}$.   We assume 
	\begin{assumption}
		\label{linearityb}
		For all $j$,
		\begin{itemize}
			\item[i)] 
			There exists a known (up to $\theta_0$ and $\eta_0$) function $\nu_j$  such that 
			$$
			\left[S^{(j)}_{\theta_0,\eta_0}b\right]\left(Z_j\right) = \mathbb{E}\left[\left.\nu_j\left(Y,\theta_0,\eta_0,b\right)\right|Z_j\right],\;\;\; \text{for all}\;\;\; b \in \mathbf{B};
			$$
			\item[ii)] $\nu_j\left(Y,\theta_0,\eta_0,b\right) = b\left(V\right)^{\prime} \tilde{\nu}_j\left(Y,\theta_0,\eta_0\right)$, for some $d_\eta-$vector of 
			known (up to $\theta_0$ and $\eta_0$) functions $\tilde{\nu}_j$;
		\end{itemize}
	\end{assumption}
	Assumption  \ref{linearityb} \textit{i)} implies that there exists a function $\nu_j$ that links the direction $b$ with the operator $S^{(j)}_{\theta_0,\eta_0}$ through the conditional expectation given $Z_j$. In the standard non-parametric IV setting \citep[e.g.,][]{newey2003instrumental}, where we have $\mathbb{E}\left[\left.Y - \eta_0\left(V\right)\right|Z_1\right] = 0$, it holds that $\nu_j(Y,\theta_0,\eta_0,b) = -b\left(V\right)$. Hence, \textit{i)} allows for these derivatives in the first step to depend generally on $\theta_0$ and $\eta_0$. Assumption  \ref{linearityb} \textit{ii)} requires $\nu_j$ to be linear in the direction $b$ in a specific sense. In the non-parametric IV case, $\tilde{\nu}_j\left(Y,\theta_0,\eta_0\right) = -1$.

	As shown in the previous section, Assumption \ref{linearityb} is not needed, as we could directly work with the generic operators $S^{(j)}_{\theta_0,\eta_0}$'s and its properties. However, in many applications, to find those operators one shall proceed by first finding functions $\nu_j's$ and then showing that their conditional expectations are the corresponding Fréchet derivatives. Additionally, it is more convenient to work with functions $\nu_j$ and $\tilde{\nu}_j$. Our strategy below relies on them to propose a practical algorithm to compute OR-IVs.  Notice that if the residual function $m_j\left(Y,\theta_0, \cdot\right): \mathbf{B} \mapsto \mathbb{R}$ is Fréchet differentiable for each fixed $Y = y$ a.s. and its derivative is bounded by some integrable function, Assumption \ref{linearityb} \textit{i)} is satisfied, provided that the conditional expectation of such  derivative, given $Z_j$, is square-integrable.\footnote{Recall that if $m_j\left(Y,\theta_0, \cdot\right): \mathbf{B} \mapsto \mathbb{R}$ is Fréchet differentiable for each fixed $Y = y$ a.s., then $\frac{d}{d\tau}m_j\left(Y,\theta_0, \eta_0 + \tau b\right) = \nu_j\left(Y,\theta_0,\eta_0,b\right)$ for all $b$, where $\nu_j\left(Y,\theta_0,\eta_0,\cdot\right): \mathbf{B} \mapsto \mathbb{R}$ is linear and continuous. If this expression is bounded by some integrable function, \textit{i)} is implied by  the Dominated Convergence Theorem, provided that $\mathbb{E}\left[\left. \nu_j\left(Y,\theta_0,\eta_0,b\right) \right|Z_j\right] \in L^2(Z_j)$.} Assumption \textit{ii)} requires that its derivative can be written in the form $ b\left(V\right)^{\prime} \tilde{\nu}_j\left(Y,\theta_0,\eta_0\right)$. Altogether, these conditions simplify our calculations below.  We can easily verify that Assumption \ref{linearityb} holds in our example.

	\bigskip \noindent \textsc{Example:} Suppose that the assumptions in Lemma \ref{fde2} hold. It is not difficult to obtain that 
	\begin{equation*}
		\begin{split}
			\left[S^{(1)}_{\theta_0,\eta_0}b\right](Z_1) & = -b_1\left(Z_1\right),\;\;\;\;\;\; \tilde{\nu}_1(Z_1,\theta_0) = -1, \\ \left[S^{(2)}_{\theta_0,\eta_0}b\right](Z_1) & = -\theta_{0\omega}b_1\left(Z_1\right),\;\;\;\;\;\; \tilde{\nu}_2(Z_1,\theta_0) = -\theta_{0\omega} \\ \left[S^{(3)}_{\theta_0,\eta_0}b\right](Z_2)& = -b_2\left(Z_2\right),\;\;\;\;\;\;\;\;\;\;\; \tilde{\nu}_3(Z_2,\theta_0) = -1\\ \left[S^{(4)}_{\theta_0,\eta_0}b\right](Z_2) & = -\theta_{0\omega}b_2\left(Z_2\right), \;\;\;\;\;\;\tilde{\nu}_4(Z_2,\theta_0) = -\theta_{0\omega}.
		\end{split}
	\end{equation*}
	In addition, $S^{(1)}_{\theta_0,\eta_0}b, S^{(2)}_{\theta_0,\eta_0}b \in L^2(Z_1)$, and $S^{(3)}_{\theta_0,\eta_0}b, S^{(4)}_{\theta_0,\eta_0}b \in L^2(Z_2)$, for any $b$. $\square$
	 
	 \bigskip Leveraging Assumption \ref{linearityb}, we can obtain an expression for $S^{*}_{\theta_0,\eta_0}$ that can be applied to various contexts: 
	\begin{proposition}
		\label{prop_adjoint}
		Suppose Assumptions \ref{gateaux} and \ref{linearityb} hold. In addition, let $\left<b_1, b_2 \right>_{\mathbf{B}} = \mathbb{E}\left[b_1(V)^{'}b_2(V)\right]$. Then, the adjoint $S^{*}_{\theta_0,\eta_0}: L^2(Z) \mapsto \mathbf{B}$ exists, is linear, continuous, and given by
		$$
		\left[S^{*}_{\theta_0,\eta_0} g\right](V) = \sum^J_{j=1} \mathbb{E}\left[\left.\tilde{\nu}_j\left(Y,\theta,\eta_0\right)g_j\left(Z_j\right)\right|V\right].
		$$
	\end{proposition}

	We shall now write \eqref{restprogram3} using matrix notation. Split $I^c_\ell$ into three mutually exclusive pieces such that $I^c_\ell = A_\ell \cup B_\ell \cup C_\ell$. Possibly, this partition depends on $j$, but we omit this dependence for simplicity. For any estimator, a subscript will indicate the part of $I^c_\ell$ that has been used to compute it. For example, $\hat{\theta}_{A_\ell}$ means that $\theta_0$ has been estimated using observations in $A_\ell$ and so on. Let $\bm{f_{j\ell}}$ be a $n_\ell-$dimensional vector containing each $f_j\left(Z_{ji}\right)$, $i \notin I_{\ell}$. Let $\bm{\hat{M}_{j\ell}}$ be a $n_\ell \times r$ design matrix such that its $(i,l)$-entry is given by
	\begin{equation}
		\begin{split}
			\label{m1}
			\left[\bm{\hat{M}_{j\ell}}\right]_{il} = \hat{\mathbb{E}}_{C_\ell}\left[\left.\left(\hat{\mathbb{E}}_{B_\ell}\left[\left.\tilde{\nu}_{j^{'}}\left(Y_i,\hat{\theta}_{A_\ell},\hat{\eta}_{A_\ell}\right)\gamma_{j^{'}k}\left(Z_{ji}\right)\right|V_i\right]\right)^{'}\tilde{\nu}_j\left(Y_i,\hat{\theta}_{B_\ell},\hat{\eta}_{B_\ell}\right) \right|Z_{ji}\right] &, \;\;\; j^{\prime},j  =1,\cdots,J, \\ k & =1,\cdots,r.
		\end{split}
	\end{equation}
	Then, we have
	$$
	\hat{\beta}_\ell = \underset{\beta \in \mathbb{R}^r}{\operatorname{arg\;min}}\; \sum^J_{j=1} \frac{1}{n-n_{\ell}} \left(\bm{f_{j\ell}} - \bm{\hat{M}_{j\ell}}\beta\right)^{'}\left(\bm{f_{j\ell}} - \bm{\hat{M}_{j\ell}}\beta\right) + 2\lambda_n \left|\left|\beta\right|\right|_1,
	$$
	where $\hat{\eta}_{A_\ell}$, $\hat{\eta}_{B_\ell}$, $\hat{\mathbb{E}}_{B_\ell}\left[\left.\cdot\right|V\right]$, and $\hat{\mathbb{E}}_{C_\ell}\left[\left.\cdot\right|Z_j\right]$ are non-parametric estimators, possibly based on some machine learning tool, and $\hat{\theta}_{A_\ell}$ and $\hat{\theta}_{B_\ell}$ are possibly non-LR estimators of $\theta_0$.\footnote{I.e., estimators based on non-LR moments.} Recall that $\tilde{\nu}_j$ is a known function, given estimators of $\theta_0$ and $\eta_0$. Thus, these conditional expectations can be evaluated. Notice that the $\ell_1-$penalization terms allows for $r>n$. As a result,  we have transformed our initial problem into a Lasso-type one for which there exist well-known and fast algorithms. Finally, the resulting residuals from such problem is an estimator of an OR-IV. 	Let $\bm{M_{j}}$ be the population analog of matrix $\bm{\hat{M}_{j\ell}}$. Let $\hat{M}_{j\ell}\left(Z_{ji}\right)$ be a $r-$dimensional vector containing the $i-$ row of $\bm{\hat{M}_{j\ell}}$. A similar definition applies to $M_{j}\left(Z_{ji}\right)$.
	 Our algorithm can be implemented as follows: 
	
	\bigskip \noindent \textbf{Algorithm to estimate OR-IVs:}
	
	\noindent \textbf{Step 0:} Choose a real-valued function $f \in L^2(Z)$. Choose a basis for each $\gamma_j(Z_j)$, e.g., exponential, Fourier, splines, or power.  In addition, specify a low-dimensional dictionary, say $\gamma^{low}(Z)$, which is a sub-vector of $\gamma(Z)$.
	
	\bigskip \noindent \textbf{Step 1:} For each $\ell = 1,\cdots L$, compute (possible) non-LR estimators $\hat{\theta}_{A_\ell}$ and $\hat{\theta}_{B_\ell}$. Moreover, using some machine learning algorithm, compute $\hat{\eta}_{A_\ell}$, $\hat{\eta}_{B_\ell}$, $\hat{\mathbb{E}}_{B_\ell}\left[\left.\cdot\right|V\right]$, and $\hat{\mathbb{E}}_{C_\ell}\left[\left.\cdot\right|Z_j\right]$.
	
	\bigskip \noindent \textbf{Step 2:} Construct design matrix $\bm{\hat{M}_{j\ell}}$ such that its $(i,l)-$entry is \eqref{m1}. 
	
	\bigskip \noindent \textbf{Step 3:} Initialize $\hat{\beta}_\ell$ using $\gamma^{low}(Z)$ such that 
	
	\begin{equation*}
		\begin{split}
			\left[\bm{\hat{M}^{low}_{j\ell}}\right]_{il} & = \hat{\mathbb{E}}_{C_\ell}\left[\left.\left(\hat{\mathbb{E}}_{B_\ell}\left[\left.\tilde{\nu}_{j^{'}}\left(Y_i,\hat{\theta}_{A_\ell},\hat{\eta}_{jA_\ell}\right)\gamma^{low}_{j^{'}k}\left(Z_{j^{'}i}\right)\right|V_i\right]\right)^{'}\tilde{\nu}_j\left(Y_i,\hat{\theta}_{B_\ell},\hat{\eta}_{jB_\ell}\right) \right|Z_{ji}\right], \\ 
			\hat{\beta}_{\ell}  & =  \begin{pmatrix}
				\left(\sum^{J}_{j=1} \bm{\hat{M}^{low'}_{j\ell}}\bm{\hat{M}^{low}_{j\ell}}\right)^{-1}\left(\sum^J_{j=1}\bm{\hat{M}^{low'}_{j\ell}} \bm{f_{j\ell}}\right) \\ 0
			\end{pmatrix}.
		\end{split}
	\end{equation*}
	
	\noindent \textbf{Step 4:} (While $\hat{\beta}_\ell$ has not converged) 
	\begin{itemize}
		\item[(a)] Update normalization 
		\begin{equation*}
			\begin{split}
				\hat{D}_{j^{'}k\ell} & = \left[\frac{1}{n-n_\ell} \sum_{i \notin I_{\ell}} \left\{\sum^J_{j=1} \hat{\mathbb{E}}_{C_\ell}\left[\left.\left(\hat{\mathbb{E}}_{B_\ell}\left[\left.\tilde{\nu}_{j^{'}}\left(Y_i,\hat{\theta}_{A_\ell},\hat{\eta}_{A_\ell}\right)\gamma_{j^{'}k}\left(Z_{j^{'}i}\right)\right|V_i\right]\right)^{'}\tilde{\nu}_j\left(Y_i,\hat{\theta}_{B_\ell},\hat{\eta}_{B_\ell}\right) \right|Z_{ji}\right]\hat{\epsilon}_{ji\ell}\right\}^2\right]^{1/2}, \\ 
				\hat{\epsilon}_{ji\ell} & = f_j\left(Z_{ji}\right) - \hat{M}_{j\ell}\left(Z_{ji}\right)^{'}\hat{\beta}_\ell.
			\end{split}
		\end{equation*}

		\item[(b)] Update $\hat{\beta}_\ell$, where
		$$
		\hat{\beta}_\ell = \underset{\beta \in \mathbb{R}^r}{\operatorname{arg\;min}}\; \sum^J_{j=1} \frac{1}{n-n_{\ell}} \left(\bm{f_{j\ell}} - \bm{\hat{M}_{j\ell}}\beta\right)^{'}\left(\bm{f_{j\ell}} - \bm{\hat{M}_{j\ell}}\beta\right) + 2\lambda_n\sum^J_{j^{\prime}=1}\sum^{r_{j^\prime}}_{k=1}\left|\hat{D}_{j^{\prime}k\ell} \beta_{j^{\prime}k}\right|,
		$$
		and
		$$
		\lambda_n = \frac{c_1}{(n-n_\ell)^{1/4}}\Phi^{-1}\left(1-\frac{c_2}{2r}\right),
		$$
		where $\Phi\left(.\right)$ is the standard normal cdf.
	\end{itemize}
	
	\noindent \textbf{Step 5:} Given the optimal $\hat{\beta}_\ell$, compute $\hat{\kappa}_{j\ell}$ as: 
   \begin{equation*}
			\hat{\kappa}_{j\ell}\left(Z_{ji}\right) = f_j\left(Z_{ji}\right) - \hat{M}_{j\ell}\left(Z_{ji}\right)^{'}\hat{\beta}_\ell.
	\end{equation*}

	\bigskip In \textbf{Step 1}, we need to specify a low-dimensional dictionary $\gamma^{\text{low}}(Z)$ such that $\sum^J_{j=1}\bm{\hat{M}_{j\ell}^{\prime}\hat{M}_{j\ell}}$ is invertible in \textbf{Step 3}. In our Monte Carlo experiment below,  we take the first 5 components of each $\gamma_j(Z_j)$. In our empirical application, we take the first $\tilde{r}_j$ of each $\gamma_j(Z_j)$, where $\tilde{r}_j = r_j/15$.\footnote{How large $\tilde{r}_j$ is depends on the particular setting. Ideally, $\tilde{r}_j$ shall be set as large as possible as long as $\sum^J_{j=1}\bm{\hat{M}_{j\ell}^{\prime}\hat{M}_{j\ell}}$ is non-singular.}  Following \cite{belloni2012sparse}, we include a normalization term, $\hat{D}_{jk\ell}$, in the $\ell_1-$norm above, which is necessary for the good properties of our Lasso estimator. For this same reason, we suggest computing $\lambda_n$ as given in \textbf{Step 4}, as recommended by \cite{belloni2012sparse} (p. 2380). However, we let $\lambda_n$ be proportional to $n^{-1/4}$ instead of $n^{-1/2}$, as required by our asymptotic theory. Regarding the constants in $\lambda_n$, our Monte Carlo exercises have suggested $c_1 = 1.1$. Several numerical experiments have indicated that a value of at least $c_2 = 0.5/\log(n \vee  r)$ works well.\footnote{\cite{belloni2012sparse} recommend $c_1 = 1.1$ and $c_2 = 0.1/\log(n \vee  r)$. We have observed that, in our context, \cite{belloni2012sparse}'s recommendation leads to ``too much" regularization. More precisely, their recommendation tends to shrink all coefficients in $\beta$ to zero, which might be explained by the fact that $\lambda_n$ has $\left(n - n_\ell\right)^{1/4}$ in the denominator instead of the larger $\left(n - n_\ell\right)^{1/2}$. To accommodate for this, we let $c_2$ to be larger as well. On strictly theoretical grounds, $c_1$ shall be greater than 1 and $c_2$ needs to be such that it converges to zero and  $\log(1/c_2) \leq C \log(n \vee  r)$, which is satisfied by our current choices.}  To improve numerical stability we follow  \cite{chernozhukov2022automatic} and cap the maximum number of iterations at 10. In addition, we use warm start. This means that in a given iteration, the initial parameter value is equal to the $\hat{\beta}_\ell$ obtained in the previous iteration.  In Section \ref{justification} of the Supplementary Appendix we provide a justification for the previous optimization. Notice that \textbf{Step 4} (b) requires solving for $\hat{\beta}_\ell$. For this, we use an extension of the coordinate descent approach for Lasso \citep[][]{Friedman07, friedman2010regularization, fu1998penalized}; see Section \ref{optimization_beta} of the Supplementary Appendix for details. An important remark is that our theory (and several numerical experiments we conducted) suggest all these choices, as long as the requirements of our asymptotic theory are met, does not play an important role as inference on $\theta_0$ is concerned, which is a result of the OR-IVs being, by construction, in the space of IVs of the model.  
	
	We point out that our previous algorithm applies to the most general case. In some applications, it will be implemented with some simplifications, depending on the particular expression of $\tilde{\nu}_j$, and how $V$ and the conditioning variables relate. For example, it might not be necessary to partition $I_\ell$ into three pieces, as it might not be necessary to compute all the estimators in \textbf{Step 1}. This occurs, for example, when $\tilde{\nu}_j$ does not depend on $\theta_0$, or $\eta_0$, or both. Another case is when $\mathbb{E}\left[\left.\tilde{\nu}_j\left(Y,\theta_0,\eta_0\right)\gamma_{jk}(Z_{ji})\right|V\right] = \tilde{\nu}_j\left(Y,\theta_0,\eta_0\right)\gamma_{jk}(Z_{ji})$ as $\left(Y,Z_j\right)$ and the variables that $\eta_0$ depends on are contained in $V$. More simplifications can emerge if $\tilde{\nu}_j$ depends only on $Z_j$.  As we will illustrate below, in our example, $\left[\bm{\hat{M}}_{j\ell}\right]_{i\ell}$ has a simpler expression than in \eqref{m1}.

	\bigskip \noindent \textsc{Example:} The user has to provide $f(Z) = \left(f_1(Z_1), f_2(Z_1), f_3(Z_2), f_4(Z_2) \right) \in L^2\left(Z\right)$ and basis $\gamma\left(Z\right) = \left(\gamma_1(Z_1), \gamma_2(Z_1), \gamma_3(Z_2), \gamma_4(Z_2)\right)$. In this example, the regressors have the following expression
	\begin{equation*}
		\begin{split}
	\hat{M}_{1\ell}\left(Z_{1i}\right) & = \left(\gamma_{11}\left(Z_{1i}\right), \cdots, \gamma_{1r_1}\left(Z_{1i}\right),  \hat{\theta}_{\omega\ell}\gamma_{21}\left(Z_{1i}\right), \cdots \hat{\theta}_{\omega\ell}\gamma_{2r_2}\left(Z_{1i}\right) \right), \\
		\hat{M}_{2\ell}\left(Z_{1i}\right) & = \hat{\theta}_{\omega \ell}	\hat{M}_{1\ell}\left(Z_{1i}\right), \\
			\hat{M}_{3\ell}\left(Z_{2i}\right) & = \left(\gamma_{31}\left(Z_{2i}\right), \cdots, \gamma_{3r_3}\left(Z_{2i}\right),  \hat{\theta}_{\omega\ell}\gamma_{41}\left(Z_{2i}\right), \cdots \hat{\theta}_{\omega\ell}\gamma_{4r_4}\left(Z_{2i}\right) \right), \\
	\hat{M}_{4\ell}\left(Z_{2i}\right) & = \hat{\theta}_{\omega \ell}	\hat{M}_{3\ell}\left(Z_{2i}\right).
	\end{split}
	\end{equation*}
	As we can see from above, no conditional expectations or estimators of $\eta_0$ appear. Hence, it is not necessary to partition $I^c_{\ell}$. That is, we use all observations in $I^c_{\ell}$ to obtain an estimator of $\theta_{0\omega}$.   $\square$

	\section{Asymptotic Properties of OR-IVs}
	\label{asymptoticskappa}
	This section provides the mean square convergence rate for $\hat{\kappa}$ based on the Lasso estimator introduced above, which is fundamental for deriving the asymptotic properties of $\hat{\theta}$. We define 
	\begin{equation*}
		\begin{split}
			\hat{F}_{j\ell} & := \frac{1}{n-n_\ell} \sum_{i \notin I_\ell} f_j\left(Z_{ji}\right) \hat{M}_{j\ell}\left(Z_{ji}\right), \;\;\;\;\;\;\; F_j := \mathbb{E}\left[f_j\left(Z_j\right) M_j\left(Z_j\right)\right], \\
			\hat{G}_{j\ell} & := \frac{1}{n-n_\ell} \sum_{i \notin I_\ell} \hat{M}_{j\ell}\left(Z_{ji}\right) \hat{M}_{j\ell}\left(Z_{ji}\right)^{'}, \;\;\;\; G_j := \mathbb{E}\left[M_j\left(Z_j\right)M_j\left(Z_j\right)^{'}\right].
		\end{split}
	\end{equation*}
	Then, $\hat{\beta}_\ell$ can equivalently be written as 
	\begin{equation}
		\label{equivalentprogram}
		\hat{\beta}_\ell = \underset{\beta \in \mathbb{R}^r}{\operatorname{arg\;min}}\; \sum^J_{j=1}\left(-2\hat{F}^{'}_{j\ell}\beta - \beta^{'}\hat{G}_{j\ell}\beta\right)  + 2\lambda_n \left|\left|\beta\right|\right|_1.
	\end{equation}
	Interestingly, the previous characterization of the program is similar to the one proposed by \cite{chernozhukov2022automatic} for automatic estimation of Riesz representers (see Equation (3.7) in this paper), but with different matrices. An important difference is that $\hat{G}_j$, the estimated Gram matrix, depends on regressors  $\hat{M}_j$'s that are estimated. This also contrasts with the formulation considered by \cite{bakhitov2022automatic}.  These previous works formulate the estimation problem in terms of known objects analogous to $M_j$'s.  On the other hand, notice that we can deal with situations where the parameter of interest is not necessarily an average of some function depending on the vector of nuisance parameters, which is the type of setting considered by \cite{chernozhukov2022automatic} and \cite{bakhitov2022automatic}. In what follows, we work with the characterization given in \eqref{equivalentprogram}. We start by assuming 
	
	\begin{assumption}
		\label{aboundM}
		There are constants $c_1,\cdots,c_J$ such that
		$$
		\underset{1\leq k \leq r}{\operatorname{max}}\;
		\left|M_{jk}\left(Z_j\right)\right| \leq c_j, \;\; \text{a.s.},\;\; j=1,\cdots,J.
		$$
	\end{assumption}
	
	Since $\hat{G}_j$ depends on the estimated $\hat{M}_j(Z_j)\hat{M}_j(Z_j)^{'}$, we require a convergence rate for them  in a specific sense such that we can assure that $\hat{G}_j$ provides a good approximation to $G_j$.  Let $F_0$ be the distribution of the data $W$, then we hypothesize
	\begin{assumption}
		For all $j$, $\ell$, $k$, there exists a constant $C$, possibly depending on $j$ and $k$, such that, with probability approaching one,
		\label{aconvergenceM}
		\begin{equation}
			\label{eqaconvergenceM}
			\left|\left|\hat{M}_{j\ell k} -  M_{jk}\right|\right|_{2} \leq  C\varepsilon^M_{jn},  
		\end{equation}
		where $\varepsilon^M_{jn} \rightarrow 0$.
	\end{assumption}

Assumptions \ref{aboundM} and \ref{aconvergenceM} imply that for any $c>0$, 
	$$
	\left|\left|\hat{G}_{j\ell} - G_j\right|\right|_{\infty} = O_p\left(\varepsilon^G_{jn}\right),\;\;\;\text{where}\;\;\; \varepsilon^G_{jn} = \max\left\{\bar{\varepsilon}_n, n^c\varepsilon^M_{jn}\right\},\; \bar{\varepsilon}_n = \sqrt{\frac{\log(r)}{n}}.
	$$
Following similar arguments to the ones in	\cite{chernozhukov2022automatic} and \cite{bakhitov2022automatic}, who work on situations where regressors are known, we would get  $	\left|\left|\hat{G}_{j\ell} - G_j\right|\right|_{\infty} = O_p\left(\bar{\varepsilon}_n\right)$. However, as we deal with unknown regressors, we pay a price in terms of a slower rate for $\hat{G}_{j\ell}$.  Hence, the better the rates we can get for $\hat{M}_{j\ell k}(Z_j)$, as required by Assumption \ref{aconvergenceM}, the better the rate for $\hat{G}_{j\ell}$. Remark that  $\hat{M}_{j\ell k}(Z_j)$ takes the form of a conditional expectation and as such could be estimated by machine learning tools.  This conditional expectation depends on other unknown objects (conditional expectations given $V_i$ and $\eta_0$) that could be estimated with machine learning algorithms as well. Under conditions to be specified below, $\varepsilon^M_{jn}$ will be no faster than the slowest convergence rate among all those other unknown objects in $\hat{M}_{j\ell k}(Z_j)$. The only requirement that we will impose is that such a convergence rate has to be faster than $n^{-1/4}$ but could be slower than $n^{-1/2}$, a reasonable assumption for several machine learners; see for example,  neural networks \citep[][]{chen1999improved, schmidt2020, farrel21}, random forests \citep[][]{syrgkanis2020estimation, biau2012analysis}, Lasso \citep[][]{bickellasso, buneatsybakovwegkamp}, and boosting \citep[][]{luo2022highdimensional}. 
	
 In our discussion to follow, let $\varepsilon_n$ be a non-negative sequence that converges to zero no faster than $\varepsilon^G_{jn}$, for any $j$. We impose a sparse approximate condition on the orthogonal projection $f^{*}$, a key assumption for us. 
	\begin{assumption}
		\label{asparse}
	There exist $C > 1$ and $\bar{\beta} \in \mathbb{R}^r$ with $s$ non-zero elements such that 
		$$
		\sum^J_{j=1}\mathbb{E}\left[\left\{f^{*}_j\left(Z_j\right) - M_j(Z_j)^{'}\bar{\beta}\right\}^2\right] \leq Cs\varepsilon^2_n,
		$$
		where $s \leq C\varepsilon_n^{-2/(2\xi + 1)}$, $\xi>0$. 
	\end{assumption}
	This assumption is arguably standard in the high-dimensional literature; see \cite{buneatsybakovwegkamp}, \cite{belloni2012sparse}, \cite{belloni2016inference}, \cite{bradic2022minimax}. It controls the squared approximation error from using the linear combination $M_j^{'}\bar{\beta}$ to approximate the orthogonal projection. Note that the rate at which (the square of) this approximation error shrinks is governed by $\varepsilon_n$, which will ultimately influence the rate of OR-IVs, as we show below. Remark that Assumption \ref{asparse} does not impose that $f^{*}$ can be written as a linear combination of $s$ terms, i.e., that the orthogonal projection is strictly sparse. Instead, Assumption \ref{asparse} only requires the existence of $\bar{\beta}$ with $s$ terms such that the mean of the squared approximation errors across the $J$ elements is bounded by $Cs\varepsilon^2_n$. Moreover, the above assumption does allow the unknown identity of the elements in $M_j$ that give a good approximation, i.e., the researcher does not have to specify which elements are important, a task that will be typically hard to accomplish as we are dealing with highly complex functional objects; see \cite{bradic2022minimax}. We will see below that a very sparse approximation, with a small number of terms $s$ (and a large $\xi$), will typically lead to faster convergence rates for the OR-IVs.  For a more detailed discussion of approximation bias conditions with sparse specifications, see \cite{belloni2012sparse}.  For the remainder of this section let us drop the dependence of random elements on $\ell$ to simplify our notation.
	
	We next impose a sparse eigenvalue condition, following the Lasso literature \citep[e.g.,][]{bickellasso,bellonicherno2013, buhlmann2011statistics}: 
	
	\begin{assumption}
		\label{aeigen}
		The largest eigenvalue of $\sum^J_{j=1} G_j$ is uniformly bounded in $n$ and there are $ C, c >0$ such that, for all $\bar{s} \approx C \varepsilon^{-2}_n$,
		$$
		\phi(\bar{s}) = \inf\left\{\frac{\delta^{'}\sum^J_j G_j \delta}{\left|\left|\delta_{S_\beta}\right|\right|^2}, \;\;\; \delta \in \mathbb{R}^r \backslash \left\{0\right\}, \left|\left|\delta_{S^c_{\beta}}\right|\right|_1 \leq 3 \left|\left|\delta_{S_{\beta}}\right|\right|_1,\;\; \# S_\beta \leq \bar{s}\right\} > c.
		$$
	\end{assumption}

	Notice that the objective function in \eqref{equivalentprogram} depends on a sample counterpart of $F_j$, $\hat{F}_j$, and thus we hypothesize a convergence rate for it. 
	\begin{assumption}
		\label{aconvF}
		For all $j$, there exists a constant $C$, possibly depending on $j$, such that, with probability approaching one, $\left|\left|\hat{F}_{j\ell} - F_j\right|\right|_{\infty} \leq C\varepsilon^F_{jn}$, where $\varepsilon^F_{jn}\rightarrow 0$.
	\end{assumption}

	Finally, following \cite{chernozhukov2022automatic}, for simplicity of the analysis, we allow the Lasso regularization parameter $\lambda_n$ to shrink slightly slower than $\varepsilon_n$. In the discussion to follow, we argue that choosing a $\lambda_n$ proportional to $n^{-1/4}$ generally suffice.  We also restrict the growth of $r$ to be slower than some power of $n$.
	
	\begin{assumption}
		\label{assespslam}
		Let $\varepsilon_n = \max_j \left\{\varepsilon^G_{jn}, \varepsilon^F_{jn}\right\} $, $\varepsilon_n = o(\lambda_n)$, $\lambda_n = o(n^c\varepsilon_n)$ for all $c>0$, and there exist $C_1>0$ and $c_2 >1$ such that $r\leq C_1n^{c_2}$. 
	\end{assumption}
	
	Now, we can estate one of the main results of the paper. 
	\begin{theorem}
		\label{tboundkappa}
		Let Assumptions \ref{aboundM}-\ref{assespslam} hold. Then, for any $c>0$, 
		$$
		\left|\left|\hat{\kappa} - \kappa_0\right|\right|_{L^2(Z)} = O_p\left(n^c\varepsilon^{2\xi/(2\xi + 1)}_n\right).
		$$
	\end{theorem}
	
	Notice that the rate depends on $\xi$, which controls the degree of approximate sparsity in $f^{*}$. The larger $\xi$, i.e., smaller $s$, the faster the rate. Additionally, the rate depends on $\varepsilon_n$, which dominates the rates of $\hat{G}_j$ and  $\hat{F}_j$. The better these rates, the faster the rate for $\hat{\kappa}$. 
	
	\cite{bakhitov2022automatic}, in an endogenous setting, finds a slower rate of convergence for his estimator of the Riesz representer than that in \cite{chernozhukov2022automatic}; see Theorem 1 in \cite{bakhitov2022automatic} and the author's comments. \cite{bakhitov2022automatic} proposes a penalized GMM estimator, which can be seen as an estimator from a Lasso extension of the standard GMM program. He employs dictionaries in both $Z$ and $V$, where $V \not \subseteq Z$. Differently, in our case, as the operator $S_{\theta_0,\eta_0}S^{*}_{\theta_0,\eta_0}$ maps elements in $L^2\left(Z\right)$ into $L^2(Z)$, we only use dictionaries in the exogenous variables of the model and the regressors $M_j$ are functions of $Z$, even when $\eta_0$ is a function of variables that are not part of the conditioning ones. This enables us to use similar theoretical devices as in \cite{chernozhukov2022automatic}, despite we do allow for endogeneity. In other words, our procedure does not pay a price for dealing with situations with endogeneity. We regard this as a theoretical advantage of the way we have formulated our problem. 

		The rate for the OR-IVs that we obtained in Section \ref{asymptoticskappa} relies crucially on Assumptions \ref{aconvergenceM} and \ref{aconvF}. We next provide a set of sufficient conditions for these. Additionally, they will conveniently establish a direct relationship between convergence rates of the estimators of all the unknown objects in the estimated regressors $\hat{M}_{j\ell k}$ and the rates in Assumptions \ref{aconvergenceM} and \ref{aconvF}. Let us define
	\begin{equation*}
		\begin{split}
			R_{j}\left(W,\theta,\eta\right) & := \tilde{\nu}_j\left(Y,\theta,\eta\right)\gamma_{jk}\left(Z_j\right),
			\\ \alpha_{0j}\left(V,\theta,\eta\right) & := \mathbb{E}\left[\left.R_{j}\left(W,\theta,\eta\right) \right|V\right],\;\;\; \alpha_{0j}\left(\cdot,\theta_0,\eta_0\right) \in \Gamma\left(V\right) \subseteq L^2(V),
			\\ \sigma_{0j}\left(Z_j,\alpha_{j^{\prime}}\left(V,\bar{\theta},\bar{\eta}\right), \theta,\eta\right) & := \mathbb{E}\left[\left.\alpha_{j^{\prime}}\left(\bar{\theta},\bar{\eta},V\right)^{\prime}\tilde{\nu}_j\left(Y,\theta,\eta\right)\right|Z_j\right],\\ & \sigma_{0j}\left(\cdot,\alpha_{j^{\prime}}\left(V,\theta_0,\eta_0\right), \theta_0,\eta_0\right) \in \Gamma\left(Z_j\right)\subseteq L^2(Z_j).
		\end{split}
	\end{equation*}
	We next assume 
	
	\begin{assumption}
		\label{lipschitzcont}
		\begin{itemize}
			\item[i)] With probability approaching one, $\hat{\theta}_{A_\ell}$, $\hat{\theta}_{B_\ell} \in \Theta$, $\hat{\eta}_{A_\ell}$, $\hat{\eta}_{B_\ell} \in \Xi$, $\hat{\alpha}_{jB_\ell}\left(\cdot,\hat{\theta}_{A_\ell}, \hat{\eta}_{A_\ell}\right) \in \Gamma(V)$, and $\hat{\sigma}_j\left(\cdot,\hat{\alpha}_{jB_\ell}\left(V,\hat{\theta}_{A_\ell}, \hat{\eta}_{A_\ell}\right),\hat{\theta}_{B_\ell}, \hat{\eta}_{B_\ell}\right) \in \Gamma(Z_j)$, for all $j$;
			\item[ii)] 	With probability approaching one, for all $k$, $\left|\left|\hat{M}_{j\ell k} - M_{jk} \right|\right|_2$ is bounded by one of the following: $\left|\left|\hat{\theta}_{A_\ell} - \theta_0\right|\right|$, $\left|\left|\hat{\theta}_{B_\ell} - \theta_0\right|\right|$, $\left|\left|\hat{\eta}_{A_\ell} - \eta_0 \right|\right|_{\Xi}$, $\left|\left|\hat{\eta}_{B_\ell} - \eta_0 \right|\right|_{\Xi}$, $\left|\left|\hat{\alpha}_{jB_\ell}\left(\cdot,\hat{\theta}_{A_\ell}, \hat{\eta}_{A_\ell}\right)-\alpha_{0j}(\cdot,\hat{\theta}_{A_\ell}, \hat{\eta}_{A_\ell})\right|\right|_{2}$, \\ $\left|\left|\hat{\sigma}_{C_\ell}\left(\cdot,\hat{\alpha}_{jB_\ell}\left(V,\hat{\theta}_{A_\ell}, \hat{\eta}_{A_\ell}\right),\hat{\theta}_{B_\ell}, \hat{\eta}_{B_\ell}\right) - \sigma_{0j}\left(\cdot,\hat{\alpha}_{jB_\ell}\left(V,\hat{\theta}_{A_\ell}, \hat{\eta}_{A_\ell}\right),\hat{\theta}_{B_\ell}, \hat{\eta}_{B_\ell}\right)\right|\right|_{2}$. Additionally, these are all bounded by $Cn^{-d_j}$, where $0<d_j<1/2$, $j = 1, \cdots,J$, and $C$ possibly depends on $j$.
		\end{itemize}
	\end{assumption}

	Assumption \ref{lipschitzcont} \textit{i)} is a standard high-level condition that will be satisfied by several estimators. The key condition is \textit{ii)}. This assumption requires Lipschitz continuity of $M_{jk}$ in their unknown components. These are $\theta_0$, $\eta_0$, and the conditional expectations given $V$ and $Z_j$. Similar conditions to \textit{ii)} are not unusual in the semi-parametric literature while dealing with non-linear functions; see, e.g., Assumption 3.9 in  \cite{ai2003efficient} and Assumption 12 in \cite{chernozhukov2022automatic}. This condition is convenient as we can obtain a connection between rates of convergence for their estimators and the rate in Assumptions \ref{aconvergenceM} and \ref{aconvF}. When machine learning approaches are employed to compute estimators for $\eta_0$ (when $\eta_0$ is, in turn, a vector of conditional expectations), $\alpha_{0j}\left(V,\theta,\eta\right)$, and $\sigma_{0j}\left(Z_j,\alpha_{j^{\prime}}\left(V,\bar{\theta},\bar{\eta}\right), \theta,\eta\right)$, their rates are known (typically in mean square norm).  Notice that, the construction of an estimator of $\theta_0$ that appears in $M_{jk}(Z_j)$ will be typically based on non-orthogonal moments, depending on machining learning estimators of $\eta_0$. Then, under some smoothness conditions (e.g., continuity of the non-orthogonal moments in $\eta$ around $\eta_0$) we can obtain that $\left|\left|\hat{\theta}-\theta_0\right|\right| = O_p\left(\left|\left|\hat{\eta}-\eta_0\right|\right|_{\Xi}\right)$. Hence we will be able to obtain bounds in probability for all the terms above. Assumption \ref{lipschitzcont} implies then, for any $k$,
	
	\begin{equation*}
		\begin{split}
		\left|\left|\hat{M}_{j\ell k} -  M_{jk}\right|\right|_{2} & = \left|\left|\hat{\sigma}_{jC_\ell}\left(\cdot, \hat{\alpha}_{jB_\ell}\left(\cdot,\hat{\theta}_{A_\ell},\hat{\eta}_{A_\ell}, \hat{\theta}_{B_\ell}, \hat{\eta}_{B_\ell}\right)\right) - \sigma_{0}\left(\cdot, \alpha_{0j}\left(\cdot,\theta_{0},\eta_{0}, \theta_{0}, \eta_{0}\right)\right)\right|\right|_2 \\ & =  O_p\left(n^{-d_j}\right).	
		\end{split}
	\end{equation*}
   Therefore, $\varepsilon^M_{jn} = n^{-d_j}$ in Assumption \ref{aconvergenceM}. By a boundness condition, we can also determine a specific rate for $\hat{F}_{j\ell}$.
   
   \begin{lemma}
   	\label{lFconvg}
   	Assume there are constants $c_1,\cdots,c_J$ such that, with probability approaching one,
   	$$
   	\underset{1\leq k \leq r}{\operatorname{max}}\;
   	\left|\hat{M}_{j k}\left(Z_j\right)\right| \leq c_j,\;\; \text{a.s.}, 
   	$$
   Additionally, let Assumption \ref{lipschitzcont} hold and $r<C_1n^{c_2}$, for some $C_1>0$ and $c_2 > 1$. Then, 
   	$$
   	\left|\left|\hat{F}_{j\ell} - F_j\right|\right|_{\infty} = O_p\left(n^{-d_j}\right).
   	$$	
   \end{lemma}
	
	\section{Estimation of the Parameter of Interest in a Two-Step Setting}
	\label{estimationparint}
	
	 Up to now, we have allowed the functions $m_j$'s to depend on $\eta_0$ and $\theta_0$ in a general fashion. As we have shown, our construction of debiased moments can handle it. Yet, we are interested in the common situation where the researcher works within a two-step setting, in which there are functions $m_j$'s that depend on one component of $\eta_{0}$ only. That is,  $m_j\left(y,\theta,\eta\right) \equiv m_j\left(y,\eta_j\right)$ for some $j$. CMRs based on those residual functions can be used to obtain an estimator of $\eta_0$, as it occurs in our example. 
	 
	 We focus on this case as many relevant scenarios in applied work present this feature \citep[see, e.g.,][Section 5 and references therein]{chen2016methods}.  Furthermore, a two-step setting might be computationally attractive in practice while estimating $\theta_0$, as pointed out by \cite{ackerberg2014asymptotic}, as it avoids a simultaneous search over infinite and finite dimensional spaces in settings where $\eta_0$ and $\theta_0$ are jointly estimated by minimizing certain objective function. Our two-step setting can be written as 
	 \begin{align*}
	 	\mathbb{E}\left[\left.m_j\left(Y,\eta_{0j}\right)\right|Z_j\right] & = 0, \;\;\; \text{for all}\;\; j \in \mathcal{J}_1, \\
	 	\mathbb{E}\left[\left.m_j\left(Y,\theta_0,\eta_{0}\right)\right|Z_j\right] & = 0, \;\;\; \text{for all}\;\; j \in \mathcal{J}_2, 
	 \end{align*}
	where $\mathcal{J}_1$ and  $\mathcal{J}_1$ are disjoint sets of indexes and $J = |\mathcal{J}_1| + |\mathcal{J}_2|$. 
	
	Recall from our previous discussion that for a given instrument $f \in L^2(Z)$, we obtain an OR-IV $\kappa_{0}(Z)$---a vector with $J$ entries. Then, for different choices of $f$'s, say $q$ of them, we can construct $J$ vectors $\bm{\kappa_{0j}(Z_j)}$, of dimension $q$, which collects the $j-$component of the $q$ OR-IVs so constructed. We use the bold notation to emphasize that $\bm{\kappa_{0j}(Z_j)} = \bm{f_j(Z_j)} - \bm{f^{*}_j(Z_j)}$ is a $q-$vector.  For the remainder of this paper, let us re-define \eqref{lrmomenteq} such that 
	$$
	\psi\left(W, \theta, \eta, \bm{\kappa}\right) = \sum^J_{j=1}m_j\left(Y_i,\theta,\eta\right)\bm{\kappa_j(Z_j)},
	$$
	where we should notice that $\psi$ is a now a $q-$vector of LR functions. Even though we keep the notation that indicates that each $m_j$ depends on both parameters $\theta$ and $\eta$, $m_j\left(y,\theta,\eta\right) \equiv m_j\left(y,\eta_j\right)$ whenever $j \in \mathcal{J}_1$.  Let $\hat{\eta}_\ell$ be an estimator of $\eta_0$, using observations in $I^c_{\ell}$. Let
	$$
	\hat{\psi}\left(\theta\right) = \frac{1}{n} \sum^L_{\ell = 1} \sum_{i \in I_\ell} \psi\left(W_i,\theta,\hat{\eta}_\ell, \hat{\bm{\kappa}}_\ell\right).
	$$
	Our proposed estimator $\hat{\theta}$ is defined as the solution to the GMM program
	\begin{equation}
		\label{estdef}
		\hat{\theta} = \underset{\theta \in \Theta}{\operatorname{arg\;min}}\; \hat{\psi}\left(\theta\right)^{\prime} \hat{\Lambda} \hat{\psi}\left(\theta\right), 
	\end{equation}
	where $\hat{\Lambda}$ is a positive semi-definite symmetric weighting matrix. A choice that asymptotically minimizes the asymptotic variance is $\hat{\Lambda} = \hat{\Psi}^{-1}$, where 
	$$
	\hat{\Psi} = \frac{1}{n} \sum^L_{\ell = 1} \sum_{i \in I_\ell} \hat{\psi}_{i\ell} \hat{\psi}^{\prime}_{i\ell}, \;\;\;\;  \hat{\psi}_{i\ell} \equiv \psi\left(W_i,\tilde{\theta}_\ell,\hat{\eta}_\ell, \hat{\bm{\kappa}}_\ell\right),
	$$
	and $\tilde{\theta}_\ell$ is a possibly non-LR estimator of $\theta_0$, based on observations in $I^c_{\ell}$. The estimated $\hat{\psi}_{i\ell}$ directly accounts for estimating $\eta_0$ and $\kappa_0$ in a previous stage. We refer to \eqref{estdef} as our \textit{Debiased GMM Estimator (DGMM)}. Let us summarize our estimation procedure with the following steps: 
	
	\bigskip \noindent \textbf{Step 1:} For each subsample $\ell = 1,\cdots,L$ , compute estimates $\hat{\eta}_\ell$ and $\hat{\kappa}_{\ell}$, using observations not in $I_\ell$.

	\bigskip	\noindent \textbf{Step 2:} Obtain our estimator $\hat{\theta}$ by means of \eqref{estdef}. The estimator of the asymptotic variance, which accounts for the estimation of the previous objects, takes the ``sandwich" form 
	
	\begin{equation}
		\label{varianceest}
		\hat{\Sigma} = \left(\hat{\Upsilon}^{\prime}\hat{\Lambda}\hat{\Upsilon}\right)^{-1}\hat{\Upsilon}^{\prime}\hat{\Lambda}\hat{\Psi}\hat{\Lambda} \hat{\Upsilon}\left(\hat{\Upsilon}^{\prime}\hat{\Lambda}\hat{\Upsilon}\right)^{-1}, \;\;\; \hat{\Upsilon} = \frac{\partial}{ \partial \theta} \hat{\psi}(\hat{\theta}).
	\end{equation}

\begin{figure}[H]%
	\centering
	\caption{Summary of Steps for Inference on Structural Parameters}
	\label{fig:diagram}
	\fbox{\includegraphics[height=0.8\textwidth]{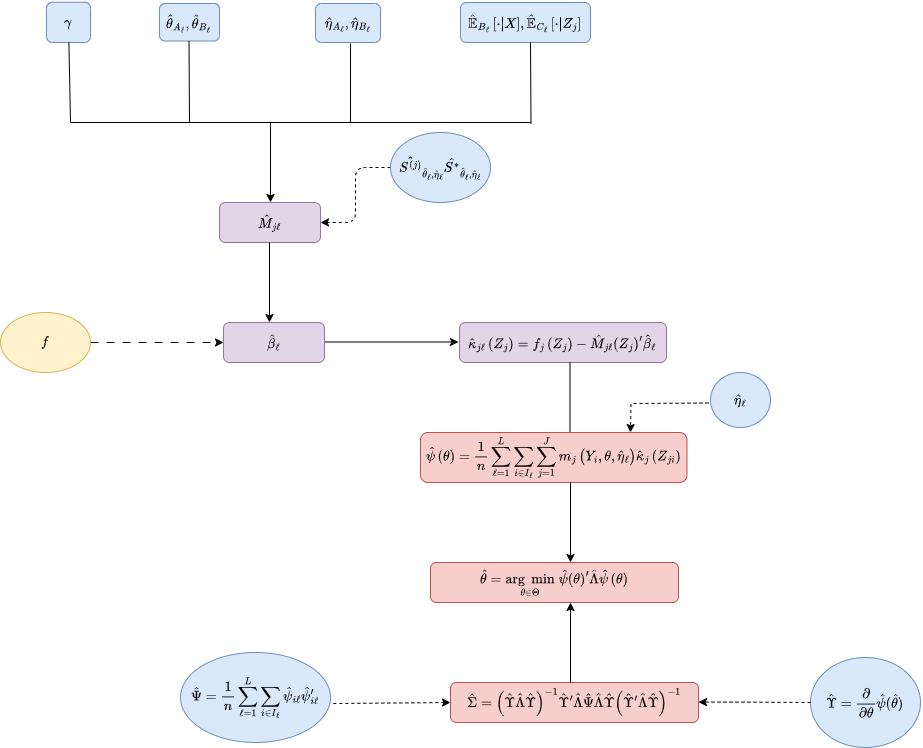}}%
\end{figure}

\bigskip Figure \ref{fig:diagram} provides a diagram that summarizes all the steps and important objects to conduct inference on structural parameters, as proposed by this paper. 

	\bigskip Before concluding this section, we would like to remark that $\eta_0$ belongs to the infinite-dimensional space $\Xi$. As such, any suitable machine learning procedure can be used to obtain an estimator  $\hat{\eta}$ by exploiting the CMRs that depend on each $\eta_{0j}$. To derive the asymptotic properties of $\hat{\theta}$, some restrictions will need to be imposed on the convergence rate of $\hat{\eta}$. Hence, it is substantial to know the behavior of such a rate asymptotically. A leading case in the semiparametric literature is when $\eta_0$ collects conditional expectations, given some $Z_j^{'}$s (as in our example). In this context, the mean-square convergence rate of $\hat{\eta}$ has been obtained for a variety of machine learners (see references in Section \ref{asymptoticskappa}). We also consider the more general case where $\eta_0$ is a function of variables different from the conditioning ones, allowing for endogeneity. In the endogenous setting, identification of $\eta_0$ is trickier and typically involves an ill-posed problem. In particular, identification of this parameter might require the so-called ``completeness condition"  \citep[][]{newey2003instrumental}. The ill-posedness results in slower rates for the mean square norm of estimators of $\eta_0$ than in the exogenous situation. However, if rates are obtained for the so-called projected mean square norm, for example, they typically improved \citep[see][]{bakhitov2022automatic, gold2020inference, singh2019kernel}. The vector $\eta_0$ might include densities as well.  In any case, we will assume that a suitable norm $\left|\left|\cdot\right|\right|_{\Xi}$ is defined in $\Xi$, and as we will show below, we will allow the rate for $\hat{\eta}$,  to be faster than $n^{-1/4}$ and slower than $n^{-1/2}$, which is arguably plausible for machine learning first steps.

	\section{Asymptotic Properties of DGMM}
	\label{apparameterinterest}
	
 We start by imposing some regularity conditions:\footnote{Recall that $\left|\left|\cdot\right|\right|$ denotes Euclidean norm.} 
	
	\begin{assumption}
		\label{contozero} For all $j$ and $\ell$, $\mathbb{E}\left[\left|\left|\psi\left(W,\theta_0,\eta_0,\bm{\kappa_0}\right)\right|\right|^2\right]$, $\left|\left|\bm{\kappa_{0j}(Z_j)}\right|\right| < \infty$ a.s., and 
		\begin{itemize}
			\item[i)] $\int \left|m_j\left(y,\theta_0,\hat{\eta}_{\ell}\right) - m_j\left(y,\theta_0,\eta_{0}\right)\right|^2 F_0\left(dw\right) \overset{p}{\to} 0$;
				\item[ii)] $\int \left|m_j\left(y,\theta_0,\hat{\eta}_{\ell}\right) - m_j\left(y,\theta_0,\eta_{0}\right)\right|^2 \left|\left|\bm{\kappa_{0j}(z_j)}\right|\right|^2 F_0\left(dw\right) \overset{p}{\to} 0$;
			\item[iii)] $\int \left|m_j\left(y,\theta_0,\eta_{0}\right)\right|^2 \left|\left|\bm{\hat{\kappa}_{j\ell}(z_j)} - \bm{\kappa_{0j}(z_j)}\right|\right|^2 F_0\left(dw\right)\overset{p}{\to} 0$.
		\end{itemize}
	\end{assumption}
	Assumptions \ref{contozero} \textit{i)} and \textit{ii)} are mean square convergence conditions for $\hat{\eta}$, while \textit{iii)} is a convergence condition for $\bm{\hat{\kappa}}$. Sufficient conditions for \textit{i)} and \textit{ii)} might involve continuity of $m_j\left(y, \theta_0,\cdot\right)$, for instance. Moreover, note that \textit{iii)} is implied, e.g., by $\left|m_j\left(Y,\theta_0,\eta_{0j}\right)\right|^2$ being bounded a.s., for all $j$. 
	
	We need to assure that the estimators of OR-IVs and $\eta_0$ are well-defined, for this, we impose
	\begin{assumption}
		\label{estspaces}
	For all $j$ and $\ell$, with probability approaching one,  $\hat{M}_{j\ell}\left(Z_{ji}\right)^{'}\hat{\beta}_\ell$ belongs to $L^2\left(Z_j\right)$, $\left|\left|\bm{\hat{\kappa}_{j\ell}(Z_j)}\right|\right| < \infty$ a.s., and $\hat{\eta}_{\ell} \in \Xi$. 
	\end{assumption}

Also, we hypothesize a rate of convergence for the estimator $\hat{\eta}$, which is required to be faster than $n^{-1/4}$ only, and the same requirement applies for our estimators of OR-IVs, as the following assumption states:
	\begin{assumption}
		\label{rates}
		\textit{i)} $\left|\left|\hat{\eta}_\ell - \eta_0 \right| \right|_{\Xi} = O_p\left(n^{-d_\eta}\right)$;
		\textit{ii)} $\frac{2\xi d + d_\eta(2\xi + 1)}{2\xi + 1} > \frac{1}{2}$, where $d = \min_j d_j$, $1/4 < d_\eta < 1/2$ and $1/4 < d < 1/2$.
	\end{assumption}
	
Lemma \ref{lFconvg} and Theorem \ref{tboundkappa} imply that the rate of the OR-IVs is as close as desired to $n^{-2\xi d/(2\xi + 1)}$. Hence, Assumption \ref{rates} is sufficient to assure that $\sqrt{n}\left|\left|\hat{\kappa}- \kappa_0\right|\right|_{L^2(Z)}\left|\left|\hat{\eta} - \eta_0\right|\right|_{\Xi} \rightarrow 0$, which is an essential condition for $\sqrt{n}-$consistency of $\hat{\theta}$.

 Unsurprisingly, this depends on the rate of the estimated regressors in our Lasso program. This contrasts with earlier related works where the conditions for $\sqrt{n}$-consistency of $\hat{\theta}$ depends on the rate of consistency of $\hat{\eta}$ only and thus $d = d_\eta$ (see, e.g., Theorem 9 in \cite{chernozhukov2022automatic}). The reason for this is that the regressors in our Lasso program, captured by $\hat{M}_{j\ell k}$, depends on unknown objects that must be estimated. In contrast, the design matrix in previous works, which have also based estimation on a Lasso-type program, depend on regressors that are known. In particular, they do not depend on $\theta_0$, $\eta_0$, or unknown conditional expectations. As a result, our proofs need to account for the presence of such unknown objects and the possibility that $d \leq d_\eta$ in obtaining a rate for OR-IVs. The practical implication of this result is that to achieve $\sqrt{n}$-consistency, in our context, we might need larger sparseness (i.e., larger $\xi$) than required by these previous papers. Nonetheless, we still allow for $d_\eta$ and $d$ being smaller than $1/2$ simultaneously.  It is important to recall that Theorem \ref{tboundkappa} requires that $\lambda_n$, the regularization parameter, to shrink slower than $\varepsilon_n$. Since $d > 1/4$, choosing a $\lambda_n$ proportional to $n^{-1/4}$ will generally suffice to achieve this requirement, as we specify in our algorithm of Section \ref{estimationoivs}.\footnote{Under our assumptions above, we let $\varepsilon_n = n^cn^{-d}$, for any $c>0$. As $c$ is sufficiently close to zero, the rate of $\varepsilon_n$ goes to zero at a rate sufficiently close to $n^{-d}$, which is faster than $n^{-1/4}$. On the other hand, $\lambda_n$ goes to zero at a rate no faster than $n^{-1/4}$.}
	
	Next, let us define the sum of product of residuals 
	$$
	\hat{\Delta}_\ell(w) := \sum^J_{j=1}\left(m_j\left(y,\theta_0,\hat{\eta}_{\ell}\right) - m_j\left(y,\theta_0,\eta_{0}\right)\right)\left(\bm{\hat{\kappa}_{j\ell}(z_j)} - \bm{\kappa_{0j}(z_j)}\right).
	$$
	 Assumptions \ref{contozero}, \ref{estspaces}, \ref{rates}, and Fréchet differentiability of the CMRs imply 
	\begin{equation}
		\label{deltadisplay}
		i)\; \int \left|\left|\hat{\Delta}_\ell(w)\right|\right|^2 F_0(dw) \overset{p}{\to} 0,  \;\;\;\; and \;\;\;\; ii)\; \sqrt{n}\int \hat{\Delta}_\ell(w) F_0(dw) \overset{p}{\to} 0.
	\end{equation}
	Expression \eqref{deltadisplay} is a $\sqrt{n}-$convergence result for object $\hat{\Delta}_\ell$, which will be key to derive the asymptotic properties of $\hat{\theta}$. Additionally, let 
	$$
	\overline{\psi}\left(\theta,\eta,\bm{\kappa}\right) = \mathbb{E}\left[\psi\left(W,\theta,\eta,\bm{\kappa}\right)\right].
	$$
	
	\begin{assumption}
		\label{twicefre}
		$\overline{\psi}\left(\theta_0,\eta,\bm{\kappa_0}\right)$ is twice continuously Fréchet differentiable in a neighborhood of $\eta_0$.  
	\end{assumption}
	\noindent Then, it can be shown that since $\psi$ leads to a debiased moment, there exists a $C >0$ such that  
	$$
	\left|\left|\overline{\psi}\left(\theta_0,\eta,\bm{\kappa_0}\right)\right|\right| \leq C \left|\left|\eta - \eta_0 \right| \right|_{\Xi}^2.
	$$
	All the previous conditions yield the most important result of this section:
	
	\begin{lemma}
		\label{keylemma1}
		Let Assumptions \ref{gateaux}, \ref{contozero}-\ref{twicefre} hold. In addition, let the assumptions in Lemma \ref{lFconvg} hold.  Then, 
		\begin{equation}
			\label{keyeq1}
			\sqrt{n} \hat{\psi}(\theta_0) = \frac{1}{\sqrt{n}} \sum^n_{i=1} \psi\left(W_i,\theta_0,\eta_0,\bm{\kappa_0}\right) + o_p(1).
		\end{equation}
	\end{lemma}
 Interestingly,  cross-fitting enables to show \eqref{keyeq1} in a simple manner, without the need to impose the so-called Donsker conditions for $\eta_0$, as discussed by \cite{chernozhukov2018double} and CEINR. Avoiding Donsker conditions is important as it is unknown if machine learners satisfy them. Furthermore, they could be restrictive in high-dimensional contexts.
	
	The next condition is relevant to show consistency of $\hat{\Psi}$, the sample second moments of the debiased functions, which appears in the asymptotic variance of our GMM estimator. 
	
	\begin{assumption}
		\label{convmall}
		$\int \left|m_j\left(y,\tilde{\theta}_\ell, \hat{\eta}_{\ell}\right) - m_j\left(y,\theta_0, \hat{\eta}_{\ell}\right)\right|^2 \left|\left|\bm{\hat{\kappa}_{j\ell}(z_j)}\right|\right|^2 F_0(dw) \overset{p}{\to} 0$.
	\end{assumption}

	Observe that the previous assumption is trivially satisfied for those $m_j$ depending on $\eta_j$ only, with probability approaching one. Finally, as in any GMM setting, we need conditions for convergence of the Jacobian: $\frac{\partial}{\partial \theta} \hat{\psi}(\bar{\theta}) \overset{p}{\to} \Upsilon = \mathbb{E}\left[\frac{\partial}{\partial \theta} \psi\left(W,\theta_0,\eta_0,\bm{\kappa_0}\right)\right]$ for any $\bar{\theta} \overset{p}{\to} 
	\theta_0$. Let $\bm{\kappa_0(Z)}$ be a $q \times J$ matrix containing all OR-IVs $\bm{\kappa_{0j}(Z_j)}$, $j = 1, \cdots, J$. A similar definition applies to $\bm{\hat{\kappa}_{\ell}(Z)}$.
	
	\begin{assumption}
		\label{ajacobian}
		$\Upsilon$ exists and there is a neighborhood $\mathcal{N}$ of $\theta_0$  such that for all $\ell$
		\begin{itemize}
			\item[i)]$\left|\left|\hat{\eta}_\ell - \eta_0 \right| \right|_{\Xi}  \int \left|\left|\bm{\hat{\kappa}_\ell(z)} - \bm{\kappa_0(z)} \right| \right|^2_{\infty}dF_0(dz)  \overset{p}{\to} 0$;
			\item[ii)] For $\left|\left|\eta - \eta_0 \right| \right|_{\Xi} \mathbb{E}\left[\left|\left|\bm{\kappa(Z)} - \bm{\kappa_0(Z)} \right| \right|^2_\mathcal{\infty}\right] $ small enough, $\psi\left(W,\theta,\eta,\bm{\kappa}\right)$ is differentiable  in $\theta$ on $\mathcal{N}$ and there are $C_1 >0$,  $C_2>0$, and $d\left(W,\eta,\bm{\kappa}\right)$ such that for $\theta \in \mathcal{N}$ 
			$$
			\left|\left|\frac{\partial\psi\left(W,\theta,\eta,\bm{\kappa}\right)}{\partial \theta} - \frac{\partial\psi\left(W,\theta_0,\eta,\bm{\kappa}\right)}{\partial \theta}\right|\right| \leq d\left(W,\eta,\bm{\kappa}\right) \left|\left|\theta - \theta_0\right|\right|^{C_1};\;\;\; \mathbb{E}\left[d\left(W,\eta,\bm{\kappa}\right)\right] < C_2;
			$$
			\item[iii)] For each s and k, $\int \left|\frac{\partial \psi_s\left(w,\theta_0,\hat{\eta}_\ell,\bm{\hat{\kappa}_\ell}\right)}{\partial \theta_k} - \frac{\partial \psi_s\left(w,\theta_0,\eta_0,\bm{\kappa_0}\right)}{\partial \theta_k}\right| F_0(dw) \overset{p}{\to} 0.$
		\end{itemize}  
	\end{assumption}
	
	Given the previous assumptions and findings, the following result shows the asymptotic normality of $\hat{\theta}$: 
	\begin{theorem}
		\label{inferencetheta}
		Let Assumptions \ref{gateaux}, \ref{contozero}-\ref{convmall} hold.  In addition, let the assumptions in Lemma \ref{lFconvg} hold. Also, let $\hat{\theta} \overset{p}{\to} \theta_0$, $\hat{\Lambda} \overset{p}{\to} \Lambda$, and $\Upsilon^{'} \Lambda \Upsilon$ be non-singular. Then,
		$$
		\sqrt{n}\left(\hat{\theta} - \theta_0\right) \overset{d}{\to}\;\;N\left(0,\Sigma\right),\;\;\; \Sigma = \left(\Upsilon^{'}\Lambda \Upsilon\right)^{-1} \Upsilon^{'}\Lambda \Psi \Lambda \Upsilon\left(\Upsilon^{'}\Lambda \Upsilon\right)^{-1}.
		$$
		If Assumption \ref{ajacobian} also holds, then $\hat{\Sigma} \overset{p}{\to} \Sigma$.
	\end{theorem}
	
	Theorem \ref{inferencetheta} implies that confidence intervals for $\hat{\theta}$ can be obtained straightforwardly in a standard way, using $\hat{\Sigma}$, and the usual quantiles of the standard normal distribution. This holds despite the convergence rates of nuisance parameters being slower than $\sqrt{n}$. Note that Theorem \ref{inferencetheta} relies on the consistency of $\hat{\theta}$. We provide sufficient conditions for this in Section \ref{consistency} of the Supplementary Appendix. 
	
	\section{Monte Carlo}
	\label{montecarlo}
	
 We have conducted several Monte Carlo experiments based on our example, the production function setting, by simulating data from a panel of $n$ firms observed over $T$ periods. Only for this section, let us distinguish between variables in logs and in levels: uppercase letters denote variables in logs (as before), while lowercase letters stand for variables in levels. 

Our baseline data generating process (DGP 1) is inspired by that in \cite{ackerberg2014asymptotic}, Section 4.3.  In our experiments, firms are observed over three periods, i.e., $T=3$. We consider a Cobb-Douglas production function in logs with only capital input: 
	$$
	Y_{it} = \theta_{01} + \theta_{0k}K_{it} + \omega_{it} + \epsilon_{it},
	$$
	where $\theta_{01} = 0$ and $\theta_{0k} = 1$. The law of motion of capital is given by 
	$$
	k_{it} = \left(1-\delta\right)k_{i,t-1} + \mu_{it}i_{i,t-1},
	$$
	where $1-\delta = 0.9$,  $\mu_{it}$ is a lognormal shock to the capital accumulation process (with the underlying normal distribution having a mean and standard deviation of 1),  and $i_{it}$ is the firm's investment decision. The log of investment is determined by
	\begin{equation}
		\label{invmc}	
		I_{it} = \gamma_0 +\gamma_1K_{it} + \gamma_2\omega_{it} + \exp\left(-0.5K_{it} + 0.5\omega_{it}\right),
	\end{equation}
	where $\gamma_0 = 0$, $\gamma_1 = -0.7$, and $\gamma_2 = 5$. We consider a large value for $\gamma_2$ to exacerbate the endogeneity bias due to the correlation between inputs and the anticipated productivity. Note that \eqref{invmc} implies that $\omega_{it} = \omega_{t}\left(I_{it},K_{it}\right)$.  We have specified an admittedly ad-hoc process for $I_{it}$ for two reasons. First, this avoids solving potentially complex firms' dynamic programs. Second, this process gives the link between productivity and the proxy variable ($I_{it}$ in this case), which is the source of the non-parametric component of the model. We intentionally model it flexibly and nonlinearly to justify the use of sophisticated machine learning tools.  
	
	One essential assumption of the proxy variable approach is the correct specification of the link between productivity and the proxy variable in the sense that $\omega_{it}$ is assumed to be a function of $K_{it}$ and $I_{it}$ only; see the discussion by \cite{ackerberg2015identification}. In other words, the proxy variable approach rules out the possibility that other unobserved variables or shocks influence $\omega_{it}$. If anticipated productivity is governed by other factors, this is problematic.  Controlling only for $I_{it}$ and $K_{it}$ in the estimation of $\eta_{0t}$ could lead to substantial bias in the first stage, resulting in poor estimation of the parameter of interest, both in terms of bias and coverage. As this cannot be tested in practice, if this situation takes place, one should prefer a less sensitive estimator to the first stage, i.e., a debiased estimator, rather than one that is not. This is what we aim to assess with our Monte Carlo simulation.  We have considered two additional data generating processes where we add a normal shock to Equation \eqref{invmc}, with mean zero and standard deviations of 0.5 (DGP 2), and 0.7 (DGP 3).\footnote{Note that in this situation both estimators are not necessarily consistent, and thus it is a deeply disadvantageous scenario for both approaches.}

	In all cases, productivity is assumed to follow a normal AR(1) process with persistence parameter $\theta_{0\omega} = 0.7$. The variance of the innovation term in this process is specified such that the standard deviation of $\omega_{it}$ is $\sigma_\omega = 0.1$. The unanticipated productivity or measurement error in output is normal and iid over firms and time.  To ensure that our DGPs do not depend on the starting values of the variables, we focus on data drawn from the steady-state distribution implied by the model. Specifically, we have simulated the data over one hundred periods and kept the last three.\footnote{The same idea has been followed by \cite{ackerberg2014asymptotic} and \cite{ackerberg2015identification}.} 
	
 Let us recall that in this case, 
 		\begin{align}
 		\mathbb{E}\left[\left.Y_{1} - \eta_{01}\left(I_{1}, K_{1}\right)\right|I_{1}, K_{1}\right] & = 0, \label{prod21} \\ \mathbb{E}\left[\left.Y_{2} - F\left(K_{2},\theta_{0p}\right) - \theta_{0\omega}\left(\eta_{01}\left(I_1,K_1\right) - F\left(K_1,\theta_{0p}\right)\right)\right|\Omega_{1}\right] & = 0, \label{prod22} \\  \mathbb{E}\left[\left.Y_{2} - \eta_{02}\left(I_{2}, K_{2}\right)\right|I_{2}, K_{2}\right] & = 0, \label{prod23} \\ \mathbb{E}\left[\left.Y_{3} - F\left(K_{3},\theta_{0p}\right) - \theta_{0\omega}\left(\eta_{02}\left(I_2,K_2\right) - F\left(K_2,\theta_{0p}\right)\right)\right|\Omega_{2}\right] & = 0\label{prod24},
 	\end{align}
 where $F$ is Cobb-Douglas. Furthermore, an LR moment is simply 
 	\begin{equation*}
 	\begin{split}
 		\psi\left(W,\theta_0,\eta_0, \kappa_0\right)  = &  \left(Y_1 - \eta_{01}\left(Z_1\right)\right)\kappa_{01}\left(Z_1\right) + \left(Y_2 - F\left(K_2,\theta_{0p}\right) - \theta_{0\omega}\left(\eta_{01}\left(Z_1\right) - F\left(K_1,\theta_{0p}\right)\right)\right)\kappa_{02}\left(Z_1\right) \\ & + \left(Y_2 - \eta_{02}\left(Z_2\right)\right)\kappa_{03}\left(Z_2\right) + \left(Y_3 - F\left(K_3,\theta_{0p}\right) - \theta_{0\omega}\left(\eta_{02}\left(Z_2\right) - F\left(K_2,\theta_{0p}\right)\right)\right)\kappa_{04}\left(Z_2\right),
 	\end{split}
 \end{equation*}
 where $Z_t = \left(I_t,K_t\right)$. 
 
 We now comment on our estimation strategy. Recall that our algorithm begins with the choice of a vector of functions depending on the conditioning variables in the CMRs \eqref{prod21}-\eqref{prod24}. Since this example has four CMRs, a initial vector function must have four entries.  A given $f(Z)$ will allow us to obtain one debiased moment.  In our simulations, we automatically construct four of these moments, and thus we specify four vectors of functions:
	\begin{equation*}
		\begin{split}
			f_1(Z) & = \left(K_{1}, K_{1}, K_{2}, K_{2}\right)^{\prime},  \\ 
			f_2(Z) & = \left(I_{1}, I_{1}, I_{2}, I_{2}\right)^{\prime}, \\
			f_3(Z) & = \left(K_{1}, K_{1}, I_{2}, I_{2}\right)^{\prime}, \\
			f_4(Z) & = \left(K_{1}, I_{1}, I_{2}, I_{2}\right)^{\prime}.
		\end{split} 
	\end{equation*}
	Each of these vectors is converted into an OR-IV to be used in estimation. For simplicity, in our experiments, the bases coincide and they are equal to $\gamma_1\left(I_{t}, K_{t}\right)$, which is obtained from the tensor product of univariate exponential bases in $I_{it}$ and $K_{it}$, that is 
	\begin{equation*}
		\begin{split}
			\gamma_1\left(I_{t}, K_{t}\right) =  & \left(\tilde{\gamma}_1(K_{t}) \times \tilde{\gamma}_1(I_{t}), \tilde{\gamma}_1(K_{t}) \times \tilde{\gamma}_2(I_{t}), \cdots, \tilde{\gamma}_{\sqrt{n}/5}(K_{t}) \times \tilde{\gamma}_{\sqrt{n}/5}(I_{t}),\right)^{\prime},
		\end{split}
	\end{equation*}
	where $\tilde{\gamma}_s$ denotes the $s-$term of the sequence of exponential bases.\footnote{A one-dimensional exponential basis is of the form $\tilde{\gamma}_k(V) = \exp\left(\alpha_k V\right)$, each with different rate parameter $\alpha_k$.  For a discussion of the use of bases with R, we refer the reader to \cite{functionaldataR}.} Hence, $r_1$, the dimension of $\gamma_1$, is $r_1 = n/25$. Recall that we have to construct the regressors: 
		\begin{equation*}
		\begin{split}
			\hat{M}_{1\ell}\left(Z_{1i}\right) & = \left(\gamma_{11}\left(Z_{1i}\right), \cdots, \gamma_{1r_1}\left(Z_{1i}\right),  \hat{\theta}_{\omega\ell}\gamma_{11}\left(Z_{1i}\right), \cdots \hat{\theta}_{\omega\ell}\gamma_{1r_1}\left(Z_{1i}\right) \right), \\
			\hat{M}_{2\ell}\left(Z_{1i}\right) & = \hat{\theta}_{\omega \ell} 	\hat{M}_{1\ell}\left(Z_{1i}\right), \\
			\hat{M}_{3\ell}\left(Z_{2i}\right) & = \left(\gamma_{11}\left(Z_{2i}\right), \cdots, \gamma_{1r_1}\left(Z_{2i}\right),  \hat{\theta}_{\omega\ell}\gamma_{11}\left(Z_{2i}\right), \cdots \hat{\theta}_{\omega\ell}\gamma_{1r_1}\left(Z_{2i}\right) \right), \\
			\hat{M}_{4\ell}\left(Z_{2i}\right) & = \hat{\theta}_{\omega \ell} 	\hat{M}_{3\ell}\left(Z_{1i}\right).
		\end{split}
	\end{equation*}
  To reduce the dimension of the setting, we have imposed a restriction among the $\beta_j$'s. We let $\beta_1 = \beta_j$ for all $j$. As far as our original algorithm is concerned, this restriction can be simply imposed by noticing that since each entry in $\beta_1$ is a common factor across the entries of  $\hat{M}_j$, the effective regressors are  
	\begin{equation*}
		\begin{split}
			\hat{M}_{1\ell}\left(Z_{1i}\right) & = \left(\gamma_{11}\left(Z_{1i}\right) + \hat{\theta}_{\omega\ell}\gamma_{11}\left(Z_{1i}\right), \cdots, \gamma_{1r_1}\left(Z_{1i}\right) + \hat{\theta}_{\omega\ell}\gamma_{1r_1}\left(Z_{1i}\right)\right), \\
					\hat{M}_{2\ell}\left(Z_{1i}\right) & = \hat{\theta}_{\omega \ell} 	\hat{M}_{1\ell}\left(Z_{1i}\right), \\
			\hat{M}_{3\ell}\left(Z_{2i}\right) & = \left(\gamma_{11}\left(Z_{2i}\right) + \hat{\theta}_{\omega\ell}\gamma_{11}\left(Z_{2i}\right), \cdots, \gamma_{1r_1}\left(Z_{2i}\right) + \hat{\theta}_{\omega\ell}\gamma_{1r_1}\left(Z_{2i}\right)\right), \\
			\hat{M}_{4\ell}\left(Z_{1i}\right) & = \hat{\theta}_{\omega \ell} 	\hat{M}_{3\ell}\left(Z_{1i}\right),
		\end{split}
	\end{equation*}
	and then we simply need to find the solution $\beta \in \mathbb{R}^{r_1}$ using our algorithm above, accordingly accommodated. This is computational appealing as the dimension of our matrices of regressors does not depend on the number of conditioning moments. The low-dimensional dictionary $\gamma^{\text{low}}(Z)$ is based on the first 5 terms of $\gamma_1$. The first stage $\eta_{0t}$ is estimated by Boosting using $L=5$ folds.\footnote{We use the R-function \texttt{gbmt} for boosting. Default options were kept. In particular, weights are equal to 1, the model offset is a vector of zeros, the number of trees is 2000, interaction depth is 3, the minimum number of observations in a node is 10, shrinkage is set at 0.001, bag fraction is 0.5, and the observations in the first half of the database  are for training the model while the remainder are used for computing out-of-sample estimates of the loss function. Predictions were conducted using the first 500 iterations of the boosting sequence.} This choice follows a recommendation  by \cite{chernozhukov2018double} (p. C24). We set $\hat{\Lambda}$ as the identity matrix. All results are based on 500 Monte Carlo repetitions.
	
   We remark a last point regarding our numerical computation. Notice that this model has three unknowns: $\theta_{01}$, $\theta_{0k}$, and $\theta_{0\omega}$. While nothing in our theory prevents us from estimating all of them simultaneously, we only focus on $\theta_{0k}$ for two important reasons. First, in the production function literature, researchers are primarily interested in marginal product estimates. Since our goal in the next section is to conduct an empirical exercise, we wanted our Monte Carlo exercise to serve as a preliminary step for applying our algorithm in a more real-world setting. Second, trying to estimate the entire vector of unknown parameters might not be advisable, as it involves a three dimensional nonlinear minimization problem. Our DGPs do not appear to generate strong variation in the instruments (non-OR-IVs and OR-IVs) to ensure a sufficiently convex objective function, potentially leading to very large standard errors, which seems to be potentially serious for $\theta_{0\omega}$, the AR(1) coefficient.\footnote{This issue has been previously noted by \cite{ackerberg2015identification} (p. 2448).} Conveniently, it is possible to write the GMM objective function in terms of $\theta_{0k}$ only, as the AR(1) process imposes a restriction on the remaining parameters, which must satisfy certain conditions given $\theta_{0k}$; for details see Section \ref{additionalmontecarlo}. At the same time, this might make our results more reliable.

   To assess the performance of DGMM, we have computed a naive plug-in estimator (PI) as our benchmark. This estimator is based on the same first stage as DGMM but uses as instruments natural choices, which are non-OR-IVs, resulting in a non-robust estimator. It is obtained by GMM using the CMRs \eqref{prod22} and \eqref{prod24}, with instruments: $ K_{t-1}, I_{t-1}, K^2_{t-1}$, and $I^2_{t-1}$. To compute its standard errors, we now have to explicitly take into account the estimation effect from the first stage, unlike DGMM, which automatically correct for this; see \cite{newey1994asymptotic}.
   
    Table \ref{table:MC1} reports our results. First, let us focus on DGP1, the one that has a correctly specified link between $\omega_{it}$ and $I_{it}$, given $K_{it}$. We observe a clear bias-variance trade-off.  Although the bias does not seem to be practically large for the naive estimator, as the sample size increases, it tends to be larger relative to DGMM. Our estimator, across the Monte Carlo repetitions, shows systematically greater standard errors. This is not surprising as it involves the estimation of the OR-IVs, introducing additional sources of uncertainty. However, this greater imprecision does not seem large. In fact, only when $n=250$ and $n=500$, DGMM reports a higher RMSE. In terms of coverage, while our estimator has coverage figures close to the expected nominal level, the naive estimator always undercover, which seems to deteriorate as the sample size increases. All these observations are coherent with our theory.  
    
    When we introduce shocks to investment, which increase the bias in the first-stage estimation, we observe that the bias in the plug-in estimator increases relative to DGP1 with no shock, as the first-stage bias becomes greater (captured by larger variability in the investment shock). In contrast, DGMM seems to be less sensitive to the size of the bias in estimating $\eta_{0t}$, which is expected as it uses OR-IVs, resulting in systematically smaller bias compared to the plug-in estimator. Again, although DGMM shows larger standard errors, this price does not seem to be large in this numerical exercise, which can be noticed by the fact that DCRMs can deliver a lower RMSE than the plug-in estimator. Possibly more concerning, the naive approach results in poor coverage, with a convergence level as low as 60$\%$. On the other hand, despite the misspespecification in the first stage, DGMM still provides satisfactory coverage. The implication of this exercise is that, even though DGMM may have higher standard errors, it might be recommendable to use a debiased estimator rather than a non-debiased one, as it is less affected by first-stage bias.
	
\begin{table}[H]
	\centering 
	\caption{Monte Carlo Results}
	\label{table:MC1}
	\begin{threeparttable}
		\begin{tabular}{ccccccccc}
			\hline \hline
		$n$ &
		\makecell{PI\\Bias} &
		\makecell{DGMM\\Bias} &
		\makecell{PI\\SE} &
		\makecell{DGMM\\SE} &
		\makecell{PI\\RMSE} &
		\makecell{DGMM\\RMSE} &
		\makecell{PI\\95Cvg} &
		\makecell{DGMM\\95Cvg} \\
			\hline
			\multicolumn{9}{l}{$\;\;\;\;\;\;\;\;\;\;\;\;\;\;\;\;\;\;\;\;\;\;\;\;\;\;\;\;\;\;\;\;\;\;\;\;\;\;\;\;\;\;\;\;\;\;\;\;\;\;\;\;\;$\textit{DGP 1: No shock to investment}} \\ 
			$250$   & 0.02 & -0.03 & 0.12 & 0.14 & 0.12 & 0.14 & 0.88 & 0.92 \\ 
			$500$   & 0.04 & -0.03 & 0.10 & 0.11 & 0.11 & 0.12 & 0.86 & 0.92 \\ 
			$750$   & 0.06 & -0.02 & 0.09 & 0.10 & 0.11 & 0.11 & 0.81 & 0.93 \\ 
			$1,000$ & 0.06 & -0.03 & 0.07 & 0.09 & 0.10 & 0.09 & 0.79 & 0.93 \\ 
			\multicolumn{9}{l}{$\;\;\;\;\;\;\;\;\;\;\;\;\;\;\;\;\;\;\;\;\;\;\;\;\;\;\;\;\;\;\;\;\;\;\;\;\;\;\;\;\;\;\;\;\;\;\;\;\;\;\;\;\;$\textit{DGP 2: Shock to investment (sd=0.5)}} \\
			$250$   & 0.04 & -0.00 & 0.12 & 0.15 & 0.13 & 0.15 & 0.89 & 0.92 \\ 
			$500$   & 0.06 & -0.00 & 0.10 & 0.12 & 0.12 & 0.12 & 0.84 & 0.93 \\ 
			$750$   & 0.08 &  0.00 & 0.09 & 0.11 & 0.12 & 0.11 & 0.77 & 0.95 \\ 
			$1,000$ & 0.09 & -0.00 & 0.08 & 0.10 & 0.12 & 0.10 & 0.70 & 0.95 \\ 
			\multicolumn{9}{l}{$\;\;\;\;\;\;\;\;\;\;\;\;\;\;\;\;\;\;\;\;\;\;\;\;\;\;\;\;\;\;\;\;\;\;\;\;\;\;\;\;\;\;\;\;\;\;\;\;\;\;\;\;\;$\textit{DGP 3: Shock to investment (sd=0.7)}} \\
			$250$   & 0.06 & 0.02 & 0.13 & 0.16 & 0.14 & 0.16 & 0.88 & 0.93 \\ 
			$500$   & 0.09 & 0.03 & 0.10 & 0.13 & 0.13 & 0.14 & 0.79 & 0.95 \\ 
			$750$   & 0.10 & 0.03 & 0.08 & 0.13 & 0.13 & 0.13 & 0.69 & 0.97 \\ 
			$1,000$ & 0.11 & 0.03 & 0.08 & 0.12 & 0.14 & 0.12 & 0.60 & 0.96 \\ 
			\hline \hline  
		\end{tabular}
		\begin{tablenotes}
			\scriptsize
			\item NOTE: The table shows the bias, standard error (SE), root mean squared error (RMSE), and 95$\%$ coverage (95Cvg) for the naive plug-in estimator (PI) and DGMM, obtained from 500 Monte Carlo repetitions.  
		\end{tablenotes}   
	\end{threeparttable}
\end{table}
	
   \section{Empirical Application}
   \label{empiricalapp}
   We now evaluate the performance of our estimator using real data to estimate production functions at the firm level. Our goal is to assess whether the use of machine learning tools makes a significant difference in practice, specifically in delivering different production function estimates, relative to other approaches.
   
   \cite{cha2023inference} use an adjusted version of the Double Debiased Machine Learning (DML) estimator of \cite{chernozhukov2018double} within the context of production functions estimation. They exploit  the orthogonal score function from \cite{robinson1988root}. Therefore,  the authors conduct inference on only one parameter at a time and treat the parameters of the rest of inputs, including capital, as nuisance ones. Moreover, they focus on a particular strategy to estimate nuisance components. As recognized by the authors (p. 24), their strategy might not lead to identification of the production function parameters as it might suffer from the functional dependence problems discussed by \cite{ackerberg2015identification}. In contrast, our two-step approach follows \cite{ackerberg2014asymptotic} and \cite{ackerberg2015identification}, avoiding functional dependence issues. In addition,  we apply debiased machine learning to production functions using off-the-shelf machine learning routines, e.g., random forest, to conduct inference on the parameters of all the inputs in the production function simultaneously.  We are not aware of previous works doing this.
   
   Generally speaking, we consider a model similar to our example. We examine a production function in two inputs: $L_{it}$ and $K_{it}$, labor and capital, respectively. The production function $F\left(\cdot,\theta_{0p}\right)$ is Cobb-Douglas and thus Equation \eqref{production} can be written as 
   \begin{equation}
   	\label{prodemp}
   	 Y_{it} = \theta_{01} + \theta_{0l}L_{it} + \theta_{0k}K_{it} + \omega_{it} + \epsilon_{it}.
   \end{equation}
   As a proxy variable we now use intermediate inputs $E_{it}$, following \cite{levinsohn2003estimating}.\footnote{The monotonicity assumption, required by the proxy variable approach, is more likely to hold with intermediate inputs than investment. Several firms in our data reports no investment for several periods, which might indicate failure of monotonicity. } Hence, $E_{it} = E_{t}\left(\omega_{it}, L_{it}, K_{it}\right)$ is the variable that allows us to control for unobserved productivity. Note that $E_{it}$ does not appear in \eqref{prodemp}, meaning that $Y_{it}$ is ``value-added" rather than output. One possible interpretation for this is that output follows a Leontief specification in $E_{it}$; for other possible interpretations see \cite{bruno1978analysis} and \cite{diewert78}. As showed by \cite{bond2005adjustment} for the Cobb-Douglas specification, and \cite{gandhi2011identification} for a non-parametric production function, under the assumptions of the proxy variable approach, outlined by \cite{olleypakes96}, \cite{levinsohn2003estimating}, and \cite{ackerberg2015identification}, production functions with gross output are not identified unless additional assumptions (e.g., observation on prices that have to vary sufficiently across periods) are imposed. As emphasized by most of the previous authors, the circumstances under which the coefficients in \eqref{prodemp} can be interpreted in a meaningful way are not trivial. Thus, we simply interpret these coefficients as a measure of the ``effect" that inputs have on value-added output, not necessarily as  marginal productivity as in the usual sense.\footnote{Still, for simplicity, we refer to these coefficients as marginal productivities/products.}
   
   Our starting CMRs are 
    \begin{equation*}
   	\begin{split}
   		\mathbb{E}\left[\right. & \left.\left.Y_{t-1} - \eta_{0,t-1}\left(E_{t-1}, L_{t-1}, K_{t-1}\right)\right|E_{t-1}, L_{t-1}, K_{t-1}\right]  = 0, \;\;\; 2 \leq t \leq T,\\ 
   		\mathbb{E}\left[ \right. & \left. \left. Y_{t} - \theta_{01} - \theta_{0l}L_{t} - \theta_{0k}K_{t} \right. \right. \\ & \left. \left. - \theta_{0\omega}\left(\eta_{0,t-1}\left(E_{t-1}, L_{t-1}, K_{t-1}\right)  - \theta_{01} - \theta_{0l}L_{t-1} - \theta_{0k}K_{t-1}\right)\right| E_{t-1}, L_{t-1}, K_{t-1}\right]  = 0, \;\;\; 2 \leq t \leq T .
   	\end{split}
   \end{equation*}
 We follow \cite{ackerberg2015identification} and let $L_{it}$ be a state variable in the firm's dynamic problem, which can be chosen at $t$, $t-1$, or at some intermediate point (potentially with dynamic implications).  Similarly as above, an LR moment can be written as follows 
   	\begin{equation*}
   	\begin{split}
   	 	\psi\left(W_i,\theta_0,\eta_0, \kappa_0\right) & = \sum^T_{t=2}\left(Y_{i,t-1} - \eta_{0,t-1}\left(E_{i,t-1}, L_{i,t-1}, K_{i,t-1}\right)\right)\kappa_{01t}\left(E_{i,t-1}, L_{i,t-1}, K_{i,t-1}\right)  \\ & + \sum^T_{t=2}\left(Y_{it} - \theta_{01} - \theta_{0l}L_{it} - \theta_{0k}K_{it} - \theta_{0\omega}\left(\eta_{0,t-1}\left(E_{i,t-1}, L_{i,t-1}, K_{i,t-1}\right) \right. \right. \\ & \left. \left.  - \theta_{01} - \theta_{0l}L_{i,t-1} - \theta_{0k}K_{i,t-1}\right)\right)  \kappa_{02t}\left(E_{i,t-1}, L_{i,t-1}, K_{i,t-1}\right),
   	\end{split}
   \end{equation*}
   where $\kappa_{0} = \left(\kappa_{012},\kappa_{013}, \cdots,\kappa_{01T}, \kappa_{022}, \kappa_{023}  \cdots, \kappa_{02T} \right)$ is a vector of OR-IVs obtained from a vector of functions of starting instruments, to be specified below. These OR-IVs are computed following the algorithm outlined in Section \ref{estimationOIV}. This setting has $2(T-1)$ CMRs and thus  $2(T-1)$ design matrices $\bm{\hat{M}_{j}}$. The $i-$ row of each of these matrices can be written as
   \begin{equation*}
   	\begin{split}
   	 \hat{M}_{1t\ell}\left(E_{it}, L_{it}, K_{it}\right)  = & \left(\gamma_{1t1}\left(E_{it}, L_{it}, K_{it}\right),\cdots, \gamma_{1tr_{1t}}\left(E_{it}, L_{it}, K_{it}\right),  \hat{\theta}_{\omega \ell}\gamma_{2t1}\left(E_{it}, L_{it}, K_{it}\right),\cdots, \right. \\ & \left. \hat{\theta}_{\omega \ell}\gamma_{2tr_{2t}}\left(E_{it}, L_{it}, K_{it}\right) \right), \\ 
   	  \hat{M}_{2t\ell}\left(E_{it}, L_{it}, K_{it}\right)  = & \; \hat{\theta}_{\omega \ell}  \hat{M}_{1t\ell}\left(E_{it}, L_{it}, K_{it}\right),
   	\end{split}
   \end{equation*}
  where $t = 1,\cdots, T-1$. In our results below we simply construct the same basis for all $t$ (so we can drop the time dependence when referring to the bases) and also we let $\gamma_{1} =  \gamma_{2}$, which in our results below is obtained from the tensor product of univariate exponential basis of dimension 5. Specifically, let $\tilde{\gamma}$ be a vector collecting the terms of the sequence of exponential basis, we have 
 \begin{equation*}
 	\begin{split}
 		\gamma_1\left(E_{t}, L_{t}, K_{t}\right) =  & \left(\tilde{\gamma}_1(L_{t}) \times \tilde{\gamma}_1(K_{t}) \times \tilde{\gamma}_1(E_{t}), \tilde{\gamma}_1(L_{t}) \times \tilde{\gamma}_1(K_{t}) \times \tilde{\gamma}_2(E_{t}), \cdots, \right. \\ & \left. \tilde{\gamma}_{5}(L_{t}) \times \tilde{\gamma}_{5}(K_{t}) \times \tilde{\gamma}_{5}(E_{t})\right),
 	\end{split}
 \end{equation*}
 where $\tilde{\gamma}_s$ denotes the $s-$term of the sequence of exponential of bases. Note that $r_1 = 125$. Our starting vectors of functions are, for $2 \leq t \leq T$, 
 	\begin{equation*}
 	\begin{split}
 		f_{1t}(Z) & = \left(K_{t-1}, K_{t-1}\right)^{\prime}, \\ 
 		f_{2t}(Z) & = \left(K_{t-1}, K_{t}\right)^{\prime}, \\
 		f_{3t}(Z) & = \left(L_{t-1},L_{t-1}\right)^{\prime},\\
 		f_{4t}(Z) & =\left(K^2_{t-1}, K^2_{t}\right)^{\prime}, \\
 		f_{5t}(Z) & = \left(K^4_{t-1}, K^4_{t}\right)^{\prime}.
 	\end{split} 
 \end{equation*}

In Section \ref{ANempiricalapp} in the Supplementary Appendix, we give specific conditions for the validity of our inference procedure, as explained in Section \ref{estimationparint}, in this setting. This result ensures $\sqrt{n}-$consistency and asymptotic normality of our debiased estimator for the entire vector $\theta_0 = \left(\theta_{01}, \theta_{0l}, \theta_{0k}, \theta_{0\omega}\right)$. However, as we have pointed out in the previous section, in practice researchers are typically interested in inference for $\theta_{0l}$ and $\theta_{0k}$ only. Moreover, estimating all parameters simultaneously might lead to significantly larger standard errors for these key coefficients. Therefore, we only focus on inference for $\theta_{0l}$ and $\theta_{0k}$. We have re-written our problem accordingly, in the same way as we did with our Monte Carlo exercise. That is, we express the GMM objective function such that it depends on $\theta_{0l}$ and $\theta_{0k}$ only. This is advantageous as we only have to proceed by minimizing the GMM objective function in two parameters instead of four, presumably improving the reliability of our results.  
      
  We use data from the Instituto Nacional de Estadistica de Chile, which conducted a census of Chilean manufacturing plants with at least ten employees during the period 1979-1987. We are specially interested in this data as it has been extensively employed in the estimation of production functions and productivity at the firm level \citep[e.g.,][]{levinsohn2003estimating, gandhi2020identification, ackerberg2015identification, pavcnik2002trade, alvarez2005exporting}.\footnote{We are indebted to David A. Rivers and Salvador Navarro for sharing the data with us and for their assistance with the variables construction details.}  The variables in our analysis  have been constructed following \cite{gandhi2020identification}, who use the conventions in \cite{greenstreet2007exploiting}. In particular, we construct real gross output as deflated revenues. Intermediate inputs are obtained as the sum of the firm's expenditure on electricity, fuels, raw materials, and services. The difference between these two yields value-added output. Capital is formed by the perpetual inventory method. This implies that in each period investment in $t$ is combined with deflated capital from period $t-1$ to produce capital. Labor is aggregated from white-collar and blue-collar workers, with the former weighted by the ratio of the (sector) average white-collar wage to the (sector) average blue-collar wage. Deflators are obtained from \cite{bergoeing2003idiosyncratic} and all variables are expressed in real 1980 Chilean thousands pesos.\footnote{See Saction D in the Supplementary Appendix of \cite{greenstreet2007exploiting} for more details; see also \cite{liu1991}.} Our sample consists of firms from the five largest three-digit ISIC manufacturing industries in Chile: food products (311), textiles (321), apparel (322), wood products (331), and fabricated metal products (381). To maximize statistical power, we pool all firms into a single sample, which has 1,069 establishments.\footnote{This sample size is the one that remains after we drop from the original data those firms with missing values for at least one of the variables in our analysis.} We focus on three subperiods: 1979-1981, 1982-1984, and 1985-1987. These coincide with macroeconomic cycles of the Chilean economy in the data, as identified by \cite{levinsohn2003estimating}. We will conduct our analysis for each of these time periods separately below. 
   
   To reduce the dimensionality of the setting, we have assumed that the coefficients across CMRs are constant as in our Monte Carlo experiments. That is, $\beta_{jk} = \beta_k$ for all $1 \leq j \leq J$ and $k = 1,\cdots, r$, where $r = 125$. 
    The low-dimensional dictionary $\gamma^{\text{low}}(Z)$ is based on the first $r_1/15$ terms of $\gamma_1$. While we could have computed different DGMM estimators using different off-the-shelf machine learning tools to estimate $\eta_{0t}\left(E_t,L_t,E_t\right)$, we focus on first stages computed by random forest, which provided us with the most precise point estimates.\footnote{We have implemented this by the R-function \textit{ranger} (default options).} We have let $c_1=2$ in our expression for $\lambda_n$, which has also helped us reduce standard errors with this particular data.
   
   We compare the results from DGMM  with other two estimators that do not rely on machine learning tools. First, an OLS estimator, obtained by regressing $Y_t$ on $L_t$, $K_t$, and a constant.\footnote{As we discussed, this estimator is inconsistent but gives a simple framework.} Second, the estimator developed by \cite{ackerberg2015identification}, which is currently implemented by the R function \textit{prodestACF}.\footnote{See the package at \texttt{https://github.com/GabrieleRovigatti/prodest/tree/master/prodest}.} There are three key differences between this estimator and ours. Firstly, the estimator by \cite{ackerberg2015identification} (ACF) treats $\eta_{0t}$ non-parametrically and is estimated by a sieve specification \citep[][]{chen2007large}. In particular, the off-the-shelf routine regresses $Y_t$ on a polynomial of degree $d$ in $E_t$, $L_t$, and $K_t$. This strategy assumes then that all the terms appearing in the polynomial are important in the first stage, which might restrict the unknown complexity of $\eta_{0t}$ and affect the estimates. The default option is $d=2$, a common choice in applied work. We consider this degree and also a more flexible specification with $d=5$.  Secondly, while \cite{ackerberg2015identification} assume that $\omega_t$ follows a first-order Markov process, as we do, they treat it non-parametrically rather than assuming an AR(1) process, and approximate this with sieves. Lastly, standard errors are computed by bootstrap.  In our estimation, the optimal GMM weighting matrix is used.

   Table \ref{table:EstResults1} presents the estimation results for the labor and capital parameters in Equation \eqref{prodemp} across the four estimation routines we have implemented.\footnote{These estimation results should be interpreted with caution. Despite their widespread use in applications, identification strategies based on timing and information set assumptions—such as those employed in the proxy variable approach to production functions—may fail and remain not well understood; see \cite{kim2019robust} and \cite{ackerberg2023under} for recent discussions.} 
    DGMM, across all time periods, suggests relatively large labor marginal products, in the range of 1.1 and 1.2 (std. err.= 0.38-0.50).\footnote{Note that these are not strictly marginal productivities as the dependent variable is not output but value-added. Yet, they are strongly correlated. We refer to them as marginal productivities to simplify the exposition. These results are intended primarily to illustrate first-stage biases in this dataset.} By contrast, ACF reports lower labor marginal products: between 0.9 and 0.7 (std. err.=0.2–0.3) when the first stage uses $d=2$, and between 0.9 and 0.65 (std. err.=0.2–0.5) when $d=5$. These figures are very similar to those obtained by the inconsistent OLS estimator.
    
    Turning to capital, all methods report systematically lower marginal products. DGMM estimates values around 0.2–0.3, with, for instance, an estimate of 0.32 (std. err.=0.16) in 1982–1984. ACF, in both specifications, yields capital marginal products close to 0.3 in 1979–1981 and 1985–1987. However, in 1982–1984 ACF delivers implausible negative marginal products, likely reflecting severe bias driven by functional-form restrictions. Relative to DGMM, OLS appears downward biased for labor, while the capital coefficient is virtually identical.\footnote{In general, the sign of the bias depends on firms’ ability to adjust inputs after observing productivity.} Taken together, the DGMM results suggest that as firms became more productive, they reduced blue-collar employment while keeping their capital stock essentially unchanged. This aligns with the extensive literature documenting high unemployment in the 1980s and the adverse short-term employment effects of trade liberalization \citep[see, e.g.,][]{edwards1997trade,riveros86,paus1994economic}. In contrast, the ACF results would imply that the transmission bias from ignoring endogeneity is negligible—a conclusion difficult to reconcile with the large structural economic reforms that Chile was undertaken during our period of analysis.

    Both DGMM and ACF support constant returns to scale at the 5\% level across all periods. For example, in 1985–1987 the sum of labor and capital marginal products is 1.03 under ACF ($d=2$), 1.12 under ACF ($d=5$), and 1.50 under DGMM, with standard errors of similar magnitude. DGMM also suggests that capital–labor intensity ranges from 0.15 (std. err.=0.06) in 1979–1981 to 0.29 (std. err.=0.13–0.17) in 1982–1987. OLS delivers larger ratios, from 0.35 (std. err.=0.02) in 1979–1981 to 0.43 (std. err.=0.03) in 1985–1987, broadly consistent with ACF in some subperiods. Therefore, conclusions differ depending on how we model the first stage.
     
  The data suggests that estimates of labor marginal products are less precise than those of capital: standard errors for labor are typically larger across methods and years. While DGMM requires the additional estimation of OR-IVs, its standard errors are not necessarily larger than those of ACF. In fact, while DGMM typically reports the largest standard errors for $\hat{\theta}_l$, the two methods produce standard errors of comparable magnitude for $\hat{\theta}_k$.

   An important observation of this table is that the approach a researcher undertakes for modeling the first stage should be chosen with caution. Failing to flexibly model the first-stage in finite samples might be misleading.\footnote{However, one should not generally observed differences asymptotically when nonparametric approaches are employed.} The point estimates that ACF yields are different from the results using more flexible machine learning tools. In addition, more restrictive specifications might lead to inaccurate results, with negative marginal products reported by ACF.   Researchers should therefore evaluate how their results depend on the specific modeling choices for the first stage while maintaining the ability to perform standard inference.  We believe that our procedure---which provides a tool to implement more general first-stage estimations---could help them in that regard.

\begin{table}[H]
	\centering 
	\caption{Production Function Estimation}
	\label{table:EstResults1}
	\begin{threeparttable}
		\begin{tabular}{lcccc}
			\hline \hline
			Coeff. &	\makecell{OLS} & \makecell{ACF \\ ($d=2$)} & \makecell{ACF \\ ($d=5$)} & \makecell{DGMM} \\ 
			\hline
			\multicolumn{5}{l}{$\;\;\;\;\;\;\;\;\;\;\;\;\;\;\;\;\;\;\;\;\;\;\;\;\;\;\;\;\;\;\;\;$\textit{1979-1981}} \\ 
			\hline
			$\hat{\theta}_l$                 & 0.85 & 0.88 & 0.86 & 1.23 \\ 
			& (0.02) & (0.31) & (0.48) & (0.38) \\ 
			$\hat{\theta}_k$                 & 0.29 & 0.37 & 0.33 & 0.18 \\ 
			& (0.01) & (0.13) & (0.18) & (0.14) \\ 
			$\hat{\theta}_l + \hat{\theta}_k$& 1.14 & 1.26 & 1.18 & 1.41 \\ 
			& (0.02) & (0.25) & (0.42) & (0.26) \\ 
			$\hat{\theta}_k/\hat{\theta}_l$  & 0.35 & 0.42 & 0.38 & 0.15 \\ 
			& (0.02) & (0.27) & (0.37) & (0.06) \\ 
			\hline
			\multicolumn{5}{l}{$\;\;\;\;\;\;\;\;\;\;\;\;\;\;\;\;\;\;\;\;\;\;\;\;\;\;\;\;\;\;\;\;$\textit{1982-1984}} \\ 
			\hline
			$\hat{\theta}_l$                 & 0.85 & 0.72 & 0.65 & 1.11 \\ 
			& (0.03) & (0.18) & (0.24) & (0.50) \\ 
			$\hat{\theta}_k$                 & 0.34 & -0.03 & -0.20 & 0.32 \\ 
			& (0.01) & (0.21) & (0.23) & (0.16) \\ 
			$\hat{\theta}_l + \hat{\theta}_k$& 1.19 & 0.68 & 0.45 & 1.44 \\ 
			& (0.02) & (0.27) & (0.33) & (0.36) \\ 
			$\hat{\theta}_k/\hat{\theta}_l$  & 0.40 & -0.05 & -0.30 & 0.29 \\ 
			& (0.03) & (0.30) & (0.37) & 0.13 \\ 
			\hline 
			\multicolumn{5}{l}{$\;\;\;\;\;\;\;\;\;\;\;\;\;\;\;\;\;\;\;\;\;\;\;\;\;\;\;\;\;\;\;\;$\textit{1985-1987}} \\ 
			\hline
			$\hat{\theta}_l$                 & 0.84 & 0.78 & 0.82 & 1.16 \\ 
			& (0.03) & (0.20) & (0.26) & (0.45) \\ 
			$\hat{\theta}_k$                 & 0.37 & 0.25 & 0.30 & 0.34 \\ 
			& (0.01) & (0.16) & (0.18) & (0.14) \\ 
			$\hat{\theta}_l + \hat{\theta}_k$& 1.21 & 1.03 & 1.12 & 1.50 \\ 
			& (0.02) & (0.23) & (0.29) & (0.33) \\ 
			$\hat{\theta}_k/\hat{\theta}_l$  & 0.43 & 0.32 & 0.36 & 0.29 \\ 
			& (0.03) & (0.23) & (0.26) & (0.17) \\ 
			\hline \hline  
		\end{tabular}
		\begin{tablenotes}
			\scriptsize
			\item NOTE: The table shows estimated coefficients in Equation \eqref{prodemp} using value-added output. The routines are OLS, \cite{ackerberg2015identification} (ACF), and our DGMM estimators using random forest in the first-stage estimation. Standard errors are reported in parenthesis below the point estimates. ACF computes (block) bootstrap standard errors based on 1,000 repetitions. 
		\end{tablenotes}   
	\end{threeparttable}
\end{table}

	\section{Final Remarks}
	\label{conclusion}
	This paper has developed an approach to estimate OR-IVs, which are the key objects to construct LR/(Neyman-)orthogonal/debiased moments in general semiparametric models defined by a finite number of CMRs, with possible different conditioning variables and endogenous regressors.  Our estimation strategy leverages the CMRs implied by the model in a general way, making it applicable to a wide variety of settings. Our approach will hopefully pave the way for the use of machine learning techniques in contexts where the construction of LR has remained unexplored, such as highly non-linear and complex settings with CMRs, prominent in structural economics.
	
	We have demonstrated that  OR-IVs can be obtained as the residuals of a Lasso-type program, which can be implemented straightforwardly. We have derived a convergence rate for these functions. Based on this construction, we have introduced an estimator of structural parameters, the DGMM, in a two-step GMM setting. This estimator has been shown to be $\sqrt{n}$-consistent and asymptotically normal, requiring only that nuisance parameters, including OR-IVs, be estimated at a rate faster than $n^{-1/4}$. Our Monte Carlo experimentation has shown that employing such an estimator in high-dimensional contexts offers advantages over naive procedures, particularly in terms of bias and coverage.
	
Building on our theoretical results and Monte Carlo experiments, we applied our methodology to estimate production functions and productivity at the firm level in our empirical application. A key implication of our analysis is that the choice of approach for estimating first stages can significantly impact empirically relevant conclusions, particularly regarding production function parameters. Consequently, it is appealing to have inference methods, such as the one presented in this paper, that enable researchers to adopt more flexible specifications in the first stage. This is important for two reasons. First, it provides a valuable tool for scholars who prefer to avoid imposing restrictive assumptions. Second, it offers a way to conduct robustness checks for those more confident in traditional (somewhat less flexible) approaches.

	We acknowledge that our paper has limitations. We have assumed that our limited number of LR moments are sufficient for identifying  $\theta_0$. Moreover, while assuming identification, we have not addressed how efficiency could be improved by selecting other orthogonal moments. These are crucial matters that should not be overlooked. Future work should focus on constructing debiased moments that ensure identification in models defined by a number of CMRs, using modern tools as discussed by \cite{muandet2020kernel} and \cite{zhang2021instrument}. Furthermore, as we have seen, several $\kappa_0$'s might exist. This raises the question of whether it is possible to characterize a suitable notion of ``optimality" among these OR-IVs and propose an algorithm to \textit{select} OR-IVs that improve efficiency. From an efficiency perspective, it is well-known that the first best is the optimal IV \citep[see, e.g.,][]{chamberlain1992efficiency, ai2003efficient}. This special IV not only yields an estimator that achieves the efficient semiparametric bound, but it is also a valid OR-IV \citep[][]{vandervaart98, newey1990semiparametric}. Such a choice, nonetheless, is difficult to implement in general settings.  It might be interesting to define a broader criterion for selecting second-best OR-IVs that improve efficiency compared to other OR-IVs while remaining computationally feasible. Additionally, a more general theory for the estimation of OR-IVs can be derived. We have performed our construction exclusively for the space of sparse functions. It might be promising to develop a general framework for different functional spaces $\mathcal{G}_n$, including Reproducing Kernel Hilbert Spaces and Neural Networks. Finally, our algorithm focuses on debiasing structural parameters $\theta_0$. It might be important to extend this approach to more general parameters that depend on both $\theta_0$ and $\eta_0$, that do not necessarily appear in the CMRs. Hopefully, these ideas will be addressed in subsequent works.

	\newpage
	\begin{center}
		{\Huge{ \textsc{Supplementary Appendix}}}
	\end{center}
	
	\begin{appendix}
		\section{Proofs of Results in Section \ref{estimationoivs}}
		\label{orthogonalityappendix}
	
		\bigskip \noindent \textbf{Proof of Lemma \ref{fde2}:} Observe that 
		\begin{equation*}
			\begin{split}
				h_1\left(Z_1,\theta_0,\eta\right) & = \mathbb{E}\left[\left.Y_{1}\right|Z_1\right] - \eta_{01}\left(Z_1\right),   \\ 	h_2\left(\Omega_1,\theta_0,\eta\right)  & = \mathbb{E}\left[\left.Y_{2} - F\left(X_{2},\theta_{0p}\right)\right|\Omega_{1}\right] - \theta_{0\omega}\left(\eta_{01}\left(Z_1\right) - F\left(X_1,\theta_{0p}\right)\right),  \\  	h_3\left(Z_2,\theta_0,\eta\right) & = \mathbb{E}\left[\left.Y_{2}\right|Z_2\right] - \eta_{02}\left(Z_2\right),  \\ 	h_4\left(\Omega_2,\theta_0,\eta\right) & =\mathbb{E}\left[\left.Y_{3} - F\left(X_{3},\theta_{0p}\right)\right|\Omega_2\right] - \theta_{0\omega}\left(\eta_{02}\left(Z_2\right) - F\left(X_2,\theta_{0p}\right)\right),
			\end{split}
		\end{equation*}
		where recall that $Z_t$ and $X_t$ are measurable functions of $\Omega_t$. In this case, 
		\begin{align}
				\frac{d}{d\tau}h_1\left(Z_1,\theta_0,\eta + \tau b\right) & = -b_1\left(Z_1\right),  \label{d1}\\ \frac{d}{d\tau}h_2\left(\Omega_1,\theta_0,\eta + \tau b\right) & = -\theta_{0\omega}b_1\left(Z_1\right),  \label{d2}\\ \frac{d}{d\tau}h_3\left(Z_2,\theta_0,\eta + \tau b\right) & = -b_2\left(Z_2\right),  \label{d3}\\ \frac{d}{d\tau}h_4\left(\Omega_1,\theta_0,\eta + \tau b\right) & = -\theta_{0\omega}b_2\left(Z_2\right),  \label{d4}
		\end{align}
		which are linear and continuous in $b$, under the assumptions of the lemma. In addition, we can check that trivially
		\begin{equation*}
			\begin{split}
			\left|\left|h_1\left(\cdot,\theta_0,\eta_0 + b\right) - h_1\left(\cdot,\theta_0,\eta_0 \right) + b_1\right|\right|_{2} & = 0,  \\ 
			\left|\left|h_2\left(\cdot,\theta_0,\eta_0 + b\right) - h_2\left(\cdot,\theta_0,\eta_0 \right) + \theta_{0\omega}b_1\right|\right|_{2} & = 0, \\ 	\left|\left|h_3\left(\cdot,\theta_0,\eta_0 + b\right) - h_3\left(\cdot,\theta_0,\eta_0 \right) + b_2\right|\right|_{2} & = 0, \\ 	\left|\left|h_4\left(\cdot,\theta_0,\eta_0 + b\right) - h_4\left(\cdot,\theta_0,\eta_0 \right) + \theta_{0\omega}b_2\right|\right|_{2} &	= 0,
			\end{split}
		\end{equation*}
		which implies that \eqref{d1}-\eqref{d4} are Fréchet derivatives of the corresponding CMRs. $\blacksquare$ 
		
		\bigskip \noindent \textbf{Proof of Proposition \ref{prop_adjoint}:} Notice that by definition, $S_{\theta_0,\eta_0}$ is a linear and bounded operator between Hilbert spaces. Then, the existence, linearity, and continuity of $S^{*}_{\theta_0,\eta_0}$ follows by Theorem 2.21 in \cite{CARRASCO20075633}. For any function $g \in L^2(Z)$, by definition, $S^{*}_{\theta_0,\eta_0}$ satisfies 
		$$
		\left<S_{\theta_0,\eta_0}b,g\right>_{L^2(Z)} = \left<b,S^{*}_{\theta_0,\eta_0}g\right>_{\mathbf{B}}.
		$$
		Then, the result of the proposition follows by writing: 
		\begin{align}
			\left<S_{\theta_0,\eta_0}b,g\right>_{L^2(Z)} & = \sum^J_{j=1} \mathbb{E}\left[\mathbb{E}\left[\left.b(V)^{'}\tilde{\nu}_j\left(Y,\theta_0,\eta_0\right)\right|Z_j\right]g_j(Z_j)\right] \nonumber \\ & =  \sum^J_{j=1} \mathbb{E}\left[b(V)^{'}\tilde{\nu}_j\left(Y,\theta_0,\eta_0\right)g_j(Z_j)\right] \nonumber \\ &  =  \mathbb{E}\left[b(V)^{'} \sum^J_{j=1} \mathbb{E} \left[\left.\tilde{\nu}_j\left(Y,\theta_0,\eta_0\right)g_j(Z_j) \right|V \right]\right] \label{expadjoint},
		\end{align}
		where the first equality holds by Assumption \ref{linearityb}, and the second and third equalities use the law of iterated expectations. Hence, we conclude
		$$
		\left[S^{*}_{\theta_0,\eta_0}g\right](V) = \sum^J_{j=1} \mathbb{E} \left[\left.\tilde{\nu}_j\left(Y,\theta_0,\eta_0\right)g_j(Z_j) \right|V \right], 
		$$
		as needed $ \blacksquare$. 
		
		\section{Justification of the Algorithm for Estimation of OR-IVs}
		\label{justification}
		The algorithm to estimate OR-IVs that we propose can be seen as a multidimensional extension of the algorithm proposed by \cite{chernozhukov2022automatic}. These authors minimize one mean squared error in their Lasso problem while in our case we have the sum of $J$ mean squared errors, as we are trying to approximate a vector of $J$ functions. Nonetheless, the objective function is essentially the same, under an appropriate construction. More specifically, as we have argued in Section \ref{asymptoticskappa}, our estimator $\hat{\beta}_\ell$ can be expressed as 
			\begin{align}
			\hat{\beta}_\ell = & \underset{\beta \in \mathbb{R}^r}{\operatorname{arg\;min}}\; \sum^J_{j=1}\left(-2\hat{F}^{'}_{j\ell}\beta - \beta^{'}\hat{G}_{j\ell}\beta\right)  + 2\lambda_n \left|\left|\beta\right|\right|_1 \nonumber \\ = & \underset{\beta \in \mathbb{R}^r}{\operatorname{arg\;min}}\; \left\{-2\hat{H}^{\prime}_{1\ell}\beta - \beta^{'}\hat{H}^{\prime}_{2\ell}\beta  + 2\lambda_n \left|\left|\beta\right|\right|_1\right\} \label{sameeq},
				\end{align}
		where $\hat{H}_{1\ell} = \sum^J_{j=1}\hat{F}_{j\ell}$ and 	where $\hat{H}_{2\ell} = \sum^J_{j=1}\hat{G}_{j\ell}$. Expression \eqref{sameeq} coincides with that in \cite{chernozhukov2022automatic} (cf. Equation (3.7)), given our definitions of $\hat{H}_{1\ell} $ and $\hat{H}_{2\ell} $. 
		
		 More broadly, our algorithm is analogous to the one introduced by \cite{belloni2012sparse} and thus is justified by a similar argument to Theorem 1 of such a paper. The analogy is given by the observation that our $\hat{D}_{jk\ell}$ is the sample second moment of the first order conditions of the unrestricted  minimization program, obtained as the product between the resulting regressors and the residuals, equivalent to $\hat{\Upsilon}_\ell$ in \cite{belloni2012sparse}. The choice of $\lambda_n$ is the same, after accounting for the fact that we have $\lambda_n$ whereas they write their Lasso problem using $\lambda/n$. The only difference is that their tuning parameter is proportional to $n^{-1/2}$. In our case, we need to choice a $\lambda_n$ proportional to $n^{-1/4}$, as required by our asymptotic theory. Although our tuning parameter is suboptimal by their theory, since our choice goes to zero at a slower rate, it still suffices to preserve good properties for variable selection, asymptotically (see p. 2380 of \cite{belloni2012sparse}). 
		 
		 \subsection{Numerical Performance using Simulated Data}
		 We assess the ability of our algorithm, as outlined in the main text, to estimate $\beta_0$ in a setting where the truth is known. In addition, we want to determine if our procedure is able to uncover the identity of relevant regressors (and estimate zeros in the corresponding entries of $\hat{\beta}$ for the non-relevant predictors).
		 
		Our design is similar to the one in \cite{chernozhukov2022automatic}. However, we accommodate for a multivariate Lasso estimation, using generated regressors. Specifically, in our experiments, the truth is $\beta_{0j} = \left(1,1,1,0,0,\cdots\right)^{\prime}$, where dim($\beta_{0j}$) = 100, $j=1,2$. The data generating process is
		$$
		Y_j = X^{\prime}_j\beta_{0j} + \epsilon_{j}, \;\;\; j=1,2,
		$$
		 where $X_j = \left(X_{1j}, \cdots,X_{100j}\right)^{\prime}$, $X_{kj} \overset{\mathrm{iid}}{\sim} N(0,1)$, and $\epsilon_j \overset{\mathrm{iid}}{\sim} N(0,1)$. Since in our case, $X_{kj}$ is unknown, our Lasso procedure applies to its generated counterpart. In this simple experimentation, generated regressors are obtained as follows:
		 $$
		 \hat{X}_{kj} = X_{kj} + n^{-0.4}.
		 $$
		 Table \ref{table:MC2} reports MSEs defined as the average of $\left|\left|\hat{\beta} - \beta_0\right|\right|^2$ across 200 Monte Carlo repetitions. Additionally, it shows the fraction of times when our algorithm was able to correctly identify relevant and non-relevant regressors, i.e., when $\hat{\beta}$ presented non-zero point estimates for its first three entries only. We observe that the performance of our algorithm is satisfactory. In particular, it is able to deliver reasonable estimations of $\beta_0$ as the sample size increases. Moreover, it could learn the identity of the important regressors systematically across all the experiments.\footnote{While our theory does not provide us with a specific $c_2$, in our numerical experiments, we have considered a larger value than $ 0.5/\log(n \vee  r)$ for $c_2$, as with it, our algorithm was able to identify the important predictors at relatively small sample sizes, allowing us to reduce computational time.  Moreover, we let $c_1 = 1.1$. Our theory and several numerical experiments have suggested that the specific values of these constants, as long as the requirements of our asymptotic theory are met, does not affect inference on structural parameters, which is our primary focus.}
		  \begin{table}[H]
		  	\centering 
		  	\caption{Monte Carlo Results}
		  	\label{table:MC2}
		  	\begin{threeparttable}
		  		\begin{tabular}{ccc}
		  			\hline \hline
		  			$n$ & MSE & Correct Selection \% \\
		  			\hline 
		  			500    & 0.94 & 100\% \\
		  			1{,}000  & 0.47 & 100\% \\
		  			5{,}000  & 0.15 & 100\% \\
		  			10{,}000 & 0.10 & 100\% \\
		  			\hline \hline
		  		\end{tabular}
		  		\begin{tablenotes}
		  			\scriptsize
		  			\item NOTE: The table shows MSEs, defined as $\lVert \hat{\beta} - \beta_0 \rVert^2$, and the fraction of times where our algorithm uncovered the identity of the relevant and non-relevant regressors correctly, across 200 Monte Carlo repetitions.
		  		\end{tablenotes}
		  	\end{threeparttable}
		  \end{table}

		\section{Optimization}
		\label{optimization_beta}
		\subsection{Algorithm}
		
		Step 4 of the iterative algorithm above requires to solve 
		\begin{equation}
			\label{program}
			\underset{\beta \in \mathbb{R}^r}{\operatorname{min}}\; \sum^J_{j=1} \frac{1}{n-n_{\ell}} \left(\bm{f_{j\ell}} - \bm{\hat{M}}_{j\ell}\beta\right)^{'}\left(\bm{f_{j\ell}} - \bm{\hat{M}}_{j\ell}\beta\right) + 2\lambda_n \left|\left|\hat{D}_{\ell}\beta\right|\right|_1,
		\end{equation}
		where $\hat{D}_{\ell}$ is a diagonal matrix with elements $\hat{D}_{jk\ell} \equiv \hat{D}_{l\ell}$ along the main diagonal, with $l = 1,\cdots,r$. Hence, the first $r_1$ entries correspond to the regressors with $\gamma_1(Z_1)$, the next $r_2$ entries are the regressors with $\gamma_2(Z_2)$, and so on. To solve \eqref{program}, we use an extension of the coordinate descent approach for Lasso \citep{fu1998penalized, Friedman07, friedman2010regularization} to our particular objective function. To be precise, we implement a coordinate-wise descent algorithm with a soft-thresholding update. Let $v_l$ denote the $l^{th}$ element of a generic vector $v$ and let $e_l$ be a $r \times 1$ unit vector with 1 in the $l^{th}$ coordinate and zeros elsewhere. This algorithm can be implemented as follows: 
		
		\bigskip \noindent For $l = 1:r$, do
		
		\noindent \textbf{Step 1:} Compute loadings (which do not depend on $\beta_k$): 
		\begin{equation*}
			\begin{split}
				A_l & = \frac{1}{n - n_\ell} \sum^J_{j=1}e^{'}_l \bm{\hat{M}_j}^{'}\left(\bm{f_{j}} - \bm{\hat{M}_j}\beta + \bm{\hat{M}_j}e_l\beta_l\right)\\
				B_l & = \frac{1}{n - n_\ell} \sum^J_{j=1} e^{'}_l\bm{\hat{M}^{'}_j} \bm{\hat{M}_j} e_l. 
			\end{split}
		\end{equation*}
		
		\noindent \textbf{Step 2:} Update coordinate $\beta_l$:
		$$
		\beta_l = \begin{cases} \frac{A_l + \hat{D}_l\lambda_n}{B_l} & \text{if}\;\;\; A_l < - \hat{D}_l\lambda_n\\ 0 &  \text{if}\;\;\;  A_l \in \left[-\hat{D}_l \lambda_n, \hat{D}_l \lambda_n\right] \\ \frac{A_l - \hat{D}_l \lambda_n}{B_l} & \text{if}\;\;\; A_l > \hat{D}_l \lambda_n.\end{cases} 
		$$
		-end-
		
		\subsection{Justification}
		In this section, we justify the previous coordinate-wise soft-thresholding update, involved in Step 1 and 2 above. Observe that
		$$
		\frac{\partial}{\partial \beta_l}\left[\sum^J_{j=1} \frac{1}{n-n_{\ell}} \left(\bm{f_{j\ell}} - \bm{\hat{M}_{j\ell}}\beta\right)^{'}\left(\bm{f_{j\ell}} - \bm{\hat{M}_{j\ell}}\beta\right)\right] = -2A_l + 2B_l\beta_l,
		$$
		where we note that neither $A_l$ nor $B_l$ depend on $\beta_l$. The subgradient of the penalty term is 
		$$
		\frac{\partial}{ \partial \beta_l} 2 \left|\left|\hat{D}\beta \right|\right|_1 = \begin{cases}
			-2\hat{D}_l\lambda_n & \text{if}\;\;\; \beta_l < 0 \\ 
			\left[-2\hat{D}_l\lambda_n, 2\hat{D}_l\lambda_n\right] & \text{if}\;\;\; \beta_l = 0 \\ 
			2\hat{D}_l\lambda_n &  \text{if}\;\;\; \beta_l > 0 
		\end{cases}
		$$
		Therefore, ($(1/2)$ of) the subgradient of the objective function of our program is 
		\begin{equation*}
			\begin{split}
				\frac{\partial}{\partial \beta_l} \frac{1}{2}\left[\sum^J_{j=1} \frac{1}{n-n_{\ell}} \left(\bm{f_{j\ell}} - \bm{\hat{M}_{j\ell}}\beta\right)^{'}\left(\bm{f_{j\ell}} - \bm{\hat{M}_{j\ell}}\beta\right) + 2\left|\left|\hat{D}\beta \right|\right|_1\right] = \begin{cases}
					-A_l + B_l \beta_l - \hat{D}_l \lambda_n & \text{if}\;\;\; \beta_l < 0 \\ \left[-A_l - \hat{D}_l\lambda_n, -A_l + \hat{D}_l\lambda_n\right] & \text{if}\;\;\; \beta_l = 0 \\  -A_l + B_l \beta_l + \hat{D}_l\lambda_n & \text{if}\;\;\; \beta_l > 0
				\end{cases}
			\end{split}
		\end{equation*}
		Hence, equalizing those terms to zero and solving for $\beta_l$ gives the element-wise update provided above. 
		
		Note that the first term of the objective function in \eqref{program} is differentiable and convex and the penalty term is the sum of convex functions. Hence, the whole objective function in \eqref{program} is a particular case of Equation 21 in \cite{Friedman07}, and thus the coordinate descent converges to the solution to \eqref{program} \citep[][]{tseng2001convergence}.

\section{Asymptotic Results of OR-IVs}
		\label{appendixoiv}
		\begin{lemma}
			\label{lboundnormM}
		Let Assumptions \ref{aboundM} and \ref{aconvergenceM} hold. Then, for any $q$ and $k$, 
		$$
		\int \left| \hat{M}_{j\ell q}\left(z_{j}\right)\hat{M}_{j\ell k}\left(z_{j}\right) - M_{j q}\left(z_{j}\right)M_{j k}\left(z_{j}\right)\right| F_0(dz_j) = O_p\left(\varepsilon^M_{jn}\right). 
		$$
		
		\end{lemma}
			\bigskip \noindent \textbf{Proof of Lemma \ref{lboundnormM}:} 
		Let $\mathcal{W}^c_{\ell}$ contain the data for each $i \in I^c_{\ell}$. By the triangle and conditional Hölder's  inequalities, 
			\begin{equation*}
				\begin{split}
					 & \mathbb{E} \left[\left.\left| \hat{M}_{j\ell q}\left(z_{j}\right)\hat{M}_{j\ell k}\left(Z_{j}\right) - M_{j q}\left(Z_{j}\right)M_{j k}\left(Z_{j}\right)\right| \right| \mathcal{W}^c_\ell\right] \\ & \leq 	\mathbb{E}\left[\left.\left| \left(\hat{M}_{j\ell q}\left(Z_{ji}\right) - M_{j q}\left(Z_{ji}\right)\right) \left(\hat{M}_{j\ell k}\left(Z_{ji}\right) - M_{j k}\left(Z_{j}\right)\right)\right| \right| \mathcal{W}^c_\ell\right] \\ & + \mathbb{E}\left[\left.\left| \left(\hat{M}_{j\ell q}\left(Z_{ji}\right) - M_{j q}\left(Z_{ji}\right)\right) M_{j k}\left(Z_{j}\right)\right| \right| \mathcal{W}^c_\ell\right] \\ & +  \mathbb{E}\left[\left.\left| \left(\hat{M}_{j\ell k}\left(Z_{ji}\right) - M_{j k}\left(Z_{ji}\right)\right) M_{j q}\left(z_{j}\right)\right| \right| \mathcal{W}^c_\ell\right] \\ & \leq  \left|\left|\hat{M}_{j\ell q} - M_{j q}\right|\right|_{2}\left|\left|\hat{M}_{j\ell k} - M_{j k}\right|\right|_{2}  + \left|\left|\hat{M}_{j\ell q} - M_{j q}\right|\right|_{2}\left|\left| M_{j k}\right|\right|_{2}  + \left|\left|\hat{M}_{j\ell k} - M_{j k}\right|\right|_{2}\left|\left| M_{j q}\right|\right|_{2} \\ & = O_p\left(\varepsilon^M_{jn}\right),
				\end{split}
			\end{equation*}
			as needed. $\blacksquare$
			
		\begin{lemma}
			\label{lG}
			Let Assumptions \ref{aboundM} and \ref{aconvergenceM} hold. Then, 
			$$
			\left|\left|\hat{G}_{j\ell} - G_j\right|\right|_{\infty} = O_p\left(\varepsilon^G_{jn}\right), \;\;\;\; \varepsilon^G_{jn} = \max\left\{\bar{\varepsilon}_n, \tilde{\varepsilon}_{jn}\right\},
			$$
			where $\bar{\varepsilon}_n = \sqrt{\frac{\log(r)}{n}}$ and $\tilde{\varepsilon}_{jn} = n^c \varepsilon^M_{jn}$, for any $c>0$.
		\end{lemma}
		
		\bigskip \noindent \textbf{Proof of Lemma \ref{lG}:} Let $\tilde{G}_{j\ell} = \frac{1}{n-n_\ell} \sum_{i \notin I_{\ell}} M_j\left(Z_{ji}\right)M_j\left(Z_{ji}\right)^{'}$. Then, by the triangle inequality, 
		$$
		\left|\left|\hat{G}_{j\ell} - G_j \right|\right|_{\infty} \leq \left|\left|\hat{G}_{j\ell} - \tilde{G}_{j\ell} \right|\right|_{\infty} + \left|\left|\tilde{G}_{j\ell} - G_j \right|\right|_{\infty}.  
		$$
		We first show that $\left|\left|\tilde{G}_{j\ell} - G_j \right|\right|_{\infty} = O_p\left(\bar{\varepsilon}_n\right)$. To prove this, we follow the proof of Lemma A10 of \cite{chernozhukov2022automatic}. Let us define 
		$$
		T^j_{iqk} := M_{jq}\left(Z_{ji}\right)M_{jk}\left(Z_{ji}\right) - \mathbb{E}\left[M_{jq}\left(Z_{ji}\right)M_{jk}\left(Z_{ji}\right)\right], \;\;\; U^{j}_{qk} := \frac{1}{n-n_\ell} \sum_{i \notin I_\ell} T^j_{iqk}. 
		$$
	 Note that in the previous display, $U^{j}_{qk}$ depends on $\ell$, but we omitted this dependence to simplify the exposition. Then, for any constant $C$, we have 
		\begin{equation*}
			\begin{split}
				\mathbb{P}\left(\left|\left|\tilde{G}_{j\ell} - G_j\right|\right|_{\infty} \geq C \bar{\varepsilon}_n\right) & \leq \sum^r_{q,k = 1}  \mathbb{P}\left(\left|U^{j}_{qk}\right| \geq C \bar{\varepsilon}_n \right) \\ & \leq r^2 \underset{k,q}{\operatorname{max}}\; \mathbb{P}\left(\left|U^{j}_{qk}\right| \geq C \bar{\varepsilon}_n \right).
			\end{split}
		\end{equation*}
		Note that $\mathbb{E}\left[T^j_{iqk}\right] = 0$ and by Assumption \ref{aboundM}, 
		\begin{equation*}
			\begin{split}
				\left|T^j_{iqk}\right| & \leq \left|M_{jq}\left(Z_{ji}\right)\right|\left|M_{jk}\left(Z_{ji}\right)\right| + \mathbb{E}\left[\left|M_{jq}\left(Z_{j}\right)\right|\left|M_{jk}\left(Z_{j}\right)\right|\right] \\  & \leq 2c^2_j.   
			\end{split}
		\end{equation*}
		The previous fact shows that $T^j_{iqk}$ is a bounded random variable. Therefore, it is sub-Gaussian. Let $\left|\left|T^j_{iqk}\right|\right|_{\Psi_2}$ denote the sub-Gaussian norm. Then, $K_j = \frac{2c^2_j}{\log 2} \geq \left|\left|T^j_{iqk}\right|\right|_{\Psi_2}$. By Hoeffding's inequality \citep[see Thereom 2.6.2 in][]{vershynin2018high}, there is a constant $c$ such that 
		\begin{equation*}
			\begin{split}
				r^2 \underset{k,q}{\operatorname{max}}\; \mathbb{P}\left(\left|U^{j}_{qk}\right| \geq C \bar{\varepsilon}_n \right) & \leq 2r^2 \exp \left(-\frac{cC^2\log(r)}{K_j^2}\right) \\ & = 2 \exp\left(\log(r) \left[2 - \frac{c C^2}{K_j^2}\right]\right) \\ & \longrightarrow 0,
			\end{split}
		\end{equation*}
		for any $C$ such that $C > K_j \sqrt{\frac{2}{c}}$. Hence, for $C$ large enough, $\mathbb{P}\left(\left|\left|\tilde{G}_{j\ell} - G_j\right|\right|_{\infty} \geq C \bar{\varepsilon}_n\right) \longrightarrow 0$ as $r \longrightarrow \infty$, as needed. 
		
		 Next, notice that each estimated element in the matrix $\hat{G}_{j\ell}$ depends on $\mathcal{W}^c_{\ell}$ only and let 
		$$
		P^j_{i\ell q k} = \hat{M}_{j\ell q}\left(Z_{ji}\right)\hat{M}_{j\ell k}\left(Z_{ji}\right) - M_{j q}\left(Z_{ji}\right)M_{j k}\left(Z_{ji}\right), \;\;\; Q^j_{\ell q k } = \frac{1}{n - n_\ell} \sum_{i \notin I_\ell} P^j_{i\ell q k}.
		$$
		Conditional on $\mathcal{W}^c_\ell$, by the  the conditional Markov's and triangle inequalities, we can write for any $C > 0$
		\begin{equation*}
			\begin{split}
				\mathbb{P}\left(\left. \left|\left|\hat{G}_{j\ell} - \tilde{G}_{j\ell}\right|\right|_{\infty} \geq C n^c\varepsilon^M_{jn}\right| \mathcal{W}^c_\ell\right) & \leq \mathbb{P}\left(\left. \underset{k,q}{\operatorname{max}}\;  \left|Q^j_{\ell q k}\right|  \geq C n^c\varepsilon^M_{jn} \right|\mathcal{W}^c_\ell\right) \\ &  \leq \mathbb{P}\left(\left.   \left|Q^j_{\ell \bar{q} \bar{k}}\right|  \geq C n^c\varepsilon^M_{jn} \right|\mathcal{W}^c_\ell\right)\\  & \leq  \frac{1}{Cn^c\varepsilon^M_{jn}} \mathbb{E}\left[\left. \left|Q^j_{\ell \bar{q} \bar{k}}\right|\right|\mathcal{W}^c_\ell\right] \\ & \leq \frac{1}{C n^c\varepsilon^M_{jn}} \mathbb{E}\left[\left. \left|P^j_{i\ell \bar{q} \bar{k}}\right|\right|\mathcal{W}^c_\ell\right] \\&  \longrightarrow 0,
			\end{split}
		\end{equation*}
		as $n$ goes to infinity, where  $\underset{k,q}{\operatorname{max}}\;  \left|Q^j_{\ell q k}\right| \equiv \left|Q^j_{\ell \bar{q} \bar{k}}\right|$, and the last display follows from Lemma \ref{lboundnormM}. We have shown then 
		$$
		\left|\left|\hat{G}_{j\ell} - G_j \right|\right|_{\infty} \leq O_p\left(\bar{\varepsilon}_n\right) + O_p\left(\tilde{\varepsilon}_{jn}\right) = O_p\left(\varepsilon^G_{jn}\right),
		$$
		as required. $\blacksquare$
		
		\bigskip Let $\beta_0$ be the slope of the linear projection of $f^{*}$ on $M = \left(M_1,\cdots,M_J\right)$. Hence,
		$$
		\sum^J_{J=1}\mathbb{E}\left[M_j\left(Z_j\right)\left(f^{*}_j\left(Z_j\right) - M_j\left(Z_j\right)^{'}\beta_0\right)\right] = 0.
		$$
		
		In addition, let us define $\beta_*$ as 
		\begin{equation}
			\label{betaast}
			\beta_* \in \underset{v\in \mathbb{R}^r}{\operatorname{arg\;min}}\;\left(\beta_0 - v\right)^{'}\sum^J_{j=1}G_j\left(\beta_0 - v\right) + 2\varepsilon_n \sum_{k \in S^{c}_{\bar{\beta}}}\left|v_k\right|.
		\end{equation}
		A maintained assumption throughout this work is that $\left|\left|\beta_*\right|\right|_1 = O_p(1)$. 
		
		\begin{lemma}
			\label{lbetaastbetazero}
			$\left|\left|\sum^J_{j=1}G_j\left(\beta_* - \beta_0\right)\right|\right|_{\infty} \leq \varepsilon_n$.
		\end{lemma}
		
		\bigskip \noindent \textbf{Proof of Lemma \ref{lbetaastbetazero}:} The first order condition (sub-gradient of the objective function) for $\beta_*$ implies that for $k \in S_{\bar{\beta}}$, we have $e^{'}_k\sum^J_{j=1}G_j\left(\beta_* - \beta_0\right) = 0$, where $e_k$ is the $k-th$ column of an identity matrix $I_r$ of dimension $r \times r$. For $k \in S^c_{\bar{\beta}}$, we have that $e^{'}_k \sum^J_{j=1} G_j \left(\beta_* - \beta_0\right) + \varepsilon_n \pi_k = 0$, where $\pi_k = sign\left(\beta_{*k}\right)$ if $\beta_{*k} \neq 0$, and $\pi_k \in \left[-1,1\right]$ if $\beta_{*k} = 0$. Hence, in any case $\left|e^{'}_k\sum^J_{j=1}G_j\left(\beta_* - \beta_0\right)\right| \leq \varepsilon_n$. $\blacksquare$
		
		\begin{lemma}
			\label{beta0matGbetaast}
			$\left(\beta_0 - \beta_*\right)\sum^J_{j=1}G_j\left(\beta_0 - \beta_*\right) \leq Cs\varepsilon^2_n$. 
		\end{lemma}
		
		\bigskip \noindent \textbf{Proof of Lemma \ref{beta0matGbetaast}:} By definition of $\beta_{*}$, 
		\begin{equation}
			\label{ineqbeta0betaast}
			\begin{split}
				\left(\beta_0 - \beta_{*}\right)^{'}\sum^J_{j=1}G_j\left(\beta_0 - \beta_{*}\right) + 2\varepsilon_n \sum_{k \in S^{c}_{\bar{\beta}}}\left|\beta_{*,k}\right| & \leq \left(\beta_0 - \bar{\beta}\right)^{'}\sum^J_{j=1}G_j\left(\beta_0 - \bar{\beta}\right) + 2\varepsilon_n \sum_{k \in S^{c}_{\bar{\beta}}}\left|\bar{\beta}_{k}\right| \\ & = \left(\beta_0 - \bar{\beta}\right)^{'}\sum^J_{j=1}G_j\left(\beta_0 - \bar{\beta}\right).
			\end{split}
		\end{equation}
		
		Next, notice that by the triangle inequality
		\begin{equation}
			\label{beta0betabar}
			\begin{split}
				\left(\beta_0 - \bar{\beta}\right)^{'}\sum^J_{j=1}G_j\left(\beta_0 - \bar{\beta}\right) & = \sum^J_{j=1} \mathbb{E}\left[\left\{M_j\left(Z_j\right)^{'}\left(\beta_0 - \bar{\beta}\right)\right\}^2\right] \\ & = \sum^J_{j=1}\left|\left|M_j^{'}\beta_0 - M_j^{'}\bar{\beta}\right|\right|^2_{2} \\ & \leq 2\sum^J_{j=1}\left(\left|\left|f^{*}_j - M_j^{'}\beta_0\right|\right|^2_{2}  + \left|\left| f^{*}_j - M_j^{'}\bar{\beta}\right|\right|^2_{2} \right) \\ & \leq 4 \sum^J_{j=1} \left|\left| f^{*}_j - M_j^{'}\bar{\beta}\right|\right|^2_{2} \\ & \leq Cs\varepsilon^2_n,
			\end{split}
		\end{equation}
		where the last inequality follows by Assumption \ref{asparse}. The result then follows by Equation \eqref{ineqbeta0betaast} and $\varepsilon_n \sum_{k \in S^{c}_{\bar{\beta}}}\left|\beta_{*,k}\right| \geq 0$. $\blacksquare$
		
		\begin{lemma}
			\label{lsast}
			Let $S_{\beta_*}$ be the vector of indices of nonzero elements of $\beta_{*}$. Then, $s_* = \# S_{\beta_*} \leq Cs$, $C>1$.
		\end{lemma}
		
		\bigskip \noindent \textbf{Proof of Lemma \ref{lsast}:} For all $k \in S_{\beta_*}\backslash S_{\bar{\beta}}$, the first order conditions to \eqref{betaast} imply $\left|e^{'}_k\sum^J_{j=1}G_j\left(\beta_{*} - \beta_0\right)\right| = \varepsilon_n$. Therefore, it holds that 
		$$
		\sum_{k \in S_{\beta_*}\backslash S_{\bar{\beta}}}\left(e^{'}_k\sum^J_{j=1}G_j\left(\beta_{*} - \beta_0\right)\right)^2 = \varepsilon^2_n \#\left[S_{\beta_*}\backslash S_{\bar{\beta}}\right]. 
		$$
		Furthermore, using Lemma \ref{beta0matGbetaast} and the fact that the largest eigenvalue of $\sum^J_{j=1}G_j$ is bounded, we obtain 
		\begin{equation*}
			\begin{split}
				\sum_{k \in S_{\beta_*}\backslash S_{\bar{\beta}}}\left(e^{'}_k\sum^J_{j=1}G_j\left(\beta_{*} - \beta_0\right)\right)^2 & \leq  \sum^r_{k =1}\left(e^{'}_k\sum^J_{j=1}G_j\left(\beta_{*} - \beta_0\right)\right)^2  \\ & = \left(\beta_{*} - \beta_0\right)^{'}\sum^J_{j=1}G_j \left(\sum^r_{k=1}e_ke^{'}_k\right)\sum^J_{j=1}G_j \left(\beta_{*} - \beta_0\right) \\ & = \left(\beta_{*} - \beta_0\right)^{'}\left(\sum^J_{j=1}G_j\right)^2\left(\beta_{*} - \beta_0\right) \\ & \leq \lambda_{max}\left(\sum^J_{j=1}G_j\right) \left\{\left(\beta_{*} - \beta_0\right)^{'}\sum^J_{j=1}G_j\left(\beta_{*} - \beta_0\right)\right\} \\ & \leq Cs\varepsilon^2_n,
			\end{split}
		\end{equation*}
		where $\lambda_{max}(A)$ denotes the maximum eigenvalue of an arbitrary matrix $A$. The previous result implies 
		$$
		\varepsilon^2_n \#\left[S_{\beta_*}\backslash S_{\bar{\beta}}\right] \leq Cs\varepsilon^2_n.
		$$
		Diving both sides of the previous expression by $\varepsilon^2_n$ yields $\# \left[S_{\beta_*}\backslash S_{\bar{\beta}}\right]  \leq Cs$. Hence, 
		$$
		s_* = \# S_{\bar{\beta}} + \# \left[S_{\beta_*}\backslash S_{\bar{\beta}}\right] \leq s + Cs \leq Cs,
		$$
		as needed $\blacksquare$. 
		
		\begin{lemma}
			\label{lfastbetaast}
			$\sum^J_{j=1}\mathbb{E}\left[\left(f^{*}_j\left(Z_j\right) - M_j\left(Z_j\right)^{'}\beta_{*}\right)^2\right] \leq Cs \varepsilon^2_n$. 
		\end{lemma}
		
		\bigskip \noindent \textbf{Proof of Lemma \ref{lfastbetaast}:} By the triangle inequality and Assumption \ref{asparse}, we can write 
		\begin{align}
			\sum^J_{j=1}\mathbb{E}\left[\left(f^{*}_j\left(Z_j\right) - M_j\left(Z_j\right)^{'}\beta_*\right)^2\right] & \leq C \sum^J_{j=1} \left|\left|f^{*}_j - M_j^{'}\bar{\beta}\right|\right|^2_{2} + C \sum^J_{j=1} \left|\left| M_j^{'}\bar{\beta} - M_j^{'}\beta_{*}\right|\right|^2_{2} \nonumber \\ & \leq Cs\varepsilon^2_n+ C \sum^J_{j=1} \left|\left| M_j^{'}\bar{\beta} - M_j^{'}\beta_{0}\right|\right|^2_{2} \label{mbeta1}\\ & + C \sum^J_{j=1} \left|\left| M_j^{'}\beta_{0} - M_j^{'}\beta_{*}\right|\right|^2_{2}. \label{mbeta2}
		\end{align}
		Notice that by the result in \eqref{beta0betabar},
		\begin{equation}
			\label{bound1}
			\sum^J_{j=1} \left|\left|M_j^{'}\bar{\beta} - M_j^{'}\beta_{0} \right|\right|^2_{2} = \left(\beta_0 - \bar{\beta}\right)^{'}\sum^J_{j=1}G_j\left(\beta_0 - \bar{\beta}\right) \leq C s\varepsilon^2_n.
		\end{equation}
		Moreover, by Lemma \ref{beta0matGbetaast}, 
		\begin{equation}
			\label{bound2}
			\sum^J_{j=1} \left|\left| M_j^{'}\beta_{0} - M_j^{'}\beta_{*}\right|\right|^2_{2} = \left(\beta_0 - \beta_*\right)\sum^J_{j=1}G_j\left(\beta_0 - \beta_*\right) \leq Cs\varepsilon^2_n,
		\end{equation}
		Plugging \eqref{bound1} into \eqref{mbeta1} and \eqref{bound2} into \eqref{mbeta2} yields the desired result. $\blacksquare$
		
		\begin{lemma}
			\label{lghatgpop}
			$\left|\left|\sum^J_{j=1}\hat{G}_j\beta_{*} - \sum^J_{j=1}G_j\beta_{*}\right|\right|_{\infty} = O_p\left(\varepsilon_n\right)$.
		\end{lemma}
		
		\bigskip \noindent \textbf{Proof of Lemma \ref{lghatgpop}:} It can be verified that
		\begin{equation*}
			\begin{split}
				\left|\left|\sum^J_{j=1}\hat{G}_j\beta_{*} - \sum^J_{j=1}G_j\beta_{*}\right|\right|_{\infty} & = \left|\left|\sum^J_{j=1}\left(\hat{G}_j-G_j\right)\beta_{*}\right|\right|_{\infty} \\ & \leq   \left|\left|\sum^J_{j=1}\left(\hat{G}_j-G_j\right)\right|\right|_{\infty}\left|\left| \beta_{*}\right|\right|_1 \\ & =  O_p\left(\varepsilon_n\right),     
			\end{split}
		\end{equation*}
		as needed. $\blacksquare$

		\begin{lemma}
			\label{ldelta}
			Let $\Delta = \hat{\beta} - \beta_{*}$. For any $\hat{S}$ such that $\beta_{*,\hat{S}^c} = 0$, with probability approaching one, 
			$$
			\Delta^{'}\sum^J_{j=1}\hat{G}_j\Delta \leq 3 \lambda_n \left|\left|\Delta\right|\right|_1,\;\;\; \left|\left|\Delta_{\hat{S}^c} \right|\right|_1 \leq 3 \left|\left|\Delta_{\hat{S}} \right|\right|_1.
			$$
		\end{lemma}
		
		\bigskip \noindent \textbf{Proof of Lemma \ref{ldelta}:} By definition of $\hat{\beta}$, we have 
		$$
		\sum^J_{j=1}\left(\hat{\beta}^{'}\hat{G}_j\hat{\beta} - 2\hat{F}^{'}_j\hat{\beta}\right) + 2\lambda_n\left|\left|\hat{\beta}\right|\right|_1 \leq \sum^J_{j=1}\left(\beta_{*}^{'}\hat{G}_j\beta_{*} - 2\hat{F}^{'}_j\beta_{*}\right) + 2\lambda_n\left|\left|\beta_{*}\right|\right|_1.
		$$
		Using $\hat{\beta} = \Delta + \beta_{*}$ in the previous expression and re-arranging terms, we obtain 
		\begin{equation}
			\label{deltaexp}
			\Delta^{'}\sum^J_{j=1}\hat{G}_j\Delta + 2\lambda_n\left|\left|\beta_{*} + \Delta\right|\right|_1 \leq 2 \lambda_n\left|\left|\beta_{*}\right|\right|_1 + 2\sum^J_{j=1}\left(\hat{F}_j - \hat{G}_j\beta_{*}\right)^{'}\Delta.
		\end{equation}
		By definition of $\beta_0$, $\sum^J_{j=1}G_j\beta_0 - \sum^J_{j=1}F_j = 0$. Then, by Assumption \ref{aconvF}, Lemma \ref{lbetaastbetazero}, Lemma \ref{lghatgpop}, and the triangle inequality, we have 
		\begin{equation*}
			\begin{split}
				\left|\left|\sum^J_{j=1}\left(\hat{G}_j\beta_{*} - \hat{F}_j\right)\right|\right|_{\infty} & \leq \left|\left|\sum^J_{j=1}\left(\hat{G}_j\beta_{*} - G_j\beta_{*}\right)\right|\right|_{\infty} +  \left|\left|\sum^J_{j=1}\left(G_j\beta_{*} - \hat{F}_j\right)\right|\right|_{\infty} \\ & \leq \left|\left|\sum^J_{j=1}\left(\hat{G}_j\beta_{*} - G_j\beta_{*}\right)\right|\right|_{\infty} +  \left|\left|\sum^J_{j=1}\left( F_j - \hat{F}_j\right)\right|\right|_{\infty} +  \left|\left|\sum^J_{j=1}\left(G_j\beta_{*} - F_j\right)\right|\right|_{\infty} \\ & \leq O_p\left(\varepsilon_n\right) + O_p\left(\varepsilon_n\right) + \left|\left|\sum^J_{j=1}\left(G_j\beta_{*} - F_j\right)\right|\right|_{\infty} \\ & \leq O_p\left(\varepsilon_n\right) + \left|\left|\sum^J_{j=1}\left(G_j\beta_{0} - F_j\right)\right|\right|_{\infty} + \left|\left|\sum^J_{j=1}G_j\left(\beta_{*} - \beta_0\right)\right|\right|_{\infty} \\ & = O_p\left(\varepsilon_n\right).
			\end{split}
		\end{equation*}
		Therefore, by the Hölder's inequality, we have that $\left|\sum^J_{j=1}\left(\hat{F}_j - \hat{G}\beta_{*}\right)^{'}\Delta\right| \leq  \left|\left|\sum^J_{j=1}\left(\hat{F}_j - \hat{G}\beta_{*}\right)\right|\right|_\infty \left|\left|\Delta\right|\right|_1 = O_p\left(\varepsilon_n\right)\left|\left|\Delta\right|\right|_1$. Recall that $\varepsilon_n = o\left(\lambda_n\right)$, and then we can write 
		\begin{align}
			\Delta^{'}\sum^J_{j=1}\hat{G}_j\Delta + 2\lambda_n\left|\left|\beta_{*} + \Delta\right|\right|_1 & \leq 2\lambda_n \left|\left|\beta_*\right|\right|_1 + O_p\left(\varepsilon_n\right)\left|\left|\Delta\right|\right|_1 \nonumber \\ & \leq 2\lambda_n \left|\left|\beta_*\right|\right|_1 + \lambda_n\left|\left|\Delta\right|\right|_1 ,\label{lambdadelta}
		\end{align}
		with probability approaching one. Moreover, by the triangle inequality, $\left|\left|\beta_*\right|\right|_1 \leq \left|\left|\beta_* + \Delta\right|\right|_1 + \left|\left|\Delta\right|\right|_1$. Plugging this into \eqref{lambdadelta}  results in $\Delta^{'}\sum^J_{j=1}\hat{G}_j\Delta \leq 3 \lambda_n\left|\left|\Delta\right|\right|_1$, and the first result of the lemma is obtained. 
		
		Furthermore, as $\Delta^{'}\sum^J_{j=1}\hat{G}_j\Delta \geq 0$, it also follows from \eqref{lambdadelta} that 
		\begin{equation}
			\label{lambdabound}
			2\lambda_n\left|\left|\beta_{*} + \Delta\right|\right|_1 \leq 2\lambda_n\left|\left|\beta_*\right|\right|_1 + \lambda_n\left|\left|\Delta\right|\right|_1. 
		\end{equation}
		By the fact that $\beta_{*,\hat{S}^c} = 0$, it follows that $\left|\left|\beta_{*} + \Delta\right|\right|_1 = \left|\left|\beta_{*,\hat{S}} + \Delta_{\hat{S}}\right|\right|_1 + \left|\left|\Delta_{\hat{S}^c}\right|\right|_1$ and $\left|\left|\beta_*\right|\right|_1 = \left|\left|\beta_{*,\hat{S}}\right|\right|_1$. Dividing both sides of \eqref{lambdabound} by $\lambda_n$ and substituting the previous conclusions yields 
		\begin{equation*}
			\begin{split}
				2\left|\left|\beta_{*,\hat{S}} + \Delta_{\hat{S}}\right|\right|_1 + 2\left|\left|\Delta_{\hat{S}^c}\right|\right|_1 & \leq 2\left|\left|\beta_{*,\hat{S}}\right|\right|_1 + \left|\left|\Delta\right|\right|_1 \\ & =2\left|\left|\beta_{*,\hat{S}}\right|\right|_1 + \left|\left|\Delta_{\hat{S}}\right|\right|_1 +   \left|\left|\Delta_{\hat{S}^c}\right|\right|_1 \\ & \leq 2\left(\left|\left|\beta_{*,\hat{S}} - \Delta_{\hat{S}}\right|\right|_1 + \left|\left| \Delta_{\hat{S}}\right|\right|_1\right) + \left|\left| \Delta_{\hat{S}}\right|\right|_1 + \left|\left| \Delta_{\hat{S}^c}\right|\right|_1 \\ & = 2\left|\left|\beta_{*,\hat{S}} - \Delta_{\hat{S}}\right|\right|_1 + 3\left|\left| \Delta_{\hat{S}}\right|\right|_1  + \left|\left| \Delta_{\hat{S}^c}\right|\right|_1,
			\end{split}
		\end{equation*}
	 Subtracting $2\left|\left|\beta_{*,\hat{S}} + \Delta_{\hat{S}}\right|\right|_1 + \left|\left|\Delta_{\hat{S}^c}\right|\right|_1$ from both sides in the previous displays yields 
		$$
		\left|\left|\Delta_{\hat{S}^c}\right|\right|_1 \leq 3 \left|\left|\Delta_{\hat{S}}\right|\right|_1, 
		$$
		as needed. $\blacksquare$
		
		Let $\phi\left(s_*\right)$ be the restricted eigenvalue condition in Assumption \ref{aeigen} evaluated at $s=s_{*}$, i.e, let 
			$$
		\phi(s_*) = \inf\left\{\frac{\delta^{'}\sum^J_j G_j \delta}{\left|\left|\delta_{S_\beta}\right|\right|^2}, \;\;\; \delta \in \mathbb{R}^r \backslash \left\{0\right\}, \left|\left|\delta_{S^c_{\beta}}\right|\right|_1 \leq 3 \left|\left|\delta_{S_{\beta}}\right|\right|_1,\;\; \# S_\beta \leq s_*\right\} > c.
		$$
		In addition, let $\hat{\phi}(s_*)$ denote its sample counterpart: 
		  	$$
		  \hat{\phi}(s_*) = \inf\left\{\frac{\delta^{'}\sum^J_j \hat{G}_j \delta}{\left|\left|\delta_{S_\beta}\right|\right|^2}, \;\;\; \delta \in \mathbb{R}^r \backslash \left\{0\right\}, \left|\left|\delta_{S^c_{\beta}}\right|\right|_1 \leq 3 \left|\left|\delta_{S_{\beta}}\right|\right|_1,\;\; \# S_\beta \leq s_*\right\} > c.
		  $$
		 The following lemma is useful as it indicates that this empirical version can be bounded from below in a convenient manner.
		\begin{lemma}
			\label{leigenemp}
			$
			 \hat{\phi}(s_*) \geq \phi(s_*) - O_p(s_*\varepsilon_n)
			$
		\end{lemma}

	\bigskip \noindent \textbf{Proof of Lemma \ref{leigenemp}:} Adding and subtracting $\sum^J_{j=1}G_j$ and the reverse triangle inequality yield 
	\begin{align}
		\left|\delta^{\prime}\sum^J_{j=1}\hat{G}_j\delta\right| & \geq \left|\delta^{\prime}\sum^J_{j=1}G_j\delta\right| - \left|\delta^{\prime}\sum^J_{j=1}\left(\hat{G}_j - G_j\right)\delta \right| \nonumber \\ & \geq   \left|\delta^{\prime}\sum^J_{j=1}G_j\delta\right| - \left|\left|\delta\right|\right|^2_1 \sum^J_{j=1} \left|\left|\hat{G}_j - G_j\right|\right|_{\infty}. \label{inemp1}
	\end{align} 
	Recall that	
	$$
 	\left|\left|\delta_{\hat{S}^c}\right|\right|_1 \leq 3 	\left|\left|\delta_{\hat{S}}\right|\right|_1 \leq 3 \sqrt{s_{*}} \left|\left|\delta_{\hat{S}}\right|\right|_2,
	$$
	where $\hat{S}$ is as defined in Lemma \ref{ldelta} and the last inequality follows by the Cauchy–Schwarz's inequality. We next add $\left|\left|\delta_{\hat{S}}\right|\right|_1$ to both sides to obtain 
	\begin{equation}
		\label{inemp2}
			\left|\left|\delta\right|\right|_1 \leq \sqrt{s_{*}}\left|\left|\delta_{\hat{S}}\right|\right| \Rightarrow \frac{\left|\left|\delta\right|\right|^2_1}{\left|\left|\delta\right|\right|^2} \leq 16s_{*}.
	\end{equation}
Then, if we divide \eqref{inemp1} by $\left|\left|\delta\right|\right|^2$,  \eqref{inemp2} implies 
$$
\frac{	\left|\delta^{\prime}\sum^J_{j=1}\hat{G}_j\delta\right|}{\left|\left|\delta\right|\right|^2} \geq  \frac{	\left|\delta^{\prime}\sum^J_{j=1}G_j\delta\right|}{\left|\left|\delta\right|\right|^2} - 16s_{*} \sum_j^J\left|\left|\hat{G}_j - G_j\right|\right|_{\infty}.
$$
Since $\frac{	\left|\delta^{\prime}\sum^J_{j=1}G_j\delta\right|}{\left|\left|\delta\right|\right|^2} \geq \phi(s_*) $ for all $\delta$ satisfying $	\left|\left|\delta_{\hat{S}^c}\right|\right|_1 \leq 3 	\left|\left|\delta_{\hat{S}}\right|\right|_1$, if we minimize the left-hand side of the expression above and use Lemma \ref{lG}, we obtain
$$
 \hat{\phi}(s_*) \geq \phi(s_*) - O_p(s_*\varepsilon_n),
$$
as needed. $\blacksquare$
	
		\begin{lemma}
			\label{ldeltabound}
			$\left|\left|\Delta\right|\right| \leq C\lambda_n\sqrt{s}$.
		\end{lemma}

	\bigskip \noindent \textbf{Proof of Lemma \ref{ldeltabound}:} We first make the following observation. By Lemma \ref{leigenemp} and the fact that $\varepsilon_n = o(\lambda_n)$, 
	\begin{align}
		\frac{\lambda_n \sqrt{s_{*}}}{\hat{\phi}(s_*)} \left|\left|\Delta_{\hat{S}}\right|\right| & \leq 	\frac{\lambda_n \sqrt{s_{*}}}{ \phi(s_*) - O_p(s_*\varepsilon_n)} \left|\left|\Delta_{\hat{S}}\right|\right| \nonumber \\ & = \frac{\lambda_n \sqrt{s_{*}}}{ \phi(s_*) - o_p(s_*\lambda_n)} \left|\left|\Delta_{\hat{S}}\right|\right| \nonumber \\ & = \frac{1}{\sqrt{s_*} \left(\frac{\phi(s_*)}{s_{*}\lambda_n} - o_p(1)\right)} \left|\left|\Delta_{\hat{S}}\right|\right| \nonumber \\ & = \frac{\sqrt{s_*}\lambda_n}{\phi(s_{*})}\left|\left|\Delta_{\hat{S}}\right|\right| \nonumber \\ & = C  \sqrt{s_*}\lambda_n\left|\left|\Delta_{\hat{S}}\right|\right|. \label{ineresult}
	\end{align}
	Second, let $N$ denote the indices corresponding to the largest $ \# S_{\beta_*}$ entries in $\Delta_{S^c_{\beta_*}}$, so that $N \subseteq S^c_{\beta_*}$, $ \# N = \# S_{\beta_*}$, and $\left|\left|\Delta_k\right|\right|_1 \geq \left| \left| \Delta_q\right| \right|_1$ for any $k \in N$ and $q \in S^c_{\beta_*}\backslash N$. For $\hat{S} = S_{\beta_*} \cup N$ it follows by Assumption \ref{aeigen}, Lemma \ref{lsast}, Lemma \ref{ldelta}, the result in \eqref{ineresult}, and the Cauchy–Schwarz's inequality that with probability approaching one,
		\begin{equation*}
			\begin{split}
				\left|\left|\Delta_{\hat{S}} \right|\right|^2 & \leq \frac{1}{\hat{\phi}(s_{*})} \Delta^{'}\sum^J_{j=1}\hat{G}_j\Delta \\ & \leq \frac{C \lambda_n}{\hat{\phi}(s_{*})}\left|\left|\Delta\right|\right|_1 \\ & = \frac{C \lambda_n}{\hat{\phi}(s_{*})}\left(\left|\left|\Delta_{\hat{S}}\right|\right|_1 + \left|\left|\Delta_{\hat{S}^c}\right|\right|_1 \right) \\ & \leq \frac{C \lambda_n}{\hat{\phi}(s_{*})} \left|\left|\Delta_{\hat{S}}\right|\right|_1 \\ & \leq \frac{C \lambda_n \sqrt{s_*}}{\hat{\phi}(s_{*})}\left|\left|\Delta_{\hat{S}}\right|\right| \\ & \leq C  \sqrt{s_*}\lambda_n\left|\left|\Delta_{\hat{S}}\right|\right|  \\ & \leq C\lambda_n \sqrt{s}\left|\left|\Delta_{\hat{S}}\right|\right|.
			\end{split}
		\end{equation*}
		Then, dividing through by $ \left|\left|\Delta_{\hat{S}} \right|\right|$ then gives, with probability approaching one, 
		\begin{equation}
			\label{deltashat}
			\left|\left|\Delta_{\hat{S}} \right|\right| \leq  C\lambda_n \sqrt{s}.  
		\end{equation}
		By Lemma 6.9 of \cite{buhlmann2011statistics}, Lemma \ref{ldelta}, \eqref{deltashat}, and the Cauchy–Schwarz's inequality
		$$
		\left|\left|\Delta_{\hat{S}^c}\right|\right| \leq \left(\# \hat{S}\right)^{-1/2} \left|\left|\Delta_{\hat{S}^c}\right|\right|_1 \leq 3 \left(\# \hat{S}\right)^{-1/2}\left|\left|\Delta_{\hat{S}}\right|\right|_1 \leq 3 \left(\# \hat{S}\right)^{-1/2} \left(\# \hat{S}\right)^{1/2}\left|\left|\Delta_{\hat{S}}\right|\right| \leq C\lambda_n \sqrt{s}.
		$$
		Hence, by the triangle inequality, with probability approaching one, 
		$$
		\left|\left|\Delta\right|\right| \leq \left|\left|\Delta_{\hat{S}}\right|\right| + \left|\left|\Delta_{\hat{S}^c}\right|\right| \leq C \lambda_n \sqrt{s}. \;\blacksquare
		$$
		
			\begin{lemma}
			\label{betahatop}
			$\left|\left|\hat{\beta}\right|\right|_1 = O_p(1).$
		\end{lemma}
		
		\bigskip \noindent \textbf{Proof of Lemma \ref{betahatop}:} We follow the proof of Lemma D.9 in \cite{bakhitov2022automatic}. Notice that expression \eqref{lambdabound} in the proof of Lemma \ref{ldelta} implies 
		$$
		2\lambda_n\left|\left|\hat{\beta}\right|\right|_1 \leq 2\lambda_n\left|\left|\beta_{*}\right|\right|_1 + \lambda_n\left|\left|\hat{\beta} - \beta_{*}\right|\right|_1.
		$$
		Next, let us divide by $2 \lambda_n$ throughout, then by the triangle inequality 
		\begin{equation*}
			\begin{split}
				\left|\left|\hat{\beta}\right|\right|_1 & \leq \left|\left|\beta_{*}\right|\right|_1 + \frac{1}{2} \left|\left|\hat{\beta} - \beta_*\right|\right|_1 \\ & \leq \left|\left|\beta_{*}\right|\right|_1 + \frac{1}{2}\left( \left|\left|\hat{\beta}\right|\right|_1 + \left|\left|\beta_*\right|\right|_1\right),
			\end{split}
		\end{equation*}
		with probability approaching one. Subtracting $\left|\left|\hat{\beta}\right|\right|_1/2$ from both sides in the previous display and multiplying by 2 yields 
		$$
		\left|\left|\hat{\beta}\right|\right|_1 \leq 3\left|\left|\beta_{*}\right|\right|_1 = O_p(1),
		$$
		as needed. $\blacksquare$
		
		\begin{lemma}
			\label{lmconv}
			$\sum^J_{j=1} \mathbb{E}\left[\left.\left(\left(M_j\left(Z_j\right) - \hat{M}_j\left(Z_j\right)\right)^{'}\hat{\beta}\right)^2\right|\mathcal{W}^c_\ell\right] = O_p\left(s\lambda^2_n\right)$. 
		\end{lemma}
		
		\bigskip \noindent \textbf{Proof of Lemma \ref{lmconv}:} By the triangle inequality, 
		
		\begin{align}
			\sum^J_{j=1} \mathbb{E}\left[\left.\left(\left(M_j\left(Z_j\right) - \hat{M}_j\left(Z_j\right)\right)^{'}\hat{\beta}\right)^2\right|\mathcal{W}^c_\ell\right] & \leq C \sum^J_{j=1} \mathbb{E}\left[\left.\left(\left(M_j\left(Z_j\right) - \hat{M}_j\left(Z_j\right)\right)^{'}\left(\hat{\beta}-\beta_*\right)\right)^2\right|\mathcal{W}^c_\ell\right] \nonumber  \\ & +  C \sum^J_{j=1} \mathbb{E}\left[\left.\left(\left(M_j\left(Z_j\right) - \hat{M}_j\left(Z_j\right)\right)^{'}\beta_*\right)^2\right|\mathcal{W}^c_\ell\right] \nonumber
		\end{align}
		Let us provide a bound for the first term on the right-hand side above. Define 
		$$
		\delta_{j} := \underset{q,k}{\operatorname{max}}\;\;\left|\int \left(\hat{M}_{jq}\left(z_j\right) - M_{jq}\left(z_j\right) \right) \left(\hat{M}_{jk}\left(z_j\right) - M_{jk}\left(z_j\right) \right) F_0(dz_j)\right|,
		$$
		and observe that, by the Cauchy–Schwarz's inequality and Assumption \ref{aconvergenceM},
		\begin{equation*}
			\begin{split}
				\delta_j & \leq \underset{q,k}{\operatorname{max}}\;\; \left|\left|\hat{M}_{jq} - M_{jq}\right|\right|_{2}\left|\left|\hat{M}_{jk} - M_{jk}\right|\right|_{2} \\ & = O_p\left(\varepsilon_{n}^2\right).
			\end{split}
		\end{equation*}
		
		Let
		$$
		\tilde{B}_j = \mathbb{E}\left[\left.\left(M_j\left(Z_j\right) - \hat{M}_j\left(Z_j\right)\right)\left(M_j\left(Z_j\right) - \hat{M}_j\left(Z_j\right)\right)^{'}\right|\mathcal{W}^c_\ell\right].
		$$
		Note that the triangle inequality and Lemma \ref{betahatop} imply that $\left|\left|\Delta\right|\right|_1 = O_p\left(1\right)$. Hence, with probability approaching one,
	\begin{align}
	 \mathbb{E}\left[\left.\left(\left(M_j\left(Z_j\right) - \hat{M}_j\left(Z_j\right)\right)^{'}\left(\hat{\beta}-\beta_*\right)\right)^2\right|\mathcal{W}^c_\ell\right] &  = \left(\hat{\beta} - \beta_{*}\right)^{\prime} \tilde{B}_j \left(\hat{\beta} - \beta_{*}\right) \\  & \leq \delta_j \left|\left|\Delta\right|\right|^2_1 \nonumber\\ & \leq C \varepsilon^2_{n} \nonumber\\ & \leq Cs\varepsilon^2_n \nonumber\\ & \leq Cs\lambda^2_n \label{m11}.
	\end{align}
	where \eqref{m11} follows from $\varepsilon_n = o\left(\lambda_n\right)$. 
		Next, let 
		
		$$
		B_j = \int \left(\hat{M}_{jS_{\beta_{*}}}\left(z_j\right) - M_{jS_{\beta_{*}}}\left(z_j\right)\right)\left(\hat{M}_{jS_{\beta_{*}}}\left(z_j\right) - M_{jS_{\beta_{*}}}\left(z_j\right)\right)^{\prime}F_0(dw).
		$$
		Then, 
		\begin{equation}
			\label{eqeigen}
			\begin{split}
				\mathbb{E}\left[\left.\left(\left(M_j\left(Z_j\right) - \hat{M}_j\left(Z_j\right)\right)^{'}\beta_{*}\right)^2\right|\mathcal{W}^c_\ell\right] \leq \lambda_{max}\left(B_j\right) \left|\left|\beta_{*}\right|\right|^2.
			\end{split}
		\end{equation}
		Let $\left|\left|\cdot\right|\right|_F$ be the Frobenius norm. We shall write, by the Cauchy–Schwarz's inequality and Assumption \ref{aconvergenceM},
		\begin{align}
		\lambda_{max}\left(B_j\right) & \leq \left|\left|B_j\right|\right|_F \nonumber	\\ & \leq s^{*} \underset{k}{\operatorname{max}}\;\; \left|\left|\hat{M}_{jS_{\beta_{*}}k} - M_{jS_{\beta_{*}}k}\right|\right|^2_{2} \nonumber \\ & \leq s^{*} \underset{k}{\operatorname{max}}\;\; \left|\left|\hat{M}_{jk} - M_{jk}\right|\right|^2_{2} \nonumber \\ & \leq Cs^{*}\varepsilon^2_{n} \nonumber\\ & \leq Cs\lambda^2_n \label{m21}.	
		\end{align}
		where \eqref{m21} follows by Lemma \ref{lsast} and the fact that $\varepsilon_{n} = o\left(\lambda_n\right)$. The results in \eqref{m11}, \eqref{eqeigen}, and \eqref{m21} imply that 
		\begin{equation*}
			\begin{split}
				\sum^J_{j=1} \mathbb{E}\left[\left.\left(\left(M_j\left(Z_j\right) - \hat{M}_j\left(Z_j\right)\right)^{'}\hat{\beta}\right)^2\right|\mathcal{W}^c_\ell\right] & \leq Cs\lambda^2_n,
			\end{split}
		\end{equation*}
		as needed. $\blacksquare$

		\begin{lemma}
			\label{lfproj}
			$\sum^J_{j=1}\mathbb{E}\left[\left.\left(f^{*}_j\left(Z_j\right) - \hat{M}_j\left(Z_j\right)^{'}\hat{\beta}\right)^2\right|\mathcal{W}^c_\ell\right] = O_p\left(s\lambda^2_n\right)$.
		\end{lemma}

		\bigskip \noindent \textbf{Proof of Lemma \ref{lfproj}:} By the triangle inequality and Lemma \ref{lfastbetaast}, 
		\begin{align}
			\sum^J_{j=1}\mathbb{E}\left[\left.\left(f^{*}_j\left(Z_j\right) - \hat{M}_j\left(Z_j\right)^{'}\hat{\beta}\right)^2\right|\mathcal{W}^c_\ell\right] & \leq C \sum^J_{j=1}\mathbb{E}\left[\left(f^{*}_j\left(Z_j\right) - M_j\left(Z_j\right)^{'}\beta_*\right)^2\right] \nonumber \\ & + C\sum^J_{j=1}\mathbb{E}\left[\left.\left(M_j(Z_j)^{'}\beta_*-\hat{M}_j\left(Z_j\right)^{'}\hat{\beta}\right)^2\right|\mathcal{W}^c_\ell\right] \nonumber \\ & \leq Cs\varepsilon^2_n + C\sum^J_{j=1}\mathbb{E}\left[\left.\left(M_j(Z_j)^{'}\beta_*-\hat{M}_j\left(Z_j\right)^{'}\hat{\beta}\right)^2\right|\mathcal{W}^c_\ell\right] \label{epsM}. 
		\end{align}
		Next, by the triangle inequality 
		\begin{align}
			\sum^J_{j=1}\mathbb{E}\left[\left.\left(M_j(Z_j)^{'}\beta_*-\hat{M}_j\left(Z_j\right)^{'}\hat{\beta}\right)^2\right|\mathcal{W}^c_\ell\right] & \leq C \sum^J_{j=1}\mathbb{E}\left[\left(M_j(Z_j)^{'}\left(\beta_*- \hat{\beta}\right)\right)^2\right] \label{mb1}\\ & + C  \sum^J_{j=1}\mathbb{E}\left[\left.\left(\left(M_j(Z_j) - \hat{M}_j(Z_j)\right)^{'} \hat{\beta}\right)^2\right|\mathcal{W}^c_\ell\right] \label{mb2}
		\end{align}
		We now find a bound for \eqref{mb1}. Since the maximum eigenvalue of $\sum^J_{j=1}G_j$ is bounded, and using Lemma \ref{ldeltabound}, we have 
		\begin{equation}
			\label{mb11}
			\sum^J_{j=1}\mathbb{E}\left[\left(M_j(Z_j)^{'}\left(\beta_*- \hat{\beta}\right)\right)^2\right]  \leq \lambda_{max}\left(\sum^J_{j=1}G_j\right) \left|\left|\Delta\right|\right|^2 \leq Cs\lambda^2_n.
		\end{equation}
		Furthermore, by Lemma \ref{lmconv} we know that 
		\begin{equation}
			\label{mb21}
			\begin{split}
				\sum^J_{j=1} \mathbb{E}\left[\left.\left(\left(M_j\left(Z_j\right) - \hat{M}_j\left(Z_j\right)\right)^{'}\hat{\beta}\right)^2\right|\mathcal{W}^c_\ell\right] & \leq Cs\lambda^2_n,
			\end{split}
		\end{equation}
		Plugging the results in \eqref{mb11} and \eqref{mb21} into \eqref{mb1} and \eqref{mb2}, respectively, yields 
		$$
		\sum^J_{j=1}\mathbb{E}\left[\left.\left(M_j(Z_j)^{'}\beta_*-\hat{M}_j\left(Z_j\right)^{'}\hat{\beta}\right)^2\right|\mathcal{W}^c_\ell\right] \leq  Cs\lambda^2_n.
		$$
		Using the last displays and that $\varepsilon_n = o\left(\lambda_n\right)$ in \eqref{epsM}  gives the desired result. $\blacksquare$
		
		\bigskip \noindent \textbf{Proof of Theorem \ref{tboundkappa}:}  By Lemma \ref{lfproj}, 
		\begin{equation*}
			\begin{split}
				\left|\left|\hat{\kappa} - \kappa_0 \right|\right|^2_{L^2(Z)} & = \sum^J_{j=1}\mathbb{E}\left[\left.\left(f^{*}_j\left(Z_j\right) - \hat{M}_j\left(Z_j\right)^{'}\hat{\beta}\right)^2\right|\mathcal{W}^c_\ell\right] \\ & \leq Cs\lambda^2_n.
			\end{split}
		\end{equation*}
		The result then follows by Assumption \ref{assespslam}. $\blacksquare$
		
	\bigskip \noindent \textbf{Proof of Lemma \ref{lFconvg}:} Let $\Gamma_{j\ell}$ be the event that $\left|\left|\hat{M}_{j\ell k} - M_{jk}\right|\right|_{2}<\epsilon$ and observe that $\mathbb{P}\left(\Gamma_{j\ell}\right) \rightarrow 1$, by Assumption \ref{lipschitzcont}, for all $j$, $\ell$, $k$. Let us define 
	$$
	T_{ij}\left(\tilde{M}_{jk}\right) := f_j(Z_{ji}) \tilde{M}_{jk}(Z_{ji}) - \mathbb{E}\left[\left.f_j(Z_{ji})\tilde{M}_{jk}(Z_{ji})\right|\mathcal{W}^c_\ell\right],\;\;\;\; U_{j\ell}\left(\tilde{M}_{jk}\right) := \frac{1}{n - n_\ell} \sum_{i \notin I_\ell} T_{ij}\left(\tilde{M}_{jk}\right).
	$$
	Observe that for any constant $C^{\prime}$ and the event $\mathcal{A} = \left\{\underset{1\leq k \leq r}{\operatorname{max}} \left|U_{j\ell}\left(\hat{M}_{j\ell k}\right)\right| \geq C^{\prime}\bar{\varepsilon}_n\right\}$, where recall that $\bar{\varepsilon}_n = \sqrt{\log(r)/n}$, we have 
	\begin{equation*}
		\begin{split}
			\mathbb{P}\left(\mathcal{A}|\mathcal{W}^c_\ell\right) = & \mathbb{P}\left(\mathcal{A}|\Gamma_{j\ell},\mathcal{W}^c_\ell\right)\mathbb{P}\left(\Gamma_{j\ell}|\mathcal{W}^c_\ell\right) + \mathbb{P}\left(\mathcal{A}|\Gamma^c_{j\ell},\mathcal{W}^c_\ell\right)\left(1-\mathbb{P}\left(\Gamma_{j\ell}|\mathcal{W}^c_\ell\right)\right) \\ \leq & \mathbb{P}\left(\left.\underset{1\leq k \leq r}{\operatorname{max}}\left|U_{j\ell}\left(\hat{M}_{j\ell k}\right)\right|\geq C^{\prime}\bar{\epsilon}_n\right|\Gamma_{j\ell},\mathcal{W}^c_\ell\right) + 1-\mathbb{P}\left(\left.\Gamma_{j\ell}\right|\mathcal{W}^c_\ell\right).  
		\end{split}
	\end{equation*}
	By Lemma C.2 of \cite{bradic2022minimax}, there is a $C^{\prime}$ sufficiently large such that for any $\delta>0$, with probability approaching one, 
	$$
	\mathbb{P}\left(\left.\underset{1\leq k \leq r}{\operatorname{max}}\left|U_{j\ell}\left(\hat{M}_{j\ell k}\right)\right|\geq C^{\prime}\bar{\varepsilon}_n\right|\Gamma_{j\ell},\mathcal{W}^c_\ell\right) < \frac{\delta}{2},
	$$
	Since $1-\mathbb{P}\left(\Gamma_{j\ell}\right) \rightarrow 0$, we obtain that $\mathbb{P}\left(\mathcal{A}|\mathcal{W}^c_\ell\right) < \delta$. Hence, 
	$$
	\left|\left|U_{j\ell}\left(\hat{M}_{j\ell }\right)\right|\right|_{\infty} = \underset{1\leq k \leq r}{\operatorname{max}}\left|U_{j\ell}\left(\hat{M}_{j\ell k}\right)\right| = O_p\left(\bar{\varepsilon}_n\right).
	$$
	Furthermore, observe that for each $\ell$ and for some $\tilde{k}$, we have, by Hölder's inequality, 
	\begin{equation*}
		\begin{split}
			\left|\left|\mathbb{E}\left[\left.f_j(Z_{j})\hat{M}_{j\ell}(Z_{j})\right|\mathcal{W}^c_\ell\right] - \mathbb{E}\left[f_j(Z_{j})M_{j}(Z_{j})\right]\right|\right|_{\infty} \leq &  \left(\mathbb{E}\left[\left|f_{j\tilde{k}}\left(Z_j\right)\right|\right]\right)^{1/2} \left|\left|\hat{M}_{j\ell \tilde{k}} - M_{j\tilde{k}}\right|\right|_{2} \\ & = O_p\left(n^{-d_j}\right),
		\end{split}
	\end{equation*}
	where the last displays follows by Assumption \ref{lipschitzcont}. The conclusion of the lemma then follows by the triangle inequality. $\blacksquare$
	
	\subsection{Sufficient conditions for Assumption \ref{lipschitzcont}}
	
	This section provide a set of sufficient conditions for Lipschitz continuity of $M_{jk}$, for all $j$ and $k$, as required by Assumption \ref{lipschitzcont}. 
	\begin{assumption}
		\label{assMconvg}
		 For $\left|\left|\theta - \theta_0\right|\right|$, $\left|\left|\bar{\theta} - \theta_0\right|\right|$, $\left|\left|\eta - \eta_0\right|\right|_{\Xi}$, $\left|\left|\bar{\eta} - \eta_0\right|\right|_{\Xi}$, $\left|\left|\alpha_{j}\left(\cdot,\theta_0,\eta_0\right) -\alpha_{0j}\left(\cdot,\theta_0,\eta_0\right)\right|\right|_2$, \\ $\left|\left|\sigma_{j}\left(\cdot,\alpha_{j^{\prime}}\left(V,\theta_0,\eta_0\right),\theta_0,\eta_0\right) - \sigma_{0j}\left(\cdot,\alpha_{j^{\prime}}\left(V,\theta_0,\eta_0\right),\theta_0,\eta_0\right)\right|\right|_2$ small enough, for all $j$ and $k$,
			\begin{align}	
				\left|\left| \sigma_{0j}\left(\cdot,\alpha_{j^{\prime}}\left(V,\bar{\theta},\bar{\eta}\right),\theta,\eta\right) - \sigma_{0j}\left(\cdot,\alpha_{j^{\prime}}\left(V,\bar{\theta},\bar{\eta}\right),\theta_0,\eta\right)\right|\right|_2  & \leq C \left|\left|\theta - \theta_0\right|\right|, \label{cont11}\\ 		
				\left|\left| \sigma_{0j}\left(\cdot,\alpha_{j^{\prime}}\left(V,\bar{\theta},\bar{\eta}\right),\theta_0,\eta\right) - \sigma_{0j}\left(\cdot,\alpha_{j^{\prime}}\left(V,\bar{\theta},\bar{\eta}\right),\theta_0,\eta_0\right)\right|\right|_2  & \leq C \left|\left|\eta - \eta_0\right|\right|_{\Xi},  \label{cont2}\\ 	
				\left|\left| \sigma_{0j}\left(\cdot,\alpha_{j^{\prime}}\left(V,\bar{\theta},\bar{\eta}\right),\theta_0,\eta_0\right) - \sigma_{0j}\left(\cdot,\alpha_{0j^{\prime}}\left(V,\bar{\theta},\bar{\eta}\right),\theta_0,\eta_0\right)\right|\right|_{2} & \leq C \left|\left|\alpha_{j^{\prime}}\left(\cdot,\bar{\theta},\bar{\eta}\right) - \alpha_{0j^{\prime}}\left(\cdot,\bar{\theta},\bar{\eta}\right)\right|\right|_{2},  \label{cont3}\\
				\left|\left| \sigma_{0j}\left(\cdot,\alpha_{0j^{\prime}}\left(V,\bar{\theta},\bar{\eta}\right),\theta_0,\eta_0\right) - \sigma_{0j}\left(\cdot,\alpha_{0j^{\prime}}\left(V,\theta_0,\bar{\eta}\right),\theta_0,\eta_0\right)\right|\right|_{2}  & \leq C \left|\left|\bar{\theta}-\theta_0\right|\right|,  \label{cont4}\\
				\left|\left| \sigma_{0j}\left(\cdot,\alpha_{0j^{\prime}}\left(V,\theta_0,\bar{\eta}\right),\theta_0,\eta_0\right) - \sigma_{0j}\left(\cdot,\alpha_{0j^{\prime}}\left(V,\theta_0,\eta_0\right),\theta_0,\eta_0\right)\right|\right|_{2}  & \leq C \left|\left|\bar{\eta}-\eta_0\right|\right|_{\Xi}.  \label{cont5}
			\end{align}
	\end{assumption}
	
	Under Assumption \ref{assMconvg}, Lipschitz continuity of $M_{jk}$ follows by the triangle inequality. Let us provide sufficient conditions for Assumption \ref{assMconvg} in the context of our example. 
	
	\bigskip \noindent \textsc{Example:} Recall that in this example, the regressors have the following expression
	\begin{equation*}
		\begin{split}
			\hat{M}_{1\ell}\left(Z_{1i}\right) & = \left(\gamma_{11}\left(Z_{1i}\right), \cdots, \gamma_{1r_1}\left(Z_{1i}\right),  \hat{\theta}_{\omega\ell}\gamma_{21}\left(Z_{1i}\right), \cdots \hat{\theta}_{\omega\ell}\gamma_{2r_2}\left(Z_{1i}\right) \right), \\
			\hat{M}_{2\ell}\left(Z_{1i}\right) & = \hat{\theta}_{\omega \ell}	\hat{M}_{1\ell}\left(Z_{1i}\right), \\
			\hat{M}_{3\ell}\left(Z_{2i}\right) & = \left(\gamma_{31}\left(Z_{2i}\right), \cdots, \gamma_{3r_3}\left(Z_{2i}\right),  \hat{\theta}_{\omega\ell}\gamma_{41}\left(Z_{2i}\right), \cdots \hat{\theta}_{\omega\ell}\gamma_{4r_4}\left(Z_{2i}\right) \right), \\
			\hat{M}_{4\ell}\left(Z_{2i}\right) & = \hat{\theta}_{\omega \ell}	\hat{M}_{3\ell}\left(Z_{2i}\right).
		\end{split}
	\end{equation*}
	 Therefore, showing that Assumption \ref{assMconvg} holds in this example is equivalent to show that the functions $\theta_\omega\gamma_{1k}$, $\theta_\omega\gamma_{2k}$, $\theta^2_\omega\gamma_{2k}$, $\theta_\omega\gamma_{3k}$, $\theta_\omega\gamma_{4k}$, $\theta^2_\omega\gamma_{4k}$ are Lipschitz-continuous in $\theta_{\omega}$ at $\theta_{0\omega}$, with $\left|\theta_\omega - \theta_{0\omega}\right| < \epsilon$. This holds as $\left|\left|\gamma_{sk}\right|\right|_2$ is bounded for all $k$ and $s=1,2,3,4$. $\square$

\section{Asymptotic Results of the Parameter of Interest}
		\label{appendixparameter}

First, we shall point out the following. Notice that Lemma \ref{lFconvg} and Theorem \ref{tboundkappa} imply that the rate for OR-IVs can be written as $\mu^\kappa_n = n^{2c}n^{-2\xi d/(2\xi + 1)}$. Let 
$$
c = \left[2\xi d/(2\xi + 1) + d_\eta - 1/2\right]/4 = \left[(2\xi d + d_\eta(2\xi+1))/(2\xi + 1) - 1/2\right]/4 >0,
$$ 
where the last inequality follows by Assumption \ref{rates}. Notice that $\left|\left|\hat{\kappa} - \kappa_0\right|\right|_{L^2(Z)} = o_p(1)$. In addition, 
$$
\sqrt{n}\left|\left|\hat{\kappa} - \kappa_0\right|\right|_{L^2(Z)}\left|\left|\hat{\eta} - \eta_0\right|\right|_{\Xi} = \sqrt{n}o_p(n^{2c} n^{-\frac{2\xi d + d_\eta(2\xi + 1)}{2\xi + 1}}) = o_p(n^{2c}n^{-4c}) = o_p(1).
$$

			\begin{lemma}
			\label{deltaresult}
			Let Assumptions \ref{gateaux},  \ref{contozero}-\ref{rates}  hold. Then, 
			$$
			i)\; \int \left|\left|\hat{\Delta}_\ell(w)\right|\right|^2 F_0(dw) \overset{p}{\to} 0,  \;\;\;\; and \;\;\;\; ii)\; \sqrt{n}\int \hat{\Delta}_\ell(w) F_0(dw) \overset{p}{\to} 0.
			$$
		\end{lemma}
		
		\bigskip \noindent \textbf{Proof of Lemma \ref{deltaresult}:} Observe that by the triangle inequality, and Assumptions \ref{contozero}-\ref{estspaces}, we have 
		\begin{equation*}
			\begin{split}
				\int \left|\left|\hat{\Delta}_\ell(w)\right|\right|^2 F_0(dw) & = \int \left|\left|\sum^J_{j=1}\left(m_j\left(y,\theta_0,\hat{\eta}_{\ell}\right) - m_j\left(y,\theta_0,\eta_{0}\right)\right)\left(\bm{\hat{\kappa}_{j\ell}(z_j)} - \bm{\kappa_{0j}(z_j)}\right)\right|\right|^2 F_0(dw) \\ & \leq C\sum^J_{j=1} \int \left|m_j\left(y,\theta_0,\hat{\eta}_{\ell}\right) - m_j\left(y,\theta_0,\eta_{0}\right)\right|^2 \left|\left|\bm{\hat{\kappa}_{j\ell}(z_j)} - \bm{\kappa_{0j}(z_j)}\right|\right|^2 F_0(dw) \\ & \leq O_p(1) \sum^J_{j=1} \int \left|m_j\left(y,\theta_0,\hat{\eta}_{\ell}\right) - m_j\left(y,\theta_0,\eta_{0}\right)\right|^2 F_0(dw) \\ & \overset{p}{\to} 0.
			\end{split}
		\end{equation*}
		Second, let us show \textit{ii)}. By the Cauchy-Schwarz's inequality,
		\begin{align}
			\left|\left|\sqrt{n} \int \hat{\Delta}_\ell(w) F_0(dw)\right|\right|  \leq & \sqrt{n} \left(\int \sum^J_{j=1}\mathbb{E}\left[\left.m_j\left(y,\theta_0,\hat{\eta}_{\ell}\right) - m_j\left(y,\theta_0,\eta_{0}\right)\right|z_j,\mathcal{W}^c_\ell\right]^2F_0(dz_j)\right)^{1/2} \nonumber \\ & \left|\left|\left(\int \sum^J_{j=1}\left(\bm{\hat{\kappa}_{j\ell}(z_j)} - \bm{\kappa_{0j}(z_j)}\right)^2F_0(dz_j)\right)^{1/2}\right|\right| \label{prodmk}
		\end{align}
		By Fréchet differentiability  of the functions $h_j$ with respect to $\eta$ at $\eta_0$ \citep[see Proposition 7.2.3 in][]{luenberger1997optimization}, with probability approaching one,
		\begin{align}
			\left(\int \sum^J_{j=1}\mathbb{E}\left[\left.m_j\left(y,\theta_0,\hat{\eta}_{\ell}\right) - m_j\left(y,\theta_0,\eta_{0}\right)\right|z_j,\mathcal{W}^c_\ell\right]^2F_0(dz_j)\right)^{1/2} & = \left(\sum^J_{j=1}\left|\left|h_j\left(\cdot,\theta_0,\hat{\eta}_\ell\right) - h_j\left(\cdot,\theta_0,\eta_0\right)\right|\right|^2_2\right)^{1/2} \\  & \leq  C \left|\left|\hat{\eta}_\ell - \eta_0\right|\right|_{\Xi} \nonumber \\& = O_p\left(n^{-d_\eta}\right). \label{opeta}
		\end{align}
		Using \eqref{opeta} and Assumption \ref{rates} in \eqref{prodmk} yields, with probability approaching one,
		$$
		\left|\left|\sqrt{n} \int \hat{\Delta}_\ell(w) F_0(dw)\right|\right| = O_p\left(\sqrt{n}n^{-d_\eta}n^c\varepsilon^{2\xi/(2\xi + 1)}_n\right)  \rightarrow 0,
		$$
	Then, the conclusion follows. $\blacksquare$
		
		\begin{lemma}
			\label{boundpsibar}
			Let Assumption \ref{twicefre} hold. Then, there is a $C>0$ such that for $\left|\left|\eta - \eta_0\right|\right|_{\Xi}$ small enough, 
			$$
			\left|\left|\overline{\psi}\left(\theta_0,\eta,\bm{\kappa_0}\right)\right|\right| \leq C \left|\left|\eta - \eta_0\right|\right|^2_{\Xi}.
			$$
		\end{lemma}
		
		\bigskip \noindent \textbf{Proof of Lemma \ref{boundpsibar}:} The result follows by Proposition 7.3.3 of \cite{luenberger1997optimization}. $\blacksquare$

		\bigskip To prove Lemma \ref{keylemma1}, let $\bm{\imath_q}$ be a $q-$dimensional vector of ones and define
		\begin{equation*}
			\begin{split}
				g\left(W_i,\theta,\eta\right) & := \sum^J_{j=1}m_j\left(Y_i,\theta,\eta\right)\bm{\imath_q},\\
				\phi\left(W_i,\theta,\eta,\bm{\kappa}\right) & := \sum^J_{j=1}m_j\left(Y_i,\theta,\eta\right)\left(\bm{\kappa_{0j}(Z_{ji})} - \bm{\imath_q}\right).
			\end{split}
		\end{equation*}
		Then, we can write 
		\begin{equation}
			\label{newrep}
			\psi\left(W_i,\theta,\eta,\bm{\kappa}\right) = g\left(W_i,\theta,\eta\right) + \phi\left(W_i,\theta,\eta,\bm{\kappa}\right).
		\end{equation}
		
		Notice that, by using representation \eqref{newrep}, we have written the LR functions as the sum of two terms $g + \phi$ as in CEINR (Equation (2.3)). Two differences are worth mentioning. First, instead of having $\alpha_0$, typically obtained as a Riesz representer, entering in $\phi$, we have the OR-IVs. Second, $g$ and $\phi$, by construction, are evaluated at the same $\theta$. In particular,  $\psi\left(W_i,\theta_0,\eta,\bm{\kappa}\right) = g\left(W_i,\theta_0,\eta\right) + \phi\left(W_i,\theta_0,\eta,\bm{\kappa}\right)$.  CEINR allow for $g$ and $\phi$ being evaluated at different $\theta$'s as $\phi$ has mean zero when evaluated at the true nuisance parameters value, for any $\theta \in \Theta$. In our case, that is true only at $\theta = \theta_0$.
		
		\bigskip \noindent \textbf{Proof of Lemma \ref{keylemma1}:} To show the result we will verify the conditions of Lemma 8 of  CEINR and restrict $g$ and $\phi$ to be always evaluated at the same $\theta$.  
		
		First, note that, by Assumption \ref{contozero}, $\mathbb{E}\left[\left|\left|\psi\left(W,\theta_0,\eta_0,\bm{\kappa_0}\right)\right|\right|^2\right] < \infty$. Moreover, by the triangle inequality, 
		\begin{equation}
			\label{a11}
			\int \left|\left| g\left(y,\theta_0,\hat{\eta}_\ell\right) - g\left(y,\theta_0,\eta_0\right) \right|\right|^2 F_0(dw) \leq C \left|\left|\bm{\imath_q}\right|\right|^2 \sum^J_{j=1} \int \left|m_j\left(y,\theta_0,\hat{\eta}_{\ell}\right) - m_j\left(y,\theta_0,\eta_{0}\right)\right|^2  F_0(dw).
		\end{equation}
		Hence, \eqref{a11} and Assumption \ref{contozero} \textit{i)} imply Assumption 1 \textit{(i)} of CEINR. Similarly, by the triangle inequality and Assumption \ref{contozero} \textit{i)}, 
		\begin{align}
			\int & \left|\left| \phi\left(w,\theta_0,\hat{\eta}_\ell, \bm{\kappa_0}\right) - \phi\left(w,\theta_0,\eta_0,\bm{\kappa_0}\right) \right|\right|^2 F_0(dw) \nonumber \leq \\ &  C\sum^J_{j=1} \int \left|m_j\left(y,\theta_0,\hat{\eta}_{\ell}\right) - m_j\left(y,\theta_0,\eta_{0}\right)\right|^2 \left|\left|\bm{\kappa_{0j}(z_{j})}\right|\right|^2 F_0(dw) + o_p(1)  \label{a12} . 
		\end{align}
		Then, Assumption \ref{contozero} \textit{ii)} implies Assumption 1 \textit{(ii)} of CEINR. By the triangle inequality, we can show
		\begin{align}
			\label{a13}
			\int \left|\left|\phi\left(w,\theta_0,\eta_0,\bm{\hat{\kappa}_\ell}\right) - \phi\left(w,\theta_0,\eta_0,\bm{\kappa_0}\right)\right|\right|^2 F_0(dw) & \leq  C\sum^J_{j=1} \int \left|m_j\left(y,\theta_0,\eta_{0}\right)\right|^2 \left|\left|\bm{\hat{\kappa}_{j\ell}(z_j)} - \bm{\kappa_{j\ell}(z_j)}\right|\right|^2 F_0(dw). \nonumber
		\end{align}
		Therefore, Assumption \ref{contozero} \textit{iii)} implies Assumption 1 \textit{(iii)} of CEINR.\footnote{Assumption 1 \textit{(iii)} of CEINR is a convergence condition for estimators  $\hat{\theta}_\ell$ and $\bm{\hat{\kappa}_\ell}$, but since we are restricting $g$ and $\phi$ to be evaluated at $\theta = \theta_0$, we only need a convergence condition for $\bm{\kappa_0}$.}
		
		Observe that 
		$$
		\hat{\Delta}_{\ell}(w) = \phi\left(w,\theta_0,\hat{\eta}_\ell, \bm{\hat{\kappa}_\ell}\right) - \phi\left(w,\theta_0,\eta_0, \bm{\hat{\kappa}_\ell}\right) - \phi\left(w,\theta_0,\hat{\eta}_\ell,\bm{\kappa_0}\right) + \phi\left(w,\theta_0,\eta_0, \bm{\kappa_0}\right).
		$$
		Then, Lemma \ref{deltaresult} implies Assumption 2 $\textit{(i)}$ of CEINR. 
		
		By Assumption \ref{estspaces}, we have $\int \phi\left(w,\theta_0,\eta_0,\bm{\hat{\kappa}_\ell}\right) F_0(dw) = 0$ with probability approaching one for any fold $\ell$. Furthermore, $\left|\left|\bar{\psi}\left(\theta_0,\eta,\bm{\kappa_0}\right)\right|\right| \leq C \left|\left|\eta - \eta_0\right|\right|^2_{\Xi}$, by Lemma \ref{boundpsibar}. Since $d_\eta > 1/4$, we obtain $\sqrt{n}\left|\left|\bar{\psi}\left(\theta_0,\hat{\eta}_\ell,\bm{\kappa_0}\right)\right|\right| \overset{p}{\to} 0 $. These results verify Assumption 3 \textit{(i)} and  \textit{(iii)} of CEINR. Then, all the conditions of Lemma 8 of CEINR hold in our context and the conclusion of our Lemma \ref{keylemma1} can be obtained. $\blacksquare$
		
		\bigskip Let $\Psi = \mathbb{E}\left[\psi\left(W,\theta_0,\eta_0,\bm{\kappa_0}\right)\psi\left(W,\theta_0,\eta_0,\bm{\kappa_0}\right)^{'}\right]$. We next show that 
		\begin{lemma}
			\label{Psiconv}
			Let Assumptions \ref{contozero} and \ref{convmall} hold. Then, $\hat{\Psi} \overset{p}{\to} \Psi$. 
		\end{lemma}
		
		\bigskip \noindent \textbf{Proof of Lemma \ref{Psiconv}:} This proof closely follows the proof of Lemma E1 of CEINR. For each $i \in I_\ell$, let $\hat{\Delta}_{\ell}\left(W_i\right)$ be as in the main text. Additionally, let
		\begin{equation*}
			\begin{split}
				\hat{R}_{1\ell i} & := g\left(W_i,\theta_0,\hat{\eta}_\ell\right) - g\left(W_i,\theta_0,\eta_0\right) = \sum^J_{j=1}\left(m_j\left(Y_i, \theta_0,\hat{\eta}_{\ell }\right) -  m_j\left(Y_i, \theta_0,\eta_{0}\right)\right)\bm{\imath_q}, \\
				\hat{R}_{2\ell i} & := \phi\left(W_i,\theta_0,\hat{\eta}_\ell,\bm{\kappa_0}\right) - \phi\left(W_i,\theta_0,\eta_0,\bm{\kappa_0}\right) = \sum^J_{j=1}\left(m_j\left(Y_i,\theta_0,\hat{\eta}_\ell\right) - m_j\left(Y_i,\theta_0,\eta_0\right)\right)\left(\bm{\kappa_{0j}(Z_{ji})} - \bm{\imath_q}\right),
				\\  \hat{R}_{3\ell i} & := \phi\left(W_i,\theta_0,\eta_0,\bm{\hat{\kappa}_\ell}\right) - \phi\left(W_i,\theta_0,\eta_0,\bm{\kappa_0}\right) = \sum^J_{j=1} m_j\left(Y_i,\theta_0,\eta_0\right) \left(\bm{\hat{\kappa}_{j\ell}\left(Z_{ji}\right)} - \bm{\kappa_{0j}\left(Z_{ji}\right)}\right),\\ 
				\hat{R}_{4\ell i} & := \sum^J_{j=1}\left(m_j\left(Y_i,\tilde{\theta}_\ell, \hat{\eta}_{\ell}\right) - m_j\left(Y_i,\theta_0,\hat{\eta}_{\ell}\right)\right)\bm{\hat{\kappa}_{j\ell}\left(Z_{ji}\right)}. 
			\end{split}
		\end{equation*}
		By Assumption \ref{contozero}, $\mathbb{E}\left[\left.\left|\left|\hat{R}_{k\ell i}\right|\right|^2\right|\mathcal{W}^c_{\ell}\right] \overset{p}{\to} 0$, $k=1,2,3$. Similarly, by Assumption \ref{convmall}, $\mathbb{E}\left[\left.\left|\left|\hat{R}_{4\ell i}\right|\right|^2\right|\mathcal{W}^c_{\ell}\right] \overset{p}{\to} 0$. Also,  by Lemma \ref{deltaresult} \textit{i)}, $\mathbb{E}\left[\left.\left|\left|\hat{\Delta}_\ell(W)\right|\right|^2\right|\mathcal{W}^c_\ell\right] \overset{p}{\to} 0$. Then, it follows for $\psi_i = \psi\left(W_i,\theta_0,\eta_0,\bm{\kappa_0}\right)$, 
		\begin{equation*}
			\begin{split}
				\mathbb{E}\left[\left.\frac{1}{n}\sum_{i \in I_\ell} \left|\left|\hat{\psi}_{i \ell} - \psi_{i}\right|\right|^2\right| \mathcal{W}^c_\ell\right] & \leq \mathbb{E}\left[\left.\left|\left|\hat{\psi}_{i \ell} - \psi_{i}\right|\right|^2\right|\mathcal{W}^c_\ell\right] \\ & \leq C \left(\sum^4_{k=1}\mathbb{E}\left[\left.\left|\left|\hat{R}_{k\ell i}\right|\right|^2\right|\mathcal{W}^c_\ell\right] + \mathbb{E}\left[\left.\left|\left|\hat{\Delta}_\ell(W_i)\right|\right|^2\right|\mathcal{W}^c_\ell\right]\right) \\ & \overset{p}{\to} 0. 
			\end{split}  
		\end{equation*}
		Hence, by the Conditional Markov's inequality, $\frac{1}{n}\sum_{i \in I_\ell} \left|\left|\hat{\psi}_{i \ell} - \psi_{i}\right|\right|^2 \overset{p}{\to} 0$. Let $\tilde{\Psi} = \frac{1}{n} \sum^n_{i=1} \psi_i \psi^{'}_i$. Then, by the triangle inequality and the Cauchy-Schwarz's inequality,\footnote{With some abuse of notation, when $A$ is a generic matrix, let $\left|\left|A\right|\right|$ denote the spectral norm of $A$.} 
		\begin{equation*}
			\begin{split}
				\left|\left|\hat{\Psi} - \tilde{\Psi}\right|\right| & \leq \sum^L_{\ell = 1} \frac{1}{n} \sum_{i \in I_\ell}\left(\left|\left|\hat{\psi}_{i\ell} - \psi_i\right|\right|^2 + 2\left|\left|\psi_i\right|\right|\left|\left|\hat{\psi}_{i\ell} - \psi_i\right|\right|\right) \\ & \leq \underbrace{\sum^L_{\ell = 1} \frac{1}{n} \sum_{i \in I_\ell}\left|\left|\hat{\psi}_{i\ell} - \psi_i\right|\right|^2}_{o_p(1)}  + \underbrace{2 \sum^L_{\ell = 1} \left(\frac{1}{n} \sum_{i \in I_\ell} \left|\left|\psi_i\right|\right|^2\right)^{1/2}}_{O_p(1)} \underbrace{\left(\frac{1}{n} \sum_{i \in I_\ell}\left|\left|\hat{\psi}_{i\ell} - \psi_i\right|\right|^2\right)^{1/2}}_{o_p(1)} \\ & = o_p(1)\left(O_p(1) + 1\right)  \overset{p}{\to} 0.
			\end{split} 
		\end{equation*}
		Moreover, by the law of large numbers, $\tilde{\Psi} \overset{p}{\to} \Psi$. Hence, the conclusion of the lemma follows by the triangle and the Conditional Markov's inequalities. $\blacksquare$
		
		\begin{lemma}
			\label{jacobianconv}
			Let Assumption \ref{ajacobian} hold and $\bar{\theta} \overset{p}{\to} \theta_0$. Then, $\frac{\partial \hat{\psi}(\bar{\theta})}{\partial \theta} \overset{p}{\to} \Upsilon$.
		\end{lemma}
		
		\bigskip\noindent \textbf{Proof of Lemma \ref{jacobianconv}:} We follow the proof of Lemma E2 of CEINR. Let $\hat{\Upsilon}_\ell = \frac{1}{n_\ell} \sum_{i \in I_\ell} \frac{\partial \psi\left(W_i,\bar{\theta},\hat{\eta}_\ell,\bm{\hat{\kappa}_\ell}\right)}{\partial \theta}$ and $\tilde{\Upsilon}_\ell = \frac{1}{n_\ell} \sum_{i \in I_\ell} \frac{\partial \psi\left(W_i,\theta_0,\hat{\eta}_\ell,\bm{\hat{\kappa}_\ell}\right)}{\partial \theta}$. Notice that by Assumption \ref{ajacobian} \textit{ii)}, 
		$$
		\mathbb{E}\left[\left.\frac{1}{n_\ell} \sum_{i \in I_\ell} d\left(W_i,\hat{\eta}_\ell, \bm{\hat{\kappa}_\ell}\right)\right|\mathcal{W}^c_\ell\right] = \mathbb{E}\left[\left.d\left(W_i,\hat{\eta}_\ell, \bm{\hat{\kappa}_\ell}\right)\right|\mathcal{W}^c_\ell\right] < C_2,
		$$
		with probability approaching one. Then, by the Conditional Markov's inequality, $\frac{1}{n_\ell} \sum_{i \in I_\ell} d\left(W_i,\hat{\eta}_\ell, \bm{\hat{\kappa}_\ell}\right) = O_p(1)$. Next, by Assumption \ref{ajacobian} \textit{i)} and \textit{ii)}, and the triangle inequality, with probability approaching one,
		\begin{equation*}
			\begin{split}
				\left|\left|\hat{\Upsilon}_\ell - \tilde{\Upsilon}_\ell\right|\right| & \leq \frac{1}{n_\ell} \sum_{i \in I_\ell} d\left(W_i,\hat{\eta}_\ell,\bm{\hat{\kappa}_\ell}\right) \left|\left|\bar{\theta} - \theta_0\right|\right|^{C_1} \\ & = O_p(1)o_p(1)  \overset{p}{\to} 0.
			\end{split}
		\end{equation*}
		Then, $\hat{\Upsilon}_\ell - \tilde{\Upsilon}_\ell \overset{p}{\to} 0$ follows by the Conditional Markov's inequality. Finally, let $\bar{\Upsilon}_\ell = \frac{1}{n_\ell} \sum_{i \in I_\ell} \frac{\partial \psi\left(W_i,\theta_0,\eta_0,\bm{\kappa_0}\right)}{\partial\theta}$. Using Assumption \ref{ajacobian} \textit{iii)}, we have that $\tilde{\Upsilon}_\ell - \bar{\Upsilon}_\ell \overset{p}{\to} 0$. What is more, by the law of large numbers, $\bar{\Upsilon}_\ell \overset{p}{\to} \Upsilon$. Hence, the conclusion follows by the triangle and the Conditional Markov's inequalities. $\blacksquare$

		\bigskip \noindent \textbf{Proof of Theorem \ref{inferencetheta}:} Based on the results of the previous lemmas, the conclusion can be derived using standard asymptotic arguments and the Central Limit Theorem as in, e.g., the proof of Proposition 21.20 in \cite{ruud2000introduction}. $\blacksquare$

		\subsection{Asymptotic Normality Conditions for the Example}
	
		\begin{corollary}
			\label{cnormalityex2}
			Let the assumptions in Lemma  \ref{fde2}, Lemma \ref{lFconvg}, Assumption \ref{estspaces},  and \ref{rates} hold. Let $\hat{\theta} \overset{p}{\to} \theta_0$, $\hat{\Lambda} \overset{p}{\to} \Lambda$, and $\Upsilon^{'} \Lambda \Upsilon$ be non-singular. In addition, assume 
			\begin{itemize}
				\item[i)] The functions  $\left|\left|\bm{\kappa_{01}(Z_1)}\right|\right|$, $\left|\left|\bm{\kappa_{02}(Z_1)}\right|\right|$, $\left|\left|\bm{\kappa_{03}(Z_2)}\right|\right|$,  $\left|\left|\bm{\kappa_{04}(Z_2)}\right|\right|$, $|Y_t - \eta_{0t}(Z_t)|$, and $|Y_{t+1} - F\left(X_{t+1},\theta_{0p}\right) - \theta_{0\omega}\left(\eta_{0t}(Z_t) - F\left(X_t,\theta_{0p}\right)\right)|$, for all $t$, are bounded a.s. In addition, $b= \left(b_1,b_2\right)$ has bounded entries a.s.; 
				\item [ii)] $\mathbb{E}\left[\left|\eta_{0t}\left(Z_t\right) \right|^2\right] < \infty$, for all $t$. Also, on a neighborhood $\mathcal{N}$ of $\theta_0$,  $\mathbb{E}\left[\left|F\left(X_t,\theta\right)\right|^2\right] < \infty$, for all $t$;
				\item[iii)] $F(X_t,\cdot)$ is differentiable on $\mathcal{N}$ a.s. Furthermore, $\Upsilon$ exists;
				\item[iv)]  There exists $C$, not depending on $X$, and $H_1$, $H_2 >0$ such that $\left|F\left(X_t,\theta_p\right) -  F\left(X_t,\theta_{0p}\right)\right| \leq C \left|\left|\theta_p - \theta_{0p}\right|\right|^{H_1}$ and $\left|\left|F_\theta\left(X_t,\theta_p\right) -  F\left(X_t,\theta_{0p}\right)\right|\right| \leq C \left|\left|\theta_p - \theta_{0p}\right|\right|^{H_2}$ on $\mathcal{N}$ and $\left|\left|F_\theta\left(X_t,\theta_{p}\right)\right|\right| < C$, where $F_\theta(X_t,\theta) = \frac{\partial}{\partial p}F(X_t,p)|_{p = \theta}$,  for all $t$. Moreover, $\eta_{1}(Z_1)$, $\eta_{2}(Z_2)$, $\left|\left|\bm{\kappa_2(Z_1)}\right|\right|$, and $\left| \left|\bm{\kappa_4(Z_2)} \right|\right|$ exists a.s., for $\left|\left|\eta - \eta_0 \right| \right|_{\Xi}  \mathbb{E}\left[\left|\left|\bm{\kappa(Z)} - \bm{\kappa_0(Z)} \right| \right|^2_{\infty}\right] $ small enough. In addition, $\mathbb{E}\left[\left|\left|\bm{\kappa_2(Z_1)}\right|\right|\right] < \infty$,  $\mathbb{E}\left[\left|\left|\bm{\kappa_4(Z_2)}\right|\right|\right] < \infty$;
				\item[v)] $\mathbb{E}\left[\left|\theta_{0\omega}F_{\theta k}\left(X_t,\theta_{0p}\right) - F_{\theta k}\left(X_{t+1},\theta_{0p}\right)\right|^2\right] < \infty$, for all $t$. 
			\end{itemize}
			Then, in the example, it follows that 
			$$
			\sqrt{n}\left(\hat{\theta} - \theta_0\right) \overset{d}{\to}\;\;N\left(0,\Sigma\right),\;\;\; \Sigma = \left(\Upsilon^{'}\Lambda \Upsilon\right)^{-1} \Upsilon^{'}\Lambda \Psi \Lambda \Upsilon\left(\Upsilon^{'}\Lambda \Upsilon\right)^{-1}.
			$$
			Moreover, $\hat{\Sigma} \overset{p}{\to} \Sigma$.	
		\end{corollary}
		
		 \bigskip \noindent \textbf{Proof of Corollary \ref{cnormalityex2}:} Recall that 
		\begin{equation*}
			\begin{split}
				\psi\left(W,\theta_0,\eta_0, \kappa_0\right) & = \underbrace{\left(Y_1 - \eta_{01}\left(Z_1\right)\right)\bm{\kappa_{01}\left(Z_1\right)}}_{a_1(W)} + \underbrace{\left(Y_2 - F\left(X_2,\theta_{0p}\right) - \theta_{0\omega}\left(\eta_{01}\left(Z_1\right) - F\left(X_1,\theta_{0p}\right)\right)\right)\bm{\kappa_{02}\left(Z_1\right)}}_{a_2(W)} \\ & + \underbrace{\left(Y_2 - \eta_{02}\left(Z_2\right)\right)\bm{\kappa_{03}\left(Z_2\right)}}_{a_3(W)} + \underbrace{\left(Y_3 - F\left(X_3,\theta_{0p}\right) - \theta_{0\omega}\left(\eta_{02}\left(Z_2\right) - F\left(X_2,\theta_{0p}\right)\right)\right)\bm{\kappa_{04}\left(Z_2\right)}}_{a_4(W)}.
			\end{split}
		\end{equation*}
	 By the triangle inequality and Assumption \textit{i)}, 
		$$
		\mathbb{E}\left[\left|\left|\psi\left(W,\theta_0,\eta_0,\bm{\kappa_0}\right)\right|\right|^2\right] \leq C\mathbb{E}\left[||a_1(W)||^2\right] + C\mathbb{E}\left[||a_2(W)||^2\right] + C\mathbb{E}\left[||a_3(W)||^2\right] + C\mathbb{E}\left[||a_4(W)||^2\right] < \infty.
		$$ 
		Consider the $l_2-$ norm for $\eta_t$ and notice that 
		$$
		\int \left|y_1 - \hat{\eta}_{1\ell}(z_1) - \left(y_1 - \eta_{01}(z_1)\right)\right|^2F_0(dw) =  \int \left| \hat{\eta}_{1\ell}(z_1) -  \eta_{01}(z_1)\right|^2F_0(dw)  \overset{p}{\to} 0.
		$$ 
		The same holds mutatis mutandi for the next period. Furthermore,  we have
		\begin{equation*}
			\begin{split}
				\int & \left|y_2 - F\left(x_2,\theta_{0p}\right) - \theta_{0\omega}\left(\hat{\eta}_{1\ell}(z_1) - F\left(x_1,\theta_{0p}\right)\right) \right. \\ &  \left.- \left(y_2 - F\left(x_2,\theta_{0p}\right) - \theta_{0\omega}\left(\eta_{01}(z_1) - F\left(x_1,\theta_{0p}\right)\right)\right) \right|^2F_0(dw) \\ &  = \left|\theta_{0\omega}\right|^2 \int \left|\hat{\eta}_{1\ell}(z_1) - \eta_{01}(z_1)\right|^2 F_0(dw) \overset{p}{\to} 0,
			\end{split}
		\end{equation*}
		and similarly for one period ahead. Then, Assumption \ref{contozero} \textit{i)} holds.  We have that 
		\begin{equation*}
			\begin{split}
				\int \left|y_1 - \hat{\eta}_{1\ell}(z_1) - \left(y_1 - \eta_{01}(z_1)\right)\right|^2 \left|\left|\bm{\kappa_{01}(z_1)}\right|\right|^2 F_0(dw) & \leq C \left(\int \left|\hat{\eta}_{1\ell}(z_1) - \eta_{01}(z_1) \right|^2F_0(dw)\right)^{1/2}\\ & \overset{p}{\to} 0. 
			\end{split}
		\end{equation*}
		The same is true for the next period. Also, 
		\begin{equation*}
			\begin{split}
				\int & \left|y_2 - F\left(x_2,\theta_{0p}\right) - \theta_{0\omega}\left(\hat{\eta}_{1\ell}(z_1) - F\left(x_1,\theta_{0p}\right)\right) \right. \\ &  \left.- \left(y_2 - F\left(x_2,\theta_{0p}\right) - \theta_{0\omega}\left(\eta_{01}(z_1) - F\left(x_1,\theta_{0p}\right)\right)\right) \right|^2 \left|\left|\bm{\kappa_{02}(z_1)}\right|\right|^2F_0(dw) \\ &  = C \left(\int \left|\hat{\eta}_{1\ell}(z_1) - \eta_{01}(z_1)\right|^2 F_0(dw)\right)^{1/2}  \overset{p}{\to} 0.
			\end{split}
		\end{equation*}
		The same conclusion holds for the next period. We obtain then that Assumption \ref{contozero} \textit{ii)} is satisfied. Also,  we can get
		\begin{equation*}
			\begin{split}
				\int \left|y_1 - \eta_{01}(z_1)\right|^2 \left|\left|\bm{\hat{\kappa}_{1\ell}(z_1)} - \bm{\kappa_{01}(z_1)}\right|\right|^2 F_0(dw) & \leq C \left(\int  \left|\left|\bm{\hat{\kappa}_{1\ell}(z_1)} - \bm{\kappa_{01}(z_1)}\right|\right|^2\right)^{1/2} \\ & \overset{p}{\to} 0,
			\end{split}
		\end{equation*}
		and similarly for period 2. Likewise, 
		\begin{equation*}
			\begin{split}
				\int \left|y_2 - F\left(x_2,\theta_{0p}\right) - \theta_{0\omega}\left(\eta_{01}(z_1) - F\left(x_1,\theta_{0p}\right)\right) \right|^2 \left|\left|\bm{\hat{\kappa}_{2\ell}(z_1)} - \bm{\kappa_{02}(z_1)}\right|\right|^2 & \leq C \left(\int \left|\left|\bm{\hat{\kappa}_{2\ell}(z_1)} - \bm{\kappa_{02}(z_1)}\right|\right|^2\right)^{1/2} \\ & \overset{p}{\to} 0.
			\end{split}
		\end{equation*}
		The same result can be obtained for the next period. As a result, Assumption \ref{contozero} \textit{iii)} holds. Let us work next on Assumption \ref{twicefre}. Observe that $b_1(Z_1)\bm{\kappa_{01}(Z_1)}$, $b_1(Z_1)\bm{\kappa_{02}(Z_1)}$, $b_2(Z_2)\bm{\kappa_{03}(Z_2)}$, and $b_2(Z_2)\bm{\kappa_{04}(Z_2)}$ all exists a.s. Furthermore, 
		\begin{equation*}
			\begin{split}
				\mathbb{E}\left[\left|\left| b_1\left(Z_1\right)\bm{\kappa_{01}(Z_1)}\right|\right|\right]  \leq C \mathbb{E}\left[\left|\left| \bm{\kappa_{01}(Z_1)}\right|\right|\right] < \infty, \\ 
				\mathbb{E}\left[\left|\left| \theta_{0\omega} b_1\left(Z_1\right)\bm{\kappa_{02}(Z_1)}\right|\right|\right]  \leq C \mathbb{E}\left[\left|\left| \bm{\kappa_{01}(Z_1)}\right|\right|\right] < \infty, \\ 
				\mathbb{E}\left[\left|\left| b_2\left(Z_2\right)\bm{\kappa_{03}(Z_2)}\right|\right|\right]  \leq C \mathbb{E}\left[\left|\left| \bm{\kappa_{03}(Z_2)}\right|\right|\right] < \infty, \\ 
				\mathbb{E}\left[\left|\left| \theta_{0\omega} b_2\left(Z_2\right)\bm{\kappa_{04}(Z_2)}\right|\right|\right]  \leq C \mathbb{E}\left[\left|\left| \bm{\kappa_{04}(Z_2)}\right|\right|\right] < \infty.
			\end{split}
		\end{equation*}
		Then, by the Dominated Convergence Theorem, we obtain 
		\begin{equation*}
			\begin{split}
				\bar{\psi^{\prime}}\left(\theta_0,\eta_0,\kappa_0\right) & = \frac{d}{d\tau} \bar{\psi}\left(\theta_0,\eta_ 0 + \tau b,\kappa_0\right)  \\ & =  \mathbb{E}\left[-b_1(Z_1)\bm{\kappa_{01}(Z_1)} - \theta_{0\omega}b_1(Z_1)\bm{\kappa_{02}(Z_1)} -b_2(Z_2)\bm{\kappa_{03}(Z_2)} - \theta_{0\omega}b_2(Z_2)\bm{\kappa_{04}(Z_2)}\right].
			\end{split}
		\end{equation*}
		The above is linear and continuous in $b$. Let us verify that it is the Fréchet derivative of $\bar{\psi}$. This holds since 
		$$
		\left|\left|\bar{\psi}\left(\theta_0,\eta_0 + b, \bm{\kappa_0}\right) - \bar{\psi}\left(\theta_0,\eta_0, \bm{\kappa_0}\right) - \bar{\psi^{\prime}}\left(\theta_0,\eta_0,\kappa_0\right)\right|\right| = 0.
		$$ 
		Moreover, note that $\bar{\psi^{\prime}}$ does not depend on $\eta_0$. Then, $\bar{\psi^{\prime \prime}}\left(\theta_0,\eta_0,\kappa_0\right) = \frac{d}{d\tau} \bar{\psi^{\prime}}\left(\theta_0,\eta_ 0 + \tau b,\kappa_0\right) = 0 $, which is trivially the Fréchet derivative of $\bar{\psi^{\prime}}$ and is trivially continuous in $\eta$ at $\eta_0$. Hence, Assumption \ref{twicefre} is verified. 
		
		Let us continue with Assumption \ref{convmall}. Trivially, 
		$$
		\int \left|y_1 - \hat{\eta}_{1\ell}(z_1) - \left(y_1 - \hat{\eta}_{1\ell}(z_1) \right)\right|^2\left|\left|\bm{\hat{\kappa}_{1\ell}(z_1)}\right|\right|^2F_0(dw) = 0.
		$$
		The same result holds mutatis mutandi for period 2. In addition, with probability approaching one,
		\begin{equation*}
			\begin{split}
				\int & \left|y_2 - F\left(x_2,\tilde{\theta}_{p\ell}\right) - \tilde{\theta}_{\omega \ell}\left(\hat{\eta}_{1\ell}(z_1) - F\left(x_1,\tilde{\theta}_{p\ell}\right)\right) - \left(y_2 - F\left(x_2,\theta_{0p}\right) - \theta_{0\omega}\left(\hat{\eta}_{1\ell}(z_1) - F\left(x_1,\theta_{0p}\right)\right)\right)\right|^2 \\ & \left|\left|\bm{\hat{\kappa}_2(z_1)}\right|\right|^2 F_0(dw) \\ \leq & \; C \int \left|F\left(x_2,\theta_{0p}\right) - F\left(x_2,\tilde{\theta}_{p}\right)\right|^2F_0(dw) + C \int \left|\hat{\eta}_{1\ell}(z_1)\left(\theta_{0\omega} - \tilde{\theta}_{\omega \ell}\right)\right|^2F_0(dw) \\ & + C\int \left|\tilde{\theta}_{\omega \ell}F\left(x_1,\tilde{\theta}_{p \ell}\right) - \theta_{0\omega}F\left(x_1,\theta_{0p}\right) \right|^2 F_0(dw) \\ \leq & o_p(1) + C \left|\theta_{0\omega} - \tilde{\theta}_{\ell \omega}\right|^2 \int \left| \hat{\eta}_{1\ell}(z_1) - \eta_{01}(z_1)\right|^2F_0(dw) + C \left|\theta_{0\omega} - \tilde{\theta}_{\ell \omega}\right|^2 \int \left| \eta_{01}(z_1)\right|^2F_0(dw) \\ & + C\left|\theta_{0\omega}\right|^2 \int \left|F\left(x_2,\theta_{0p}\right) - F\left(x_2,\tilde{\theta}_{p}\right)\right|^2F_0(dw) + C \left|\theta_{0\omega} - \theta_{\omega \ell}\right|^2 \int \left|F\left(X,\tilde{\theta}_{p\ell}\right)\right|^2 F_0(dw) \\  \leq & o_p(1) + o_p(1)o_p(1) + o_p(1)O_p(1) + O_p(1)o_p(1) + o_p(1)O_p(1) \\ \leq & o_p(1),
			\end{split}
		\end{equation*} 
		where $\int \left|F\left(x_2,\theta_{0p}\right) - F\left(x_2,\tilde{\theta}_{p}\right)\right|^2F_0(dw) = o_p(1)$ by the Dominated Convergence Theorem and continuity of $F\left(X,\cdot\right)$ a.s. The same conclusion can be derived for the next period. Hence, Assumption \ref{convmall} holds.
		
		It remains to show that Assumption \ref{ajacobian} is satisfied. By Assumption \textit{iii)}, we know that 
		\begin{equation*}
			\begin{split}
				\Upsilon = - \mathbb{E} & \left[   \frac{\partial}{\partial \theta}\left(F\left(X_2,\theta_{0p}\right) + \theta_{0\omega}\left(\eta_{01}(Z_1) - F\left(X_1,\theta_{0p}\right)\right)\right)\bm{\kappa_{02}(Z_1)} \right. \\ & \left. + \frac{\partial}{\partial \theta}\left(F\left(X_3,\theta_{0p}\right) + \theta_{0\omega}\left(\eta_{02}(Z_2) - F\left(X_2,\theta_{0p}\right)\right)\right)\bm{\kappa_{04}(Z_2)}\right]
			\end{split}
		\end{equation*}
		exists. Furthermore, $\psi\left(W,\theta,\eta,\bm{\kappa}\right)$ is differentiable in $\theta$ on $\mathcal{N}$, for $\left|\left|\eta - \eta_0 \right| \right|_{\Xi}  \mathbb{E}\left[\left|\left|\bm{\kappa(Z)} - \bm{\kappa_0(Z)} \right| \right|^2_{\infty}\right]$ small enough. By the triangle inequality,
		\begin{equation*}
			\begin{split}
				\left|\left|\frac{\partial \psi\left(W,\theta,\eta,\bm{\kappa}\right)}{\partial \theta} - \frac{\partial \psi\left(W,\theta_0,\eta,\bm{\kappa}\right)}{\partial \theta}\right|\right| \leq  & \left|\left|\theta_\omega F_\theta\left(X_1,\theta_p\right) - \theta_{0\omega} F_\theta\left(X_1,\theta_{0p}\right)\right|\right|\left|\left|\bm{\kappa_2(Z_1)}\right|\right|  \\ & +  \left|\left|F_\theta(X_2,\theta_{p}) - F_\theta(X_2,\theta_{0p})\right|\right|\left|\left|\bm{\kappa_2(Z_1)}\right|\right| \\ & + \left|\left|\theta_\omega F_\theta\left(X_2,\theta_p\right) - \theta_{0\omega} F_\theta\left(X_2,\theta_{0p}\right)\right|\right|\left|\left|\bm{\kappa_4(Z_2)}\right|\right|  \\ & +  \left|\left|F_\theta(X_3,\theta_{p}) - F_\theta(X_3,\theta_{0p})\right|\right|\left|\left|\bm{\kappa_4(Z_2)}\right|\right| \\ & + \left|F\left(X_1,\theta_p\right) - F\left(X_1,\theta_{0p}\right) \right| \left|\left|\bm{\kappa_2(Z_1)}\right|\right| \\ & + \left|F\left(X_2,\theta_p\right) - F\left(X_2,\theta_{0p}\right)\right| \left|\left|\bm{\kappa_4(Z_2)}\right|\right| \\ \leq & \left(\left|\theta_\omega - \theta_{0\omega}\right| \left|\left|F_\theta\left(X_1,\theta_p\right)\right|\right| + |\theta_{0\omega}| \left|\left|F_\theta\left(X_1,\theta_p\right) - F_\theta\left(X_1,\theta_{0p}\right) \right|\right|\right) \left|\left|\bm{\kappa_2(Z_1)}\right|\right| \\ & + \left|\left|F_\theta(X_2,\theta_{p}) - F_\theta(X_2,\theta_{0p})\right|\right|\left|\left|\bm{\kappa_2(Z_1)}\right|\right| \\ & + \left(\left|\theta_\omega - \theta_{0\omega}\right| \left|\left|F_\theta\left(X_2,\theta_p\right)\right|\right| + |\theta_{0\omega}| \left|\left|F_\theta\left(X_2,\theta_p\right) - F_\theta\left(X_2,\theta_{0p}\right) \right|\right|\right) \left|\left|\bm{\kappa_4(Z_2)}\right|\right| \\ & + \left|\left|F_\theta(X_3,\theta_{p}) - F_\theta(X_3,\theta_{0p})\right|\right|\left|\left|\bm{\kappa_4(Z_2)}\right|\right| \\ & + \left|F\left(X_1,\theta_p\right) - F\left(X_1,\theta_{0p}\right) \right| \left|\left|\bm{\kappa_2(Z_1)}\right|\right| \\ & + \left|F\left(X_2,\theta_p\right) - F\left(X_2,\theta_{0p}\right)\right| \left|\left|\bm{\kappa_4(Z_2)}\right|\right| \\ \leq & \left(C\left|\theta_\omega - \theta_{0\omega}\right| + C \left|\left|\theta_p - \theta_{0p}\right|\right|^{H_2} + C \left|\left|\theta_p - \theta_{0p}\right|\right|^{H_2}\right) \\ & \left(\left|\left|\bm{\kappa_2(Z_1)}\right|\right| + \left|\left|\bm{\kappa_4(Z_2)}\right|\right|\right),
			\end{split}
		\end{equation*}
	 The last display suffices to show Assumption \ref{ajacobian} \textit{ii)}. Lastly, by the triangle and Hölder's inequalities, 
		\begin{align}
			\int & \left|\left(\theta_{0\omega}F_{\theta k}\left(x_1,\theta_{0p}\right) - F_{\theta k}\left(x_2,\theta_{0p}\right)\right) \bm{\hat{\kappa}_{2\ell s}(z_1)} + \left(\theta_{0\omega}F_{\theta k}\left(x_2,\theta_{0p}\right) - F_{\theta k}\left(x_3,\theta_{0p}\right)\right) \bm{\hat{\kappa}_{4\ell s}(z_2)} \right. \nonumber\\ & \left. - \left(\left(\theta_{0\omega}F_{\theta k}\left(x_1,\theta_{0p}\right) - F_{\theta k}\left(x_2,\theta_{0p}\right)\right) \bm{\kappa_{02s}(z_1)} + \left(\theta_{0\omega}F_{\theta k}\left(x_2,\theta_{0p} - F_{\theta k}\left(x_3,\theta_{0p}\right)\right)\right) \bm{\kappa_{04s}(z_2)}\right)\right|F_0(dw) \nonumber \\  \leq & \left(\int \left|\theta_{0\omega}F_{\theta k}\left(x_1,\theta_{0p}\right) - F_{\theta k}\left(x_2,\theta_{0p}\right)\right|^2 F_0(dw)\right)^{1/2}  \left(\int \left|\bm{\hat{\kappa}_{2\ell s}(z_1)} - \bm{\kappa_{02s}(z_1)} \right|^2 F_0(dw)\right)^{1/2} \nonumber\\ & + \left(\int \left|\theta_{0\omega}F_{\theta k}\left(x_2,\theta_{0p}\right) - F_{\theta k}\left(x_3,\theta_{0p}\right)\right|^2 F_0(dw)\right)^{1/2}  \left(\int \left|\bm{\hat{\kappa}_{4\ell s}(z_2)} - \bm{\kappa_{04s}(z_2)} \right|^2 F_0(dw)\right)^{1/2} \nonumber\\ \leq & O_p(1)o_p(1) \nonumber \\  = & o_p(1). \label{zerolimit1}
		\end{align}
		Observe that, for a given row $s$ of $\frac{\partial}{\partial \theta}\psi\left(W,\theta_0,\eta,\bm{\kappa}\right)$, the elements in $p_1$ of its columns satisfies \eqref{zerolimit1} (these are the derivatives corresponding to $\theta_{p}$).\footnote{Recall that $\theta_{0p} \in \mathbb{R}^{p_1}$.} For the derivatives corresponding to $\theta_{\omega}$, by the triangle and Hölder's inequalities, we can establish the following
		\begin{align}
			\int & \left|\left(F\left(x_1,\theta_{0p}\right) - \hat{\eta}_{1\ell}(z_1)\right)\bm{\hat{\kappa}_{2\ell q}(z_1) }+ \left(F\left(x_2,\theta_{0p}\right) - \hat{\eta}_{2\ell}(z_2)\right)\bm{\hat{\kappa}_{4\ell q}(z_2)} \right.  \nonumber\\ & - \left.  \left(\left(F\left(x_1,\theta_{0p}\right) - \eta_{01}(z_1)\right)\bm{\kappa_{02q}(z_1)} + \left(F\left(x_2,\theta_{0p}\right) - \eta_{02}(z_2)\right)\bm{\kappa_{04q}(z_2)} \right)\right| F_0(dw) \nonumber \\ \leq & \left(\int \left|F\left(x_1,\theta_{0p}\right)\right|^2 F_0(dw)\right)^{1/2} \left(\int \left| \bm{\hat{\kappa}_{2\ell q}(z_1)} - \bm{\kappa_{02q}(z_1)}\right|^2F_0(dw)\right)^{1/2}	\nonumber \\ & +  \left(\int \left|F\left(x_2,\theta_{0p}\right)\right|^2 F_0(dw)\right)^{1/2} \left(\int \left| \bm{\hat{\kappa}_{4\ell q}(z_2)} - \bm{\kappa_{04q}(z_2)}\right|^2F_0(dw)\right)^{1/2} \nonumber \\ & + \int \left|\eta_{01}(z_1)\bm{\kappa_{02q}(z_1)} - \hat{\eta}_{1\ell}(z_1)\bm{\hat{\kappa}_{2\ell q}(z_1)}\right|F_0(dw)	+ \int \left|\eta_{02}(z_2)\bm{\kappa_{04q}(z_2)} - \hat{\eta}_{2\ell}(z_2)\bm{\hat{\kappa}_{4\ell q}(z_2)}\right|F_0(dw) \nonumber \\ \leq &  o_p(1) + \left(\int \left|\eta_{01}(z_1) - \hat{\eta}_{1\ell}(z_1)\right|^2F_0(dw)\right)^{1/2} \left(\int \left|\bm{\kappa_{02q}(z_1)} - \bm{\hat{\kappa}_{2\ell q}(z_1)}\right|^2F_0(dw)\right)^{1/2} \nonumber \\ &  + C \left(\int \left|\eta_{01}(z_1) - \hat{\eta}_{1\ell}(z_1)\right|^2F_0(dw)\right)^{1/2} + C \left(\int \left|\bm{\kappa_{02q}(z_1)} - \bm{\hat{\kappa}_{2\ell q}(z_1)}\right|^2F_0(dw)\right)^{1/2} \nonumber \\ & + \left(\int \left|\eta_{02}(z_2) - \hat{\eta}_{2\ell}(z_2)\right|^2F_0(dw)\right)^{1/2} \left(\int \left|\bm{\kappa_{04q}(z_2)} - \bm{\hat{\kappa}_{4\ell q}(z_2)}\right|^2F_0(dw)\right)^{1/2} \nonumber \\ &  + C \left(\int \left|\eta_{02}(z_2) - \hat{\eta}_{2\ell}(z_2)\right|^2F_0(dw)\right)^{1/2} + C \left(\int \left|\bm{\kappa_{04q}(z_2)} - \bm{\hat{\kappa}_{4\ell q}(z_2)}\right|^2F_0(dw)\right)^{1/2} \nonumber \\ \leq & o_p(1) + o_p(1)o_p(1) + O_p(1)o_p(1) + O_p(1)o_p(1) + o_p(1)o_p(1) + O_p(1)o_p(1) + O_p(1)o_p(1) \nonumber \\  = & o_p(1) \label{zerolimit2}
		\end{align}
		Then, the results in \eqref{zerolimit1} and \eqref{zerolimit2} show that Assumption \ref{ajacobian} \textit{iii)} is satisfied for the Production Function example. We have thus verified all Assumptions \ref{contozero}-\ref{ajacobian}, as needed. Hence, the conclusions of the corollary is obtained by Theorem \ref{inferencetheta}. $\blacksquare$
		
		\subsection{Asymptotic Normality in the Empirical Application}
		\label{ANempiricalapp}
		 \begin{corollary}
			\label{cnormalityex3}
			Let the assumptions in Lemma  \ref{fde2}, Lemma \ref{lFconvg}, Assumption \ref{estspaces},  and \ref{rates} hold. Let $\hat{\theta} \overset{p}{\to} \theta_0$, $\hat{\Lambda} \overset{p}{\to} \Lambda$, and $\Upsilon^{'} \Lambda \Upsilon$ be non-singular. In addition, assume 
			\begin{itemize}
				\item[i)] The functions  $\left|\left|\bm{\kappa_{01t}\left(E_{t-1}, L_{t-1}, K_{t-1}\right)}\right|\right|$,$\left|\left|\bm{\kappa_{02t}\left(E_{t-1}, L_{t-1}, K_{t-1}\right)}\right|\right|$, $|Y_t - \eta_{0t}(E_t,L_t,K_t)|$, and \\ $\left|Y_{t+1} - \theta_{01} - \theta_{0l}L_{t+1} - \theta_{0k}K_{t+1} - \theta_{0\omega}\left(\eta_{0t}\left(E_{t}, L_{t}, K_{t}\right) - \theta_{01} - \theta_{0l}L_{t} - \theta_{0k}K_{t}\right)\right|$, for all $t$, are bounded a.s. In addition, $b= \left(b_1,\cdots, b_{T-1}\right)$ has bounded entries a.s.; 
				\item [ii)] $\mathbb{E}\left[\left|\eta_{0t}\left(Z_t\right) \right|^2\right] < \infty$, for all $t$;
				\item[iii)]  $\Upsilon$ exists;
				\item[iv)]   $L_{it}$ and $K_{it}$ have bounded support a.s., for all $t$. Moreover, $\eta_{t-1}(E_{t-1}, L_{t-1}, K_{t-1})$ and \\ $\left|\left|\bm{\kappa_{2t}\left(E_{t-1}, L_{t-1}, K_{t-1}\right)}\right|\right|$, $2 \leq t \leq T$,  exists a.s., for $\left|\left|\eta - \eta_0 \right| \right|_{\Xi}\mathbb{E}\left[\left|\left|\bm{\kappa(Z)} - \bm{\kappa_0(Z)} \right| \right|^2_{\infty}\right] $ small enough. In addition, $\mathbb{E}\left[\left|\left|\bm{\kappa_{2t}\left(E_{t-1}, L_{t-1}, K_{t-1}\right)}\right|\right|\right] < \infty$.
			\end{itemize}
			Then, in the empirical application setting, it follows that 
			$$
			\sqrt{n}\left(\hat{\theta} - \theta_0\right) \overset{d}{\to}\;\;N\left(0,\Sigma\right),\;\;\; \Sigma = \left(\Upsilon^{'}\Lambda \Upsilon\right)^{-1} \Upsilon^{'}\Lambda \Psi \Lambda \Upsilon\left(\Upsilon^{'}\Lambda \Upsilon\right)^{-1}.
			$$
			Moreover, $\hat{\Sigma} \overset{p}{\to} \Sigma$.	
		\end{corollary}

		\bigskip \noindent \textbf{Proof of Corollary \ref{cnormalityex3}:} The results follows by a straightforward extension of  Corollary \ref{cnormalityex2} from three periods of observed data to an arbitrary number $T$, and focusing on a Cobb-Douglas production function. $\blacksquare$

		\section{Consistency}
		\label{consistency}
		
		\bigskip The result in Theorem \ref{inferencetheta} relies on the consistency of $\hat{\theta}$. We now proceed by establishing the consistency of our estimator.

		\begin{theorem}
			\label{consistencytheta}
			If i) $\hat{\Lambda} \overset{p}{\to} \Lambda$, where $\Lambda$ is a positive definite matrix; ii) $\mathbb{E}\left[\psi\left(W,\theta,\eta_0,\bm{\kappa_0}\right)\right] = 0$ if and only if $\theta = \theta_0$; iii) $\Theta$ is compact; iv) For all $j$ and $\ell$, $\int \left|\left| m_j\left(y,\theta,\hat{\eta}_{\ell}\right)\bm{\hat{\kappa}_{j\ell}(z_j)} - m_j\left(y,\theta,\eta_{0}\right)\bm{\kappa_{0j}(z_j)}\right|\right|F_0(dw) \overset{p}{\to} 0$ and $\mathbb{E}\left[\left|\left| m_j\left(Y,\theta,\eta_0\right) \bm{\kappa_{0j}(Z_j)} \right|\right|\right] < \infty$ for all $\theta \in \Theta$; v) There is $C_1 > 0$, $C_2 > 0$, and $d\left(W,\eta,\bm{\kappa}\right)$ such that for each $\left|\left|\eta - \eta_0 \right| \right|_{\Xi} \mathbb{E}\left[\left|\left|\bm{\kappa(Z)} - \bm{\kappa_0(Z)} \right| \right|^2_\mathcal{\infty}\right] $ small enough and all $\tilde{\theta}$, $\theta \in \Theta$, 
			$$
			\left|\left|\psi\left(W,\tilde{\theta}, \eta, \bm{\kappa}\right) - \psi\left(W,\theta, \eta, \bm{\kappa}\right)\right|\right| \leq d\left(W,\eta,\bm{\kappa}\right) \left|\left|\tilde{\theta} - \theta\right|\right|^{C_1},\;\;\; \mathbb{E}\left[d\left(W,\eta,\bm{\kappa}\right)\right] < C_2.
			$$
			Then, $\hat{\theta} \overset{p}{\to} \theta$. 
		\end{theorem}
		
		\bigskip \noindent \textbf{Proof of Theorem \ref{consistencytheta}:} We follow the proof of Theorem A3 of CEINR. Observe that, by the triangle inequality and Assumption \textit{iv)}, 
		\begin{equation*}
			\begin{split}
				\int \left|\left|\psi\left(w,\theta,\hat{\eta}_\ell, \bm{\hat{\kappa}_\ell}\right) - \psi\left(w,\theta,\eta_0, \bm{\kappa_0}\right)\right|\right|F_0(dw) & \leq \sum^J_{j=1} \int \left|\left| m_j\left(y,\theta,\hat{\eta}_{\ell}\right)\bm{\hat{\kappa}_{j\ell}(z_j)} - m_j\left(y,\theta,\eta_{0}\right)\bm{\kappa_{0j}(z_j)}\right|\right|F_0(dw) \\ & \overset{p}{\to} 0.
			\end{split}
		\end{equation*}
		It follows then that $\hat{\psi}(\theta) \overset{p}{\to} \bar{\psi}(\theta) = \mathbb{E}\left[\psi\left(W,\theta,\eta_0,\bm{\kappa_0}\right)\right]$ for all $\theta \in \Theta$, by the Conditional Markov's inequality, the law of large numbers, and the triangle inequality. Next, by \textit{v)}, with probability approaching one, 
		\begin{equation*}
			\begin{split}
				\left|\left|\hat{\psi}\left(\tilde{\theta}\right) - \hat{\psi}\left(\theta\right) \right|\right| & \leq \frac{1}{n}\sum^L_{\ell = 1} \sum_{i \in I_\ell} \left|\left|\psi\left(W_i,\tilde{\theta},\hat{\eta}_\ell,\bm{\hat{\kappa}_\ell}\right) - \psi\left(W_i,\theta,\hat{\eta}_\ell,\bm{\hat{\kappa}_\ell}\right)\right|\right| \\ & \leq \frac{1}{n}\sum^L_{\ell = 1} \sum_{i \in I_\ell} d\left(W_i,\hat{\eta}_\ell,\bm{\hat{\kappa}_\ell}\right) \left|\left|\tilde{\theta} - \theta\right|\right|^{C_1} \\ &  = \hat{A} \left|\left|\tilde{\theta} - \theta\right|\right|^{C_1}. 
			\end{split}
		\end{equation*}
		Note, by the conditional Markov's inequality, $\hat{A} = O_p(1)$. Then, by Corollary 2.2 of \cite{newey1991uniform}, we have $\sup_{\theta \in \Theta} \left|\left|  \hat{\psi}(\theta) - \bar{\psi}(\theta)\right| \right| \overset{p}{\to} 0$. Moreover, observe that condition \textit{v)} also implies that $\bar{\psi}(\theta)$ is continuous on $\Theta$. Finally, note that the second condition in \textit{iv)} implies that $\mathbb{E}\left[\left|\left|\psi\left(W,\theta,\eta_0,\bm{\kappa_0}\right)\right|\right|\right] < \infty$ for all $\theta \in \Theta$, by the triangle inequality. The conclusion then follows similarly to the proof of Theorem 2.6 of \cite{newey1994large}. $\blacksquare$

		\subsection{Consistency Conditions for the Example}

	\begin{corollary}
		\label{consistencyex2}
		Suppose that i) $\hat{\Lambda} \overset{p}{\to} \Lambda$, where $\Lambda$ is a positive definite matrix; ii) $\mathbb{E}\left[\psi\left(W,\theta,\eta_0,\bm{\kappa}_0\right)\right] = 0$ if and only if $\theta = \theta_0$; iii) $\Theta$ is compact. In addition, let 
		\begin{itemize}
			\item[iv)] $\mathbb{E}\left[\left|Y_t - \eta_{0t}(Z_t)\right|^2\right]$, $\mathbb{E}\left[\left|Y_{t+1} - F\left(X_{t+1},\theta_p\right) - \theta_\omega\left(\eta_{0t}(Z_t) - F\left(X_t,\theta_p\right)\right)\right|^2\right]$, $\mathbb{E}\left[\left|\eta_{0t}(Z_t)\right|^2\right] < \infty$, for all $t$, $\mathbb{E}\left[\left|\left|\bm{\kappa_{01}(Z_1)}\right|\right|^2\right]$, $\mathbb{E}\left[\left|\left|\bm{\kappa_{02}(Z_1)}\right|\right|^2\right]$, $\mathbb{E}\left[\left|\left|\bm{\kappa_{03}(Z_2)}\right|\right|^2\right]$, $\mathbb{E}\left[\left|\left|\bm{\kappa_{04}(Z_2)}\right|\right|^2\right] < \infty$. Also, for all $\theta \in \Theta$,  $\left|F(X_t,\theta_{p})\right|$ is bounded uniformly in $X_t$ a.s., for all $t$; 
			\item[v)] There exists a $C$, not depending on $X$, such that $\left|F(X_t,\tilde{\theta}) -  F(X_t,\theta_0) \right| \leq C \left|\left|\tilde{\theta} - \theta\right|\right|^H$ for any $\tilde{\theta}, \theta \in \Theta$ for all $t$, with $H>0$. Moreover, $\left|\left|\eta_1(Z_1)\bm{\kappa_2(Z_1)}\right|\right|$, $\left|\left|\eta_2(Z_2)\bm{\kappa_4(Z_2)}\right|\right|$, $\left|\left|\bm{\kappa_2(Z_1)}\right|\right|$, $\left|\left|\bm{\kappa_4(Z_2)}\right|\right|$ are bounded a.s., for $\left|\left|\eta - \eta_0 \right|\right|_{\Xi}\mathbb{E}\left[\left|\left|\bm{\kappa(Z)} - \bm{\kappa_0(Z)} \right|\right|^2_{\infty}\right]$ small enough.
		\end{itemize}
		Then, $\hat{\theta} \overset{p}{\to} \theta$, in the example.
	\end{corollary}
	
	\bigskip \noindent \textbf{Proof of Corollary \ref{consistencyex2}:} We only need to show Assumptions \textit{iv)} and \textit{v)} in Theorem \ref{consistencytheta}. By the triangle and Hölder inequalities, 
	\begin{equation*}
		\begin{split}
			\int \left|\left|\left(y_1 - \hat{\eta}_{1\ell}(z_1)\right) \bm{\hat{\kappa}_{1\ell}(z_1)} - \left(y_1 - \eta_{01}(z_1)\right)\bm{\kappa_{01}(z_1)} \right|\right| F_0(dw)  = &  \int \left|\left|\hat{\eta}_{1\ell}(z_1)\bm{\hat{\kappa}_{1\ell}(z_1)} -  \eta_{01}(z_1)\bm{\kappa_{01}(z_1)}\right|\right| F_0(dw) \\ \leq & \left(\int \left|\hat{\eta}_{1\ell}(z_1) - \eta_{01}(z_1)\right|^2\right)^{1/2} \\ & \left(\int \left|\left|\bm{\hat{\kappa}_{1\ell}(z_1)} - \bm{\kappa_{01}(z_1)}\right|\right|^2\right)^{1/2} \\ & + C \left(\int \left|\left|\bm{\hat{\kappa}_{1\ell}(z_1)} - \bm{\kappa_{01}(z_1)}\right|\right|^2\right)^{1/2} \\ & + C \left(\int \left|\hat{\eta}_{1\ell}(z_1) - \eta_{01}(z_1)\right|^2\right)^{1/2} \\  = & o_p(1)o_p(1) + O_p(1)o_p(1) + O_p(1)o_p(1) \\  =  & o_p(1).
		\end{split}
	\end{equation*}
	Furthermore, $\mathbb{E}\left[\left|\left|\left(Y_1 - \eta_{01}(Z_1)\right)\bm{\kappa_{01}(Z_1)}\right|\right|\right] < \mathbb{E}\left[\left|Y_1 - \eta_{01}(Z_1)\right|^2\right] \mathbb{E}\left[\left|\left| \bm{\kappa_{01}(Z_1)}\right|\right|^2\right] < \infty$. The same  findings can be obtained for period 2 mutatis mutandi. Additionally, remark that, by the triangle and Hölder's inequalities,
	\begin{equation*}
		\begin{split}
			\int & \left|\left| \left(y_2 - F\left(x_2,\theta_p\right) - \theta_\omega\left(\hat{\eta}_{1\ell}(z_1) - F\left(x_1,\theta_p\right)\right)\bm{\hat{\kappa}_{2\ell}(z_1)}\right) \right.\right. \\ & -  \left. \left. \left(y_2 - F\left(x_2,\theta_p\right) - \theta_\omega\left(\eta_{01}(z_1) - F\left(x_1,\theta_p\right)\right)\bm{\kappa_{02}(z_1)}\right) \right|\right| F_0(dw) \\ \leq & C \int \left|\left|\hat{\eta}_{1\ell}(z_1)\bm{\hat{\kappa}_{1\ell}(z_1)} -  \eta_{01}(z_1)\bm{\kappa_{01}(z_1)}\right|\right| F_0(dw) + C \int \left|\left|\bm{\kappa_{2\ell}(z_1)} - \bm{\kappa_{02}(z_1)}\right|\right| F_0(dw) \\ \leq & o_p(1) + C \left(\int \left|\left|\bm{\kappa_{2\ell}(z_1)} - \bm{\kappa_{02}(z_1)}\right|\right|^2 F_0(dw)\right)^{1/2} \\ = & o_p(1).
		\end{split}
	\end{equation*}  
	Also, observe that $\mathbb{E}\left[\left|\left|  \left(Y_2 - F\left(X_2,\theta_p\right) - \theta_\omega\left(\eta_{01}(Z_1) - F\left(X_1,\theta_p\right)\right)\bm{\kappa_{02}(Z_1)}\right) \right|\right|\right] \leq $ \\ $ \mathbb{E}\left[\left|Y_2 - F\left(X_2,\theta_p\right) - \theta_\omega\left(\eta_{01}(Z_1) - F\left(X_1,\theta_p\right)\right)\right|^2\right]^{1/2} \mathbb{E}\left[\left|\left|\bm{\kappa_{02}(Z_1)}\right|\right|^2\right]^{1/2} < \infty$. The same conclusions can be obtained for one period ahead. Therefore, Assumption \textit{iv)} in Theorem \ref{consistencytheta} holds. Next, for a fixed $W$ and arbitrary $\theta$, $\tilde{\theta} \in \Theta$, by the triangle inequality, 
 \begin{align}
   		\left|\left|\psi\left(W,\tilde{\theta},\eta,\bm{\kappa}\right) - \psi\left(W,\theta,\eta,\bm{\kappa}\right)\right|\right|  \leq & \left|F\left(X_2,\tilde{\theta}_p\right) - F\left(X_2,\theta_p\right)\right|\left|\left|\bm{\kappa_{2}(Z_1)}\right|\right| + |\tilde{\theta}_\omega - \theta_\omega|\left|\left|\eta_1(Z_1)\bm{\kappa_2(Z_1)}\right|\right| \nonumber \\ &  + \left|\left|\left(\tilde{\theta}_\omega F\left(X_1,\tilde{\theta}_p\right) - \theta_\omega F\left(X_1,\theta_p\right)\right)\bm{\kappa_2(Z_1)} \right|\right| \label{diff1} \\ &  + \left|F\left(X_3,\tilde{\theta}_p\right) - F\left(X_3,\theta_p\right)\right|\left|\left|\bm{\kappa_{4}(Z_2)}\right|\right| + \left|\tilde{\theta}_\omega - \theta_\omega\right|\left|\left|\eta_2(Z_2)\bm{\kappa_4(Z_2)}\right|\right| \nonumber \\ &   + \left|\left|\left(\tilde{\theta}_\omega F\left(X_2,\tilde{\theta}_p\right) - \theta_\omega F\left(X_2,\theta_p\right)\right)\bm{\kappa_4(Z_2)} \right|\right|. \label{diff2}
    \end{align}
Note that by the triangle inequality and Hölder continuity of $F(X_t,\cdot)$, 
\begin{align}	
\left|\left|\left(\tilde{\theta}_\omega F\left(X_1,\tilde{\theta}_p\right) - \theta_\omega F\left(X_1,\theta_p\right)\right)\bm{\kappa_2(Z_1)}\right|\right| &  \leq C \left|\left|\bm{\kappa_2(Z_1)}\right|\right| \left|\left|\tilde{\theta}_p - \theta_p \right| \right|^H + C \left|\tilde{\theta}_\omega - \theta_\omega\right| \left|\left| \bm{\kappa_2(Z_1)} \right|\right|, \label{ineq1}\\ \left|\left|\left(\tilde{\theta}_\omega F\left(X_2,\tilde{\theta}_p\right) - \theta_\omega F\left(X_2,\theta_p\right)\right)\bm{\kappa_4(Z_2)}\right|\right| &  \leq C \left|\left|\bm{\kappa_4(Z_2)}\right|\right| \left|\left|\tilde{\theta}_p - \theta_p \right| \right|^H + C \left|\tilde{\theta}_\omega - \theta_\omega\right| \left|\left| \bm{\kappa_4(Z_2)} \right|\right| \label{ineq2}.
\end{align}
Plugging \eqref{ineq1} and \eqref{ineq2} in	  \eqref{diff1} and \eqref{diff2}, respectively, and again using Hölder continuity of $F(X_t,\cdot)$  yields 
\begin{equation*}
	\begin{split}
		\left|\left|\psi\left(W,\tilde{\theta},\eta,\bm{\kappa}\right) - \psi\left(W,\theta,\eta,\bm{\kappa}\right)\right|\right|  \leq & C \left|\left|\tilde{\theta}_p - \theta_p \right|\right|^H\left(\left|\left|\bm{\kappa_2(Z_1)}\right|\right| + \left|\left|\bm{\kappa_4(Z_2)}\right|\right|\right) \\ & + C\left|\tilde{\theta}_\omega - \theta_\omega\right|\left(\left|\left|\bm{\kappa_2(Z_1)}\right|\right| + \left|\left|\bm{\kappa_4(Z_2)}\right|\right|\right) \\ & + \left|\tilde{\theta}_\omega - \theta_\omega\right|\left(\left|\left|\eta_1(Z_1)\bm{\kappa_2(Z_1)}\right|\right| + \left|\left|\eta_2(Z_2)\bm{\kappa_4(Z_2)}\right|\right|\right),
	\end{split}
\end{equation*}	
The last displays is sufficient to show that \textit{v)} is satisfied, as the norms above that depend on the data are bounded a.s. Therefore, the conclusion of the corollary follows by Theorem \ref{consistencytheta}. $\blacksquare$

\subsection{Consistency Conditions in the Empirical Application}

\begin{corollary}
	\label{consistencyex3}
	Suppose that i) $\hat{\Lambda} \overset{p}{\to} \Lambda$, where $\Lambda$ is a positive definite matrix; ii) $\mathbb{E}\left[\psi\left(W,\theta,\eta_0,\bm{\kappa}_0\right)\right] = 0$ if and only if $\theta = \theta_0$; iii) $\Theta$ is compact. In addition, let 
	\begin{itemize}
		\item[iv)] $\mathbb{E}\left[\left|Y_t - \eta_{0t}(E_t,L_t,K_t)\right|^2\right]$, \\ $\mathbb{E}\left[\left|Y_{t+1} - \theta_{01} - \theta_{0l}L_{t+1} - \theta_{0k}K_{t+1} - \theta_{0\omega}\left(\eta_{0t}\left(E_{t}, L_{t}, K_{t}\right) - \theta_{01} - \theta_{0l}L_{t} - \theta_{0k}K_{t}\right)\right|^2\right]$, \\ $\mathbb{E}\left[\left|\eta_{0t}(E_t,L_t,K_t)\right|^2\right] < \infty$, for all $t$, \\ $\mathbb{E}\left[\left|\left|\bm{\kappa_{01t}\left(E_{t-1}, L_{t-1}, K_{t-1}\right)}\right|\right|^2\right]$, $\mathbb{E}\left[\left|\left|\bm{\kappa_{02t}\left(E_{t-1}, L_{t-1}, K_{t-1}\right)}\right|\right|^2\right] < \infty$, $2 \leq t \leq T$. Also, $L_t$ and $K_t$ have bounded support a.s., for all $t$; 
		\item[v)] $\left|\left|\eta_{t-1}(E_{t-1},L_{t-1},K_{t-1})\bm{\kappa_{2t}\left(E_{t-1}, L_{t-1}, K_{t-1}\right)}\right|\right|$, $\left|\left|\bm{\kappa_{2t}\left(E_{t-1}, L_{t-1}, K_{t-1}\right)}\right|\right|$ are bounded a.s., $2\leq t \leq T$, for $\left|\left|\eta - \eta_0 \right|\right|_{\Xi}\mathbb{E}\left[\left|\left|\bm{\kappa(Z)} - \bm{\kappa_0(Z)} \right|\right|^2_{\infty}\right]$ small enough.
	\end{itemize}
	Then, $\hat{\theta} \overset{p}{\to} \theta$, in the empirical application setting.
\end{corollary}

\bigskip \noindent \textbf{Proof of Corollary \ref{consistencyex3}:} The results follows by a straightforward extension of  Corollary \ref{consistencyex2} from three periods of observed data to an arbitrary number $T$, and focusing on a Cobb-Douglas production function. $\blacksquare$

		\section{Additional Monte Carlo Details}
		\label{additionalmontecarlo}
		
		As we stated in the main text, we use GMM based on four debiased moments, which can be written as  
		\begin{equation*}
			\begin{split}
				\psi\left(W,\theta_0,\eta_0\right) & = \left(Y_1 - \eta_{01}\left(I_1,K_1\right)\right)\bm{\kappa_{01}\left(Z_1\right)} + \left(Y_2 - \theta_{01} - \theta_{0k}K_2 - \theta_{0\omega}\left(\eta_{01}\left(Z_1\right) - \theta_{01} - \theta_{0k}K_1\right)\right)\bm{\kappa_{02}\left(Z_1\right)} \\ 
				& + \left(Y_2 - \eta_{02}\left(I_2,K_2\right)\right)\bm{\kappa_{03}\left(Z_2\right)} + \left(Y_3 - \theta_{01} - \theta_{0k}K_3 - \theta_{0\omega}\left(\eta_{02}\left(Z_2\right) - \theta_{01} - \theta_{0k}K_2\right)\right)\bm{\kappa_{04}\left(Z_2\right)}.
			\end{split}
		\end{equation*}
		Notice that our GMM program involves a three-dimensional non-linear search. To increase the reliability of our results, we have reduced the dimension of the problem such that we see $\theta_{01}$ and $\theta_{0\omega}$ as functions of $\theta_{0k}$. In this way, we only search over the dimension $\theta_{0k}$. We have accomplished this as follows.   Notice 
		$$
		\eta_{0t}\left(Z_t\right) = \theta_{01} + \theta_{0k}K_t + \omega_t\left(I_t,K_t\right),
		$$
		which implies that 
		\begin{equation}
			\label{omegatheta1}
			\theta_{01} + \omega_t\left(I_t,K_t\right) = \eta_{0t}\left(Z_t\right) - \theta_{0k}K_t.
		\end{equation}
		As $\omega_t$ follows an AR(1) process, we have
		\begin{equation}
			\label{ar1}
			\omega_t = \theta_{0\omega}\omega_{t-1} + \epsilon^{\omega}_t,\;\;\;\;\; \mathbb{E}\left[\left.\epsilon^{\omega}_t\right|\omega_{t-1}\right] = 0. 
		\end{equation}
		Plugging \eqref{omegatheta1} into \eqref{ar1} and re-arranging terms yields 
		$$
		\eta_{0t}\left(Z_t\right) - \theta_{0k}K_{t} = \tilde{c} + \theta_{0\omega}\left(\eta_{0,t-1}\left(Z_{t-1}\right) - \theta_{0k}K_{t-1}\right) + \epsilon^{\omega}_t, \;\;\; \tilde{c} = \theta_{01}\left(1-\theta_{0\omega}\right). 
		$$
		Hence, for a given value of $\theta_{0k}$, we can identify $\theta_{0\omega}$ as the slope in a linear regression of $\eta_{0t} - \theta_{0k}K_t$ on $\eta_{0,t-1} - \theta_{0k}K_{t-1}$. The parameter $\theta_{01}$ can also be identified from this regression equation by using the equality $\theta_{01} = \tilde{c}/(1-\theta_{0\omega})$, provided that $\theta_{0\omega} \neq 1$. As $\theta_{01}= 0$ in our Monte Carlo experiments, we directly consider $\tilde{c} = \theta_{01}$. Then, in our non-linear search, we impose these restrictions and minimize the GMM objective function based on $\psi$, treating it as a function of $\theta_{0k}$ only.

	\end{appendix}

	\clearpage
	\phantomsection
	\bibliographystyle{ectabib}
	\bibliography{references}
	
\end{document}